\documentclass[a4paper,fleqn,usenatbib]{mnras}

\usepackage[T1]{fontenc}
\usepackage{ae,aecompl}

\usepackage{graphicx}	% Including figure files
\usepackage{amsmath}	% Advanced maths commands
\usepackage{amssymb}	% Extra maths symbols

\title[RELATIVISTIC ELECTRONS IN 3C 279]{On the Injection of Relativistic Electrons in the Jet of 3C 279}
\author[Hu et al.]{
Wen Hu,$^{1}$
Dahai Yan,$^{2}$\thanks{E-mail: yandahai@ynao.ac.cn}
Benzhong Dai,$^{3}$
Wei Zeng$^{3}$
and Qianglin Hu$^{1}$
\\
$^{1}$College of Mathematics and Physics, Jinggangshan University, Jiangxi Province, Jian 343009, People's Republic of China\\
$^{2}$Key Laboratory for the Structure and Evolution of Celestial Objects, Yunnan Observatory, Chinese Academy of Sciences,\\
Kunming 650011, People's Republic of China\\
$^{3}$Department of Astronomy, Key Laboratory of Astroparticle Physics,Yunnan Province, Yunnan University, Kunming 650091,\\ People's Republic of China
}

\date{Accepted 2020 January 23. Received 2020 January 21; in original form 2020 January 8}

\pubyear{2020}

\begin{document}
\label{firstpage}
\pagerange{\pageref{firstpage}--\pageref{lastpage}}
\maketitle

\begin{abstract}
The acceleration of electrons in 3C 279 is investigated through analyzing the injected electron energy distribution (EED) in
a time-dependent synchrotron self-Compton + external Compton emission model.
In this model, it is assumed that relativistic electrons are continuously injected into the emission region, 
and the injected EED [$Q_e^\prime(\gamma^\prime)$] follows a single power-law form with low- and high-energy cutoffs 
$\rm \gamma_{min}'$ and $\rm \gamma_{max}'$, respectively, and the spectral index $n$, i.e, $Q_e^\prime(\gamma^\prime)\propto\gamma^{\prime-n}$.
This model is applied to 14 quasi-simultaneous spectral energy distributions (SEDs) of 3C 279.
The Markov Chain Monte Carlo fitting technique is performed to obtain the best-fitting parameters and the uncertainties on the parameters.
The results show that the injected EED is well constrained in each state.
The value of $n$ is in the range of 2.5 to 3.8, which is larger than that expected by the classic non-relativistic shock acceleration.
However, the large value of $n$ can be explained by the relativistic oblique shock acceleration.
%The cooling of the relativistic electrons in the emission region is limited in the fast-cooling regime.
%Namely, $\rm \gamma_{min}'$ is the break Lorenz factor of the emitting EED.
%This indicates that the acceleration efficiency is low, which is consistent with the feature of the relativistic oblique shock acceleration.
The flaring activity seems to be related to an increased acceleration efficiency, reflected in an increased $\gamma'_{\rm min}$ and electron injection power.
%The correlations between the injection parameters and the observed $\gamma$-ray fluxes indicate that the $\gamma$-ray flares could be caused by the acceleration.
\end{abstract}

\begin{keywords}
galaxies: jets --- gamma rays: galaxies --- radiation mechanisms: non-thermal
\end{keywords}

%%%%%%%%%%%%%%%%% BODY OF PAPER %%%%%%%%%%%%%%%%%%

\section{Introduction}\label{sec:intro}

Blazars are a subclass of active galactic nuclei (AGNs), with their relativistic jets pointing very close to our line of sight \citep{Urry1995, Ulrich1997}.
The non-thermal radiation produced in the relativistic jet covers from radio up to $\gamma$-ray bands.
The jet emission is highly variable, with variability timescales from years to several minutes.
%Their broadband emission generally exhibits a two-bump structure
Blazar's spectral energy distribution (SED) presents two humps.
The low-energy hump which is believed to be produced by synchrotron radiation of relativistic electrons, peaks between infrared and X-ray bands.
The high-energy bump which could be produced by inverse-Compton (IC) scattering of the relativistic electrons, peaks at gamma-ray energies.

Blazars are divided into flat spectrum radio
quasars (FSRQs) and BL Lacertae objects (BL Lacs) based on the rest-frame equivalent width (EW) of their broad optical emission lines \citep{Stocke1991,Stickel1991}.
FSRQs have strong broad emission lines with $\rm EW>5$\AA, while BL Lacs have weak or no emission lines.
The synchrotron peak frequencies of FSRQ are usually $<10^{14}$ Hz, due to the strong cooling of relativistic electrons in
intense external photon fields \citep[][]{Ghisellini2008}.
%\textbf{For some sources, the synchrotron peak can even lie in the radio band \citep{Ackermann2015}.}

$\gamma$-rays from FSRQs could be ascribed to IC scattering of the relativistic electrons.
The seed photons could be from $\sim$sub-parsec (pc) size
broad-line-region (BLR) \citep[e.g.,][]{Sikora1994,Zhang2012,Bottcher2013,Hu2015} and/or $\sim$pc-scale size dust torus (DT) \citep[e.g.,][]{Blazejowski2000,Dermer2014,Hu2017b,Wu2018},
depending on the location of the $\gamma$-ray emitting region \citep[e.g.,][]{Ghisellini2010}.
%However, it is still an unresolved issue.

The particle acceleration mechanism in blazar jets is still a hot question.
By means of numerical simulations, particle accelerations in blazar jets were explored \citep[e.g.,][]{Sironi2009,Sironi2015,Summerlin2012,Guo}.
The studies of numerical simulation focus on micro-physics and acceleration efficiency.
However, there is a gap between the numerical simulations and the observations.

Diffusive shock acceleration is the mostly discussed particle acceleration mechanism.
The power-law form of particle distribution is a key feature of this mechanism.
%and is widely used to explain non-thermal radiation observed in varieties of astrophysical environments.
For non-relativistic shock acceleration, the index of the particle distribution only depends on the shock compression ratio $r$,
i.e,  the power-law index $n=(r+2)/(r-1)$ \citep[e.g.,][]{Drury1983,Jones1991}.
For strong shock with $r=4$, the canonical $n\simeq2$ is obtained \citep[e.g.,][]{Drury1983,Jones1991}.
For the accelerations at relativistic shocks,
%a power-law index $n=2.23$ can be found analytically \citep{Kirk2000} and numerically \citep{Baring1999,Ellison2004}.
a wide variety of power-law indices is feasible, depending on the properties of the shock and the magnetic field \citep{Kirk1989,Ellison1990,Ellison2004,Summerlin2012,Baring2017}.

By fitting observed data with a proper emission model, one can obtain emitting EED.
It is the result of the competition between acceleration/injection and cooling,
and it can be used to investigate the acceleration mechanism \citep[e.g.,][]{Massaro2006,Yan2013,Zhou2014}.
However, this tactic is only suitable for the blazars in which the cooling effect does not significantly re-shape the accelerated/injected electron distribution,
like Mrk 421 and Mrk 501\textbf{ \citep{Ushio2010,Tramacere,Yan2013,Peng2014}} (i.e., the high-synchrotron-peaked BL Lacs).

In FSRQs, the strong radiative cooling of electrons due to the IC scattering off external photons has a big impact on the evolution of the emitting EED.
Hence, the emitting EED cannot be directly connected to acceleration process.

%Here, we provide a method to investigate the acceleration process in FSRQs through fitting observed SEDs with a radiative model.
Here, we investigate the acceleration process of the electrons in the FSRQ 3C 279 through analyzing the injected EEDs
in a time-dependent radiative model.
%\textbf{In the model}, we solve the \textbf{continuity} equation that governs the evolution of electron to obtain emitting EED.
%The \textbf{continuity} equation includes an injection term, a cooling term and an escape term.
%Using the spectra calculated with the emitting EED to match observed data, the injected EED can be constrained.
%The injected EED can be directly used to investigate acceleration process.
%Moreover, a Markov chain Monte Carlo (MCMC) fitting technique is adopted to explore high-dimensional parameter space systematically,
%which enables us to obtain reliable model parameters \citep{Yan2013,Yan2015}.
%\textbf{In this work, our model is characterized by an six-dimensional parameter space.}
% The method is applied to 3C 279, a famous FSRQ having intensive multiwavelength observations, to constrain the injected EEDs in flaring states.
Throughout the paper, we adopt the cosmological parameters $\rm H_0=69.6 ~km~s^{-1}~Mpc^{-1}$, $\Omega_M=0.286$, and $\Omega_\Lambda = 0.714$.
This results in the luminosity distance $d_L=3113.6 ~\rm Mpc$ for 3C 279 with redshift $z=0.536$.

\section{Method}\label{model}

We adopt a one-zone homogeneous leptonic jet model.
It is assumed that emissions are produced in a spherical blob of radius $R^\prime$ filled with a uniform magnetic field $B^\prime$.
The blob moves with a relativistic speed $\beta_\Gamma c= c(1-1/\Gamma^2 )^{1/2}$ and an angle $\theta$ with respect to the line of sight,
where $c$ is the speed of light and $\Gamma$ is the bulk Lorentz factor of the blob.
The observed radiations are strongly boosted by the relativistic Doppler factor
$\delta_D=1/[\Gamma(1-\beta_\Gamma\cos\theta)]$. It is assumed $\theta\sim1/\Gamma$, resulting in $\delta_D\sim\Gamma$.
Here and throughout this paper, primed quantities refer to the frame comoving with the blob and unprimed quantities refer to the observer's frame.

\subsection{Solving emitting EED}

In the model, we assume that the accelerated electrons are continuously
injected into the blob. %which can be identified with the downstream region of the shock. % a spherical blob of radii $R^\prime$ filled with a uniform magnetic field ($B^\prime$).
The isotropic electrons loss energy through synchrotron radiation and IC scattering, and may also escape out of the blob.
%When continuous injection is balanced by the cooling and escape, an equilibrium electron spectrum can be established
%and the resulting electrons spectrum and produced photons are isotopically distributed throughout the blob.
%The relativistic electrons bathing in the magnetic field radiate nonthermal synchrotron, observed as the low-energy component in
%blazar SEDs, while the second component is contributed to SSC and EC process.
The evolution of the electrons in the comoving frame of the blob is governed by \citep[e.g.,][]{Coppi1990,Chiaberge1999}
\begin{equation}\label{kineticEQ}
\frac{\partial{N_e^\prime(\gamma')}}{\partial{t}}+\frac{\partial}{\partial{\gamma^\prime}}\left[\dot{\gamma}^\prime N_e^\prime(\gamma')\right]+\frac{N_e^\prime(\gamma')}{t'_{\rm esc}}=\dot{Q}_e^\prime\ ,
\end{equation}
where $N_e^\prime(\gamma')$ is the number of the electrons per unit $\gamma^\prime$,
$\dot{\gamma}'$ is the total energy-loss rate of the electrons,
$t'_{\rm esc}$ is the escape timescale of the electrons, and $\dot{Q}_e$ is the source term describing the injection rate of the electrons in units of $s^{-1}$.

%In the comoving frame of the blob, the electrons rapidly accelerated at a shock front are continuously
The injected EED is assumed to be a single power-law distribution,
\begin{equation}\label{eq2}
Q_e^\prime(\gamma^\prime)=Q_0'\gamma^{\prime-n},~\gamma_{\rm min}'\le\gamma'\le\gamma_{\rm max}',
\end{equation}
with
\begin{equation}\label{eq3}
Q_0' =\left\{
             \begin{tabular}{l}
             $\frac{P'_e}{m_ec^2}\frac{2-n}{{\gamma_{\rm max}'}^{2-n}-{\gamma_{\rm min}'}^{2-n}};~n\neq2$ \\
             $\frac{P'_e}{m_ec^2\ln\left(\gamma_{\rm max}'/\gamma_{\rm min}'\right)};~n=2$ \\
             \end{tabular}
            \right. ,
\end{equation}
where $\gamma_{\rm min}'$ and $\gamma_{\rm max}'$ are respectively the low and high energy cutoffs,
and $P'_e$ is the injection power in the units of $\rm erg/s$, and $n$ is the spectral index \citep{Bottcher2002}.

Three radiative energy losses of  the electrons are considered:

(1) synchrotron radiation cooling
\begin{equation}
-\dot\gamma'_{\rm syn}=\frac{4\sigma_T}{3m_ec}{U_B'\gamma'}^2,
\end{equation}
where $U_B'={B'}^2/8\pi$ is the magnetic field energy density.

(2) synchrotron self-Compton radiation (SSC) cooling \citep[e.g.,][]{Jones1968,Blumenthal1970,Finke2008}
\begin{equation}
-\dot\gamma'_{\rm ssc}=\frac{4\sigma_T}{3m_ec}{\gamma'}^2\int_0^\infty d\epsilon' u_{syn}'(\epsilon')f_{kn}(\epsilon',\gamma'),
\end{equation}
where $u_{\rm syn}'(\epsilon')\simeq(\sigma_{\rm T}U_B')/(2\pi R'^2\epsilon')\gamma_s'^3N'_e(\gamma_s')$ is the spectral energy density of the synchrotron radiation.
Here, $\gamma_s'=\sqrt{\epsilon'B_{\rm cr}/B'}$ is a synchrotron-emitting electron's Lorentz factor where $B_{\rm cr}\simeq4.414\times10^{13}$ G is the critical magnetic field.

\begin{equation}
f_{kn}(\epsilon',\gamma')=\frac{9}{16}\int_{\gamma'_{\rm low}}^{\gamma'}d\gamma''F_c(x,q)\frac{\gamma'-\gamma''}{{\epsilon'}^2{\gamma'}^4},
\end{equation}
where the lower limit for the integration is $\gamma'_{\rm low}\simeq\gamma'+\epsilon'-\frac{4{\gamma'}^2\epsilon'}{1+4\gamma'\epsilon'}$, and
\begin{equation}
\label{fc}
F_{c}(x,q) = \Big[2q\ln{q}+q+1-2q^2 + \frac{(xq)^2}{2(1+xq)}(1-q)\Big]\ .
\end{equation}
%is scattered photon distribution function derived by \cite{Jones1968}\citep[see also][]{Blumenthal1970}.
Here,  $x=4\epsilon'\gamma'$, $q=\frac{\epsilon_\gamma'/\gamma'}{x(1-\epsilon_\gamma'/\gamma')}$,
and $\epsilon_\gamma'=\gamma'+\epsilon'-\gamma''$ is the scattered photon energy required by the conservation of energy.
The limits on $q$ are $\frac{1}{4\gamma'^2}\le q\le1$.
%The integral in Eq are quickly solved using routine of Press 1992.

(3) external-Compton (EC) cooling
\begin{equation}
-\dot\gamma'_{\rm ec}=\frac{4\sigma_T}{3m_ec}\gamma^2\int_0^\infty d\epsilon u_{ext}(\epsilon)f_{kn}(\epsilon,\gamma)\ ,
\end{equation}
where $u_{\rm ext}(\epsilon)$ is the spectral energy density of the external photon field.
The quantities $\gamma=\delta_D\gamma'$ and $\epsilon$ refer to the stationary frame with respect to the black hole (BH).
%Here all primed quantities in $x$ and $q$ are replaced with the ones referring to the stationary frame.

%%Usually, the $\gamma$-ray emitting region is either in BLR or in DT.
For the EC processes, we consider the seed photons from BLR and DT.
In this work, BLR and IR DT radiations are assumed to be a dilute blackbody \citep[e.g.,][]{Liu2006,Tavecchio2008},
\begin{equation}
u_{ext}(\epsilon)=\frac{15U_0}{(\pi \Theta)^4}\frac{\epsilon^3}{\exp\left(\epsilon/\Theta\right)-1}\ ,
\end{equation}
where $\Theta$ and $U_0$ are the dimensionless temperature and energy density of the BLR/DT radiation field, respectively.
We consider the BLR radiation with $\Theta\simeq9.6\times10^4$ K/($5.93\times10^9$ K) (corresponding to $\sim2\times10^{15}$ Hz) 
and $\rm U_{0}\simeq2.7\times10^{-2}\ erg\ cm^{-3}$ \citep[e.g.,][]{Ghisellini2008},
and the IR DT radiation with $\Theta\simeq1.4\times10^3$ K/($5.93\times10^9$ K) (corresponding to $\sim3\times10^{13}$ Hz) 
and $\rm U_{0}\simeq2.1\times10^{-4}\ erg\ cm^{-3}$ \citep[e.g.,][]{Ghisellini2009}.

Therefore, the total cooling rate of the electrons is $\dot{\gamma}'=\dot\gamma'_{\rm syn}+\dot\gamma'_{\rm ssc}+\dot\gamma'_{\rm ec}$.
We simply assume an energy-independent escape for the electrons, i.e., $t'_{\rm esc}=\eta_{\rm esc}R^\prime/c$, where it is required that $\eta>1$ \citep[e.g.,][]{Bottcher2002}.
With the above information, Equation~(\ref{kineticEQ}) is solved by using the iterative scheme described by \cite{Graff2008} to obtain the steady-state EED.

\subsection{Calculation of emission spectra}

The spectra of synchrotron radiation, SSC and EC are calculated with the formulas in \citet{Finke2008,Dermer2009}.
We here give the key formulas.
The synchrotron spectrum is
\begin{equation}
\nu f_{\nu}^{\rm syn}=\frac{ \delta_D^4\sqrt{3}e^3B^\prime}{4\pi hd_L^2}\chi(\tau)\epsilon^\prime\int_1^\infty d\gamma^\prime N_e^\prime(\gamma^\prime) R_s(\epsilon^\prime/\epsilon^\prime_c),
\end{equation}
where $\epsilon'm_ec^2=(1+z)h\nu/\delta_D$, $e$ is the fundamental charge and $h$ is the Planck constant.
%and $d_L$ is the luminosity distance of the source at redshift $z$.
In the spherical approximation, the factor $\chi(\tau)\equiv3u(\tau)/\tau$, where $\tau=2\kappa_{\epsilon'} R'$ is the synchrotron self-absorption (SSA) opacity and
 $u(\tau)=\frac{1}{2}\Big(1-\frac{2}{\tau^2}[1-(1+\tau)\exp(-\tau)]\Big)$. The SSA coefficient is given by
%%The dimensionless form of the synchrotron self-absorption Coefficient
\begin{equation}
\kappa_{\epsilon'}=-\frac{\sqrt{3}B'e^3\lambda_c^3}{8\pi hm_ec^3{\epsilon'}^2}\int_1^\infty d\gamma' R_s(\frac{\epsilon'}{\epsilon'_c})\Big[{\gamma'}^2\frac{\partial}{\partial\gamma'}\Big(\frac{N_e'(\gamma')}{{\gamma'}^2}\Big)\Big]\ ,
\end{equation}
where $m_e$ is the rest mass of electron and $\lambda_c=h/m_ec=2.43\times10^{-10}~\rm cm$ is the electron Compton wavelength.
%and $N_e'(\gamma')=n_e'(\gamma')/V_b'$, $V_b^\prime=4\pi R_b^\prime/3$ is the intrinsic volume of the blob.
Here, $\epsilon_c'=\frac{3eB'h}{4\pi m_e^2c^3}{\gamma^\prime}^2$ is the characteristic energy of synchrotron radiation in the units of $m_ec^2$, and $R_s(x)=(x/2)\int_0^\pi{d\theta}\sin\theta\int_{x/\sin\theta}^\infty{dtK_{5/3}(t)}$.

%In a standard one-zone model for blazars,
The SSC/EC spectrum is given by
\begin{equation}
\nu f_{\nu}^{\rm SSC/EC}=f_L\epsilon_\gamma'^2\int_0^\infty{}d\epsilon' \frac{u_{\rm syn/ext}'(\epsilon')}{\epsilon'^2}\int_{1}^{\infty} {}d\gamma^\prime{}\frac{N_e'(\gamma')}{\gamma'^2}F_{c}(x,q),
\end{equation}
where $\epsilon_\gamma'm_ec^2=(1+z)h\nu/\delta_D$, $f_L=(3c\sigma_T\delta_D^4)/(16\pi d_L^2)$,
and $u_{\rm ext}'(\epsilon')=\delta_{\rm D}^3u_{\rm ext}(\epsilon'/\delta_{\rm D})$ \citep[e.g.,][]{Dermer2009,Ghisellini2009}.

The model is characterized by eight parameters, i.e.,  $B', \delta_D, P'_e, n, \gamma_{\rm min}', \gamma_{\rm max}', \eta_{\rm esc}$ and $R'$.
%In our calculations, we set $\gamma'_{max} = 3\times10^{4}$ which will not affect our results,
The radius of the emission region can be estimated from the minimum variability timescale $t_{\rm var}$, i.e., $R'=c\delta_D t_{\rm var}/(1+z)$.
%%which is related to the size of the emission region and its Doppler factor according to the formula $R_b'=c\delta_D t_{var}/(1+z)$.
%%For the sources with no reported minimum variability timescales, the typical value of 2 hr will be adopted.
%Therefore, there are six free parameters in our SED fittings.

\subsection{MCMC fitting technique}

In order to unbiasedly constrain the model parameters,
we adopt MCMC technique which is based on Bayesian statistics to perform fitting.
The MCMC fitting technique is a powerful tool to explore the multi-dimensional parameter space in blazar science \citep{Yan2013,Yan2015}.
The details on MCMC technique can be found in \cite{Lewis2002,Yuan2011,Liu2012}.
%In the fitting, a relative systematic uncertainty of 5\% was added in quadrature to the statistical error of the
%IR-optical-UV and X-rays data, as usually adopted the literatures \citep[e.g.,][]{Poole2008,Abdo2011a}.

\section{Application to 3C 279}\label{results}

3C 279 is one of the best studied FSRQs.
It has been intensively monitored from radio band to $\gamma$-ray energies \citep[e.g.,][]{Wehrle1998,Hartman1996,Bottcher2007,Collmar2010,Abdo2010,Hayashida2012,Hayashida2015,Pacciani2014, Aleksic2015}. 
3C 279 shows rapid variabilities at all wavelengths. 
The radio and optical emissions are highly-polarized.
The correlations between the optical polarization level/angle and $\gamma$-ray variabilities
provide strong evidence for the SSC+EC model \citep[e.g.,][]{Abdo2010,Paliya2015,Hayashida2012}.
%Recently, tens of simultaneous and high-quality SEDs of the source have been constructed \citep[e.g.,][]{Hayashida2012,Hayashida2015,Pacciani2014, Aleksic2015}.

\cite{Hayashida2012,Hayashida2015} and \cite{Paliya2015} have constructed 16 high-quality SEDs for 3C 279
 from (quasi-)simultaneous observations by {\it Fermi} satellite together with many other facilities.
%%It is interesting to note that
Note that there is a temporal overlap of Period H in \cite{Hayashida2012} with the low-activity state in \cite{Paliya2015},
and the X-ray data are lacking in period B in \cite{Hayashida2015}.
We therefore do not consider the SED in the low-activity state in \cite{Paliya2015} and  the one in period B in \cite{Hayashida2015}.
We apply the method described in Section~\ref{model} to the rest of 14 high-quality SEDs.
In our fitting,  the radio data of $\lesssim200\ $GHz are neglected,
due to the fact that the low-frequency radio emission comes from the large-scale jet.

 \cite{Paliya2015} showed that the $\gamma$-ray variability timescale $t_{\rm var}$ can be down to $\sim1-2$ hours,
and \cite{Hayashida2015} reported $t_{\rm var} \sim2$ hours in the flare state of Period D.
In addition, variabilities down to the timescale of a few hours were also reported in \cite{Hayashida2012}.
%Hence, it is interesting to check whether few-hour-long variability timescales can be accommodated in the framework of the SSC+EC model applied here for the collected SED sets.
Hence, to reduce the number of model parameters, we take $\rm t_{\rm var}$ = 2 hours in the fittings.
%although close inspection of the light curves from radio through $\gamma$-rays in \cite{Hayashida2012} indicates variability
%%on a timescale as short as a fraction of a day to a few days.
In the process of testing our method, it is found that the observed data is insensitive to $\gamma'_{\rm max}$.
We then fix it to a large value, $\gamma'_{\rm max} = 3\times10^{4}$.
There are finally six free parameters in the fittings.

Following \cite{Poole2008} and \cite{Abdo2011}, a relative systematic uncertainty, namely 5\% of the data,
is added in quadrature to the statistical error of the
IR-optical-UV and X-rays data.
This is due to the fact that the errors of these data are dominated by the systematic errors.

\begin{table*}
%%\tiny \scriptsize
\setlength{\tabcolsep}{3.0pt}
\def\arraystretch{1.0}
\caption{Mean values and marginalized 95\% CI of the parameters for the SED fittings with the DT photons.}
\label{MT_para}
\begin{tabular}{lccccccc}
\hline
	           & $B^\prime$ (G)	  & $\delta_{\rm D}$  (10)	&$\eta_{\rm esc}$ (10)  & $P_e'\ (10^{41}\ \rm erg/s)$	 & $\gamma_{\rm min}'\ (10^2)$ & $n$   & $\chi_{\rm DT}^2\ (dof)$  \\
\hline
$\rm Period~ A$&      $ 1.02 \pm 0.02 $&$ 3.78 \pm 0.08 $&$ 7.67          $&$ 4.27 \pm 0.19 $&$ 2.48 \pm 0.21 $&$ 2.82 \pm 0.04 $	  & 0.69(34)\\
95\% CI        &      0.97 - 1.08      & 3.64 - 3.94     &$\ge2.72$        & 3.97 - 4.71     & 2.04 - 2.86     & 2.74 - 2.90          &\\
$\rm Period~ B$&      $ 0.67 \pm 0.06 $&$ 4.10 \pm 0.14 $&$ 2.30          $&$ 10.21\pm 1.46 $&$ 2.89 \pm 0.44 $&$ 2.49 \pm 0.10 $     & 1.47(17)\\
95\% CI        &      0.56 - 0.80      & 3.84 - 4.41     & $\ge0.18$       & 7.73 - 13.99    & 1.95 - 3.80     & 2.31 - 2.70          &\\
$\rm Period~ C$&      $ 1.40 \pm 0.07 $&$ 4.68 \pm 0.13 $&$ 2.78 \pm 0.77 $&$ 4.62 \pm 0.27 $&$ 2.66 \pm 0.20 $&$ 3.16 \pm 0.06 $	  & 3.10(36)\\
95\% CI        &      1.27 - 1.52      & 4.45 - 4.93     & 1.24 - 4.26     & 4.13 - 5.16     & 2.26 - 3.04     & 3.03 - 3.28          &\\
%$\rm Period~ C^a\dag$&$ 0.79 \pm 0.04 $&$ 2.67 \pm 0.07 $&$ 0.46 \pm 0.09 $&$ 5.72 \pm 0.30 $&$ 3.60 \pm 0.24 $&$ 3.13 \pm 0.06 $	  & 3.06(36)\\
%95\% CI        &      0.72 - 0.87      & 2.54 - 2.81     & 0.27 - 0.64     & 5.18 - 6.35     & 3.13 - 4.08     & 3.00 - 3.26          &\\
%$\rm Period~ C^b$&    $ 2.17 \pm 0.15 $&$ 4.57 \pm 0.17 $&$ 5.91          $&$ 4.18 \pm 0.33 $&$ 2.22 \pm 0.18 $&$ 3.46 \pm 0.06 $	  & 0.93(35)\\
%%95\% CI        &      1.89 - 2.48      & 4.23 - 4.91     & $\ge1.87$       & 3.59 - 4.87     & 1.86 - 2.57     & 3.33 - 3.58          &\\
$\rm Period~ D$&      $ 1.19 \pm 0.05 $&$ 4.75 \pm 0.15 $&$ 4.75          $&$ 5.17 \pm 0.56 $&$ 4.66 \pm 0.40 $&$ 3.62 \pm 0.09 $     & 1.97(16)\\
95\% CI        &      1.09 - 1.29      & 4.48 - 5.07     & $\ge0.83$       & 4.14 - 6.30     & 3.93 - 5.55     & 3.45 - 3.80          &\\
$\rm Period~ E$&      $ 1.46 \pm 0.19 $&$ 3.91 \pm 0.15 $&$ 5.90          $&$ 3.74 \pm 0.38 $&$ 3.83 \pm 0.23 $&$ 3.44 \pm 0.04 $     & 2.82(21)\\
95\% CI        &      1.23 - 2.03      & 3.50 - 4.15     & $\ge1.31$       & 3.04 - 4.56     & 3.34 - 4.28     & 3.35 - 3.53          &\\
$\rm Period~ F$&      $ 1.44 \pm 0.17 $&$ 3.46 \pm 0.24 $&$ 5.73          $&$ 5.31 \pm 0.65 $&$ 3.73 \pm 0.51 $&$ 3.49 \pm 0.24 $     & 0.26(12)\\
95\% CI        &      1.14 - 1.80      & 3.01 - 3.95     & $\ge1.08$       & 4.26 - 6.86     & 2.89 - 4.85     & 3.01 - 3.95          &\\
$\rm Period~ G$&      $ 0.97 \pm 0.11 $&$ 3.81 \pm 0.29 $&$ 5.82          $&$ 9.13 \pm 1.86 $&$ 6.33 \pm 1.44 $&$ 3.31 \pm 0.11 $     & 0.66(13)\\
95\% CI        &      0.76 - 1.21      & 3.29 - 4.41     &$\ge1.32 $       & 6.25 - 13.73    & 4.03 - 9.42     & 3.10 - 3.52          &\\
$\rm Period~ H$&      $ 0.86 \pm 0.15 $&$ 3.57 \pm 0.28 $&$ 5.29          $&$ 5.21 \pm 0.78 $&$ 2.84 \pm 0.33 $&$ 3.53 \pm 0.24 $     & 0.45(16)\\
95\% CI        &      0.61 - 1.20      & 3.08 - 4.18     & $\ge0.98$       & 3.94 - 6.94     & 2.21 - 3.49     & 3.07 - 4.00          &\\
$\rm Flare1$   &      $ 1.06 \pm 0.08 $&$ 3.72 \pm 0.15 $&$ 5.91          $&$ 10.15\pm 1.59 $&$ 8.77 \pm 1.41 $&$ 3.35 \pm 0.09 $     & 1.88(19)\\
95\% CI        &      0.91 - 1.23      & 3.44 - 4.02     & $\ge1.72$       & 7.51 - 13.75    & 6.34 - 11.88    & 3.19 - 3.55          &\\
$\rm Flare2\dag$&     $ 0.87 \pm 0.04 $&$ 4.39 \pm 0.09 $&$ 1.05 \pm 0.54 $&$ 1.28 \pm 0.10 $&$ 6.48 \pm 0.61 $&$ 3.26 \pm 0.05 $	  & 2.00(20)\\
95\% CI        &      0.79 - 0.95      & 4.23 - 4.57     & 0.43 - 2.53     & 1.10 - 1.49     & 5.35 - 7.80     & 3.17 - 3.36          &\\
$\rm Post-flare$&     $ 2.00 \pm 0.32 $&$ 4.67 \pm 0.58 $&$ 5.07          $&$ 4.07 \pm 1.10 $&$ 3.31 \pm 1.48 $&$ 3.25 \pm 0.05 $     &0.92(18)\\
95\% CI         &     1.39 - 2.64      & 3.59 - 5.82     & $\ge0.58$       & 2.56 - 7.12     & 1.48 - 7.42     & 3.14 - 3.36          &\\
$\rm Period~A15$&     $ 1.52 \pm 0.12 $&$ 4.33 \pm 0.22 $&$ 20.01         $&$ 4.14 \pm 0.41 $&$ 2.76 \pm 0.53 $&$ 3.39 \pm 0.06 $     & 0.68(34)\\
95\% CI         &     1.28 - 1.78      & 3.78 - 4.64     & $\ge8.97$       & 3.50 - 5.14     & 2.12 - 4.11     & 3.28 - 3.51          &\\
$\rm Period~C15$&     $ 1.09 \pm 0.06 $&$ 3.90 \pm 0.10 $&$ 0.68 \pm 0.30 $&$ 10.96\pm 0.60 $&$ 4.82 \pm 0.43 $&$ 3.41 \pm 0.06 $     & 1.82(30)\\
95\% CI         &     0.99 - 1.22      & 3.71 - 4.08     & 0.29 - 1.42     & 9.86 - 12.22    & 4.00 - 5.71     & 3.28 - 3.54          &\\
$\rm Period~D15$& $ 0.52 \pm 0.05 $&$ 4.34 \pm 0.18 $&$ 0.25 \pm 0.10 $&$ 38.7 \pm 7.3 $&$ 8.30 \pm 1.32 $&$ 3.28 \pm 0.05 $     &1.31(20)\\
95\% CI         &     0.44 - 0.62      & 4.01 - 4.70     & 0.12-0.52       & 2.68 - 5.50     & 5.98 - 11.25    & 3.19 - 3.39          &\\
\hline
\end{tabular}
\newline
\begin{flushleft}
%Notes: \dag denoting the powers of the injected electrons in units of $10^{42}$ ergs/s.
%Notes: \dag denotes the powers of the injected electrons in units of $10^{41}$ ergs/s. We fix $\rm \gamma_{max}'=3\times10^4$ and $\rm t_{var}=2 hr$ in all fittings.  \\
\end{flushleft}
\end{table*}

\begin{table*}
%%\tiny \scriptsize
\setlength{\tabcolsep}{3.0pt}
\def\arraystretch{1.0}
\caption{Mean values and marginalized 95\% CI of the parameters for the SED fittings with the BLR photons.}
\label{BLR_para}
\begin{tabular}{lccccccc}
\hline
	           & $B^\prime$ (G)	  & $\delta_{\rm D}$ (10)	&$\eta_{\rm esc}$  (10) & $P_e'\ (10^{42}\rm\ erg/s)$	 & $\gamma_{\rm min}'\ (10^2)$ & $n$   & $\chi_{\rm BLR}^2\ (dof)$  \\
\hline
$\rm Period ~A$ & $5.57 \pm 0.15 $&$ 2.04 \pm 0.02 $&$ 5.86          $&$ 2.50 \pm 0.09 $&$ 1.64 \pm 0.08 $&$ 3.18 \pm 0.04 $  & 2.10(34)\\
95\% CI         & 5.28 - 5.88     & 2.00 - 2.08     & $\ge0.20$       & 2.36 - 2.76     & 1.48 - 1.80     & 3.10 - 3.27       &\\
$\rm Period ~B$ & $2.11 \pm 0.70 $&$ 2.25 \pm 0.12 $&$ 4.71          $&$ 6.55 \pm 1.78 $&$ 1.92 \pm 0.38 $&$ 2.69 \pm 0.26 $  & 1.95(17)\\
95\% CI         & 1.36 - 3.83     & 2.10 - 2.52     & $\ge0.16$       & 3.53 - 9.57     & 1.02 - 2.50     & 2.32 - 3.26       &\\
$\rm Period ~C$ & $7.06 \pm 0.32 $&$ 2.46 \pm 0.02 $&$ 5.62          $&$ 2.58 \pm 0.08 $&$ 2.19 \pm 0.09 $&$ 3.62 \pm 0.06 $  & 2.71(36)\\
95\% CI         & 6.41 - 7.69     & 2.42 - 2.51     & $\ge1.03$       & 2.42 - 2.75     & 2.02 - 2.34     & 3.49 - 3.74       &\\
$\rm Period~D$  & $6.13 \pm 0.32 $&$ 2.79 \pm 0.04 $&$ 5.29          $&$ 2.62 \pm 0.12 $&$ 2.35 \pm 0.18 $&$ 4.02 \pm 0.09 $  & 2.45(16)\\
95\% CI         & 5.52 - 6.79     & 2.72 - 2.86     & $\ge0.50$       & 2.40 - 2.85     & 2.02 - 2.72     & 3.85 - 4.19       &\\
$\rm Period~E$  & $8.54 \pm 0.66 $&$ 2.26 \pm 0.06 $&$ 5.00          $&$ 2.01 \pm 0.13 $&$ 1.83 \pm 0.09 $&$ 3.61 \pm 0.05 $  & 2.91(21)\\
95\% CI         & 7.37 - 9.84     & 2.14 - 2.38     & $\ge0.37 $      & 1.76 - 2.27     & 1.66 - 2.01     & 3.52 - 3.71       &\\
$\rm Period~F$  & $7.93 \pm 0.96 $&$ 1.77 \pm 0.08 $&$ 5.19          $&$ 3.64 \pm 0.34 $&$ 2.54 \pm 0.28 $&$ 3.62 \pm 0.22 $  & 0.40(12)\\
95\% CI         & 6.15 - 9.78     & 1.59 - 1.93     & $\ge0.54$       & 3.11 - 4.45     & 2.08 - 3.18     & 3.20 - 4.06       &\\
$\rm Period~G$  & $5.07 \pm 0.50 $&$ 2.36 \pm 0.07 $&$ 5.13          $&$ 3.72 \pm 0.28 $&$ 2.67 \pm 0.25 $&$ 3.56 \pm 0.11 $  & 1.18(13)\\
95\% CI         & 4.16 - 6.16     & 2.23 - 2.50     & $\ge0.37 $      & 3.22 - 4.31     & 2.23 - 3.19     & 3.35 - 3.77       &\\
$\rm Period~H$  & $3.34 \pm 1.23 $&$ 1.58 \pm 0.11 $&$ 4.59          $&$ 4.53 \pm 1.42 $&$ 1.91 \pm 0.28 $&$ 3.34 \pm 0.35 $  & 0.64(16)\\
95\% CI         & 1.74 - 6.50     & 1.41 - 1.82     & $\ge0.12 $      & 2.56 - 8.06     & 1.24 - 2.37     & 2.82 - 4.16       &\\
$\rm Flare1$    & $6.19 \pm 0.48 $&$ 2.68 \pm 0.05 $&$ 5.21          $&$ 3.11 \pm 0.20 $&$ 2.58 \pm 0.24 $&$ 3.58 \pm 0.05 $  &2.28(19)\\
95\% CI         & 5.33 - 7.20     & 2.58 - 2.79     & $\ge0.55 $      & 2.74 - 3.54     & 2.13 - 3.11     & 3.48 - 3.68       &\\
$\rm Flare2$    & $4.95 \pm 0.23 $&$ 3.14 \pm 0.03 $&$ 5.81 $         &$ 3.80 \pm 0.16 $&$ 2.24 \pm 0.11 $&$ 3.74 \pm 0.04 $  &7.33(20)\\
95\% CI         & 4.51 - 5.40     & 3.07 - 3.20     & $\ge1.04 $      & 3.49 - 4.13     & 2.02 - 2.46     & 3.66 - 3.82       &\\
$\rm Post-flare$& $10.13\pm 1.46 $&$ 2.55 \pm 0.11 $&$ 5.13          $&$ 2.41 \pm 0.20 $&$ 1.92 \pm 0.44 $&$ 3.44 \pm 0.06 $  &0.85(18)\\
95\% CI         & 7.37 - 13.05    & 2.32 - 2.76     & $\ge0.40 $      & 2.12 - 2.89     & 1.26 - 2.97     & 3.32 - 3.55       &\\
$\rm Period~A15$& $5.61 \pm 0.38 $&$ 2.03 \pm 0.05 $&$ 6.14          $&$ 3.36 \pm 0.12 $&$ 3.10 \pm 0.16 $&$ 3.62 \pm 0.07 $  & 1.35(34)\\
95\% CI         & 4.95 - 6.45     & 1.92 - 2.13     & $\ge1.46 $      & 3.15 - 3.59     & 2.79 - 3.43     & 3.48 - 3.76       &\\
$\rm Period~C15$& $4.49 \pm 0.28 $&$ 2.27 \pm 0.05 $&$ 4.65          $&$ 5.21 \pm 0.22 $&$ 3.47 \pm 0.14 $&$ 3.76 \pm 0.07 $  & 2.24(30)\\
95\% CI         & 3.98 - 5.08     & 2.16 - 2.37     & $\ge0.11 $      & 4.81 - 5.70     & 3.20 - 3.71     & 3.61 - 3.90       &\\
$\rm Period~D15$& $2.86 \pm 0.22 $&$ 3.46 \pm 0.07 $&$ 5.24          $&$ 6.69 \pm 0.51 $&$ 2.81 \pm 0.25 $&$ 3.89 \pm 0.05 $  &3.55(20)\\
95\% CI         & 2.46 - 3.34     & 3.32 - 3.60     &$\ge 0.54 $      & 5.78 - 7.75     & 2.36 - 3.33     & 3.80 - 3.98       &\\
\hline
\end{tabular}
%\newline
%\begin{flushleft}
%\end{flushleft}
\end{table*}

\begin{table}
%%\tiny \scriptsize
\setlength{\tabcolsep}{3.0pt}
\def\arraystretch{1.0}
\caption{Mean values and marginalized 95\% CI of the derived parameters for the SED fittings with the DT photons.}
\label{t3}
\begin{tabular}{lccc}
\hline
state& $\log_{10}\gamma_c'$ & $\log_{10}L_B$ (erg/s)  & $\log_{10}L_r$ (erg/s)\\
\hline
$\rm Period~ A$&     $ 0.82 \pm 0.16 $&$ 44.19 \pm 0.04 $&$ 44.17 \pm 0.02 $\\
95\% CI        &       $\le$ 1.29     & 44.11 $-$ 44.29  & 44.14 $-$ 44.20  \\
$\rm Period~ B$&     $ 1.40 \pm 0.39 $&$ 43.97 \pm 0.13 $&$ 44.59 \pm 0.02 $\\
95\% CI        &       $\le$ 2.22     & 43.71 $-$ 44.25  & 44.54 $-$ 44.64  \\
$\rm Period~ C$&     $ 0.98 \pm 0.16 $&$ 44.84 \pm 0.08 $&$ 44.38 \pm 0.01 $\\
95\% CI        &       0.74 $-$ 1.36  & 44.68 $-$ 45.00  & 44.36 $-$ 44.40  \\
%$\rm Period~ C^a$&   $ 1.40 \pm 0.10 $&$ 45.53 \pm 0.08 $&$ 44.97 \pm 0.01 $\\
%95\% CI        &       1.23 $-$ 1.65  & 45.37 $-$ 45.70  & 44.95 $-$ 44.99  \\
%$\rm Period~ C^b$&   $ 0.27          $&$ 45.18 \pm 0.10 $&$ 44.36 \pm 0.05 $\\
%95\% CI        &       $\ge0.00$      & 45.00 $-$ 45.37  & 44.26 $-$ 44.45  \\
$\rm Period~ D$&     $ 0.82 \pm 0.32 $&$ 44.72 \pm 0.08 $&$ 44.45\pm0.02   $\\
95\% CI        &       $\le$ 1.53     & 44.56 $-$ 44.90  & $44.41-44.49 $   \\
$\rm Period~ E$&     $ 0.88 \pm 0.24 $&$ 44.56 \pm 0.06 $&$ 44.13 \pm0.06  $\\
95\% CI        &       $\le$ 1.50     & 44.45 $-$ 44.69  & $43.96-44.23 $   \\
$\rm Period~ F$&     $ 1.01 \pm 0.27 $&$ 44.33 \pm 0.19 $&$ 44.17 \pm 0.04 $\\
95\% CI        &          $\le$ 1.67  & 43.95 $-$ 44.69  & 44.10 $-$ 44.24  \\
$\rm Period~ G$&     $ 0.96 \pm 0.25 $&$ 44.15 \pm 0.22 $&$ 44.51 \pm 0.03 $\\
95\% CI        &          $\le$ 1.51  & 43.71 $-$ 44.60  & 44.45 $-$ 44.58  \\
$\rm Period~ H$&     $ 1.12 \pm 0.29 $&$ 43.94 \pm 0.28 $&$ 44.18 \pm 0.04 $\\
95\% CI        &          $\le$ 1.84  & 43.41 $-$ 44.50  & 44.11 $-$ 44.26  \\
$\rm Flare1$   &     $ 0.94 \pm 0.21 $&$ 44.20 \pm 0.13 $&$ 44.54\pm0.04   $\\
95\% CI        &          $\le$ 1.44  & 43.94 $-$ 44.46  & $44.48-44.62 $   \\
$\rm Flare2\dag$&    $ 1.54 \pm 0.20 $&$ 44.32 \pm 0.07 $&$ 44.76 \pm 0.02 $\\
95\% CI        &       1.10 $-$ 1.88  & 44.18 $-$ 44.46  & 44.73 $-$ 44.80  \\
$\rm Post-flare$&    $ 0.73 \pm 0.33 $&$ 45.12 \pm 0.35 $&$ 44.30 \pm 0.03 $\\
95\% CI         &         $\le$ 1.57  & 44.36 $-$ 45.77  & 44.24 $-$ 44.38  \\
$\rm Period~A15$&    $ 0.21 \pm 0.17 $&$ 44.78 \pm 0.16 $&$ 44.27\pm0.02   $\\
95\% CI         &      $\le$0.62      & 44.42 $-$ 45.02  & $44.24-44.32    $\\
$\rm Period~C15$&    $ 1.83 \pm 0.16 $&$ 44.31 \pm 0.06 $&$ 44.55 \pm 0.03 $\\
95\% CI         &      1.50 $-$ 2.14  & 44.19 $-$ 44.43  & 44.48 $-$ 44.60  \\
$\rm Period~D15$&$ 2.20 \pm 0.18 $&$ 43.85 \pm 0.14 $&$ 45.16 \pm 0.04 $\\
95\% CI         &      1.81 $-$ 2.49  & 43.59 $-$ 44.13  & 45.07 $-$ 45.25  \\
\hline
\end{tabular}
%\newline
%\begin{flushleft}
%Column 2 is the critical Lorentz factor of the emitting electrons.
%Column 3-4 are the jet power carried by magnetic field and radiation, respectively.
%All the values are in logarithmic space. %The table are arranged as the following Table \ref{MT_para}.
%\end{flushleft}
\end{table}

\begin{figure*}
  \centering
  \includegraphics[width=\textwidth]{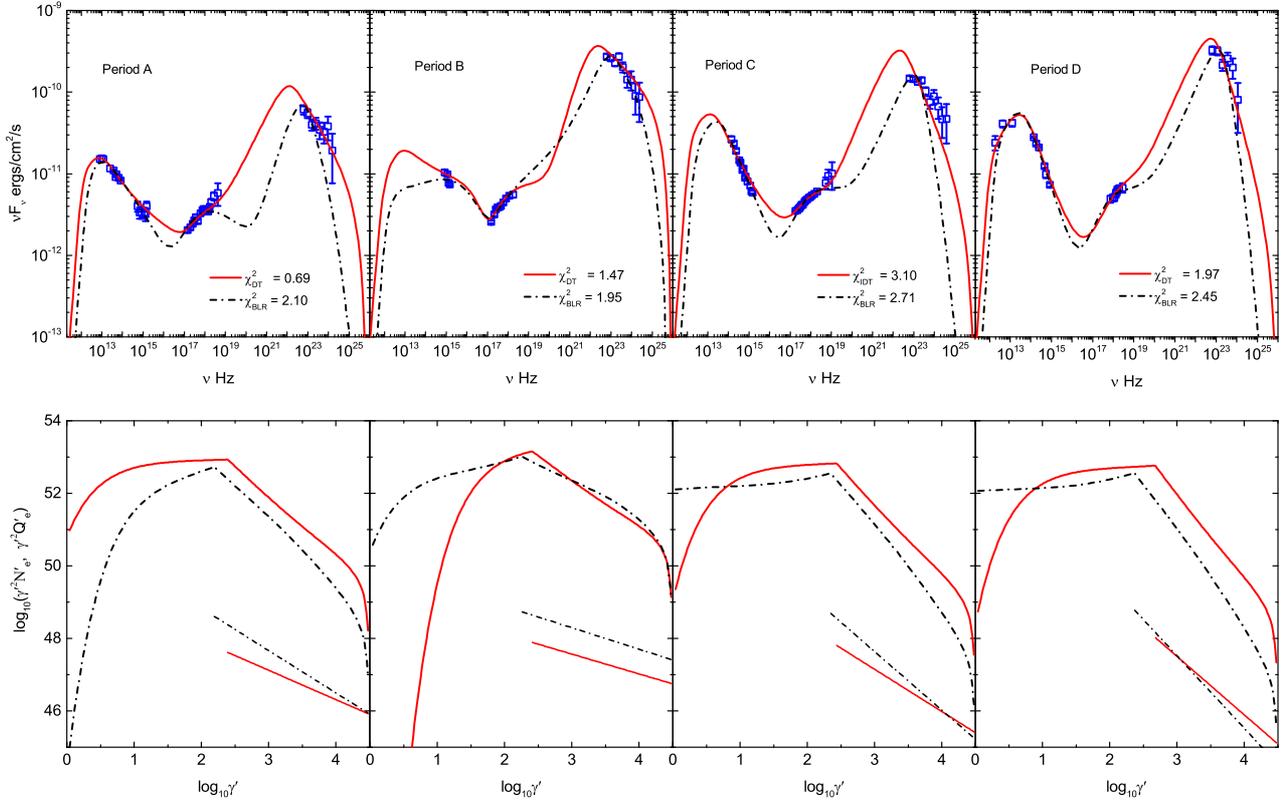}
  \caption{Upper panels: Best-fitting results for the SEDs of 3C 279 during the observations of Periods A, B, C and D reported in Hayashida et al. (2012). %\cite{Hayashida2012}.
   The red solid line and black dash-dotted line represent the fittings with the seed photons from the DT and the BLR, respectively.
   Lower panels: The steady-state emitting EED and injected EED corresponding to the best-fitting model in each state.
   The red solid thick and thin lines are respectively the steady-state emitting EED and the injected EED obtained from the fitting with the seed photons from the DT;
   and the black dash-dotted thick and thin lines are respectively the steady-state emitting EED and injected EED obtained from the fitting with the seed photons from the BLR.
   }\label{figure1}
\end{figure*}

\begin{figure*}
  \centering
  \includegraphics[width=\textwidth]{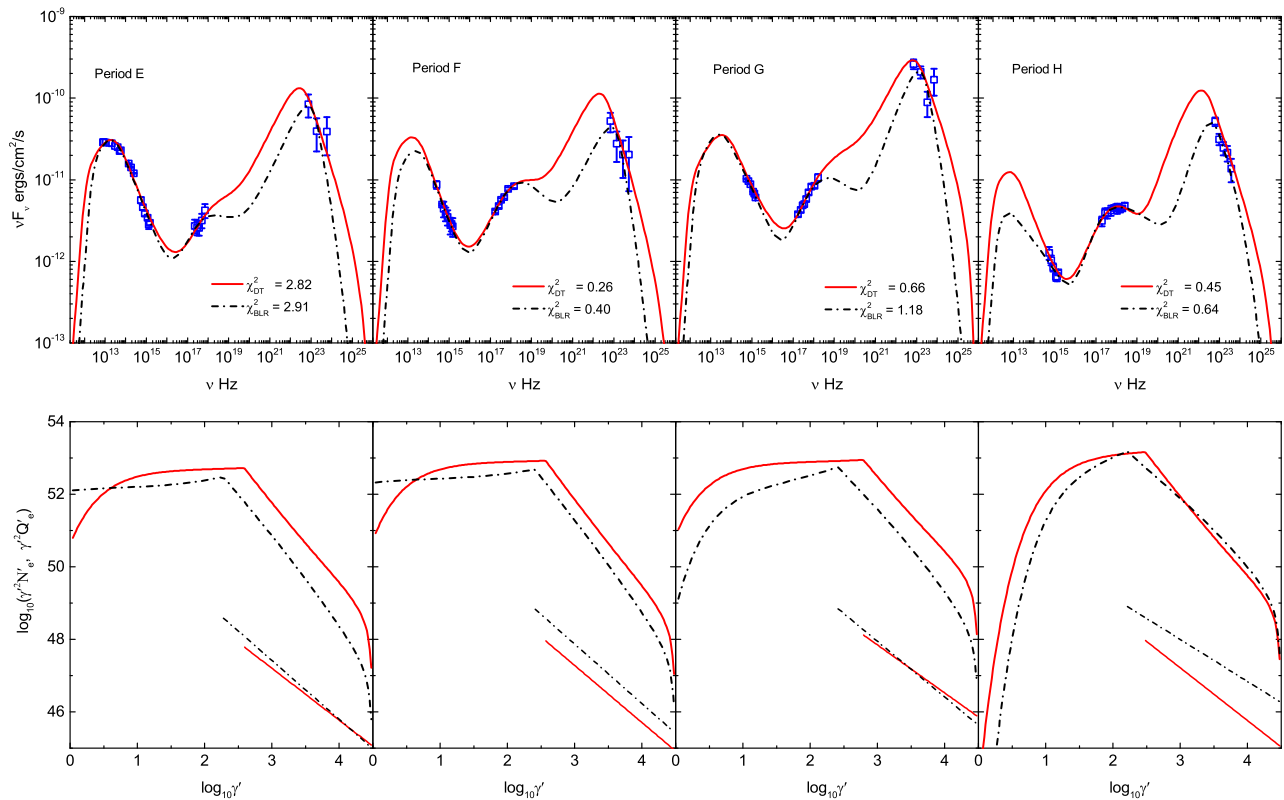}
  \caption{Same as Figure \ref{figure1}, but for the SEDs in Periods E, F, G, H reported by Hayashida et al. (2012). %\cite{Hayashida2012}.
 }\label{figure2}
\end{figure*}

\begin{figure*}
  \centering
  \includegraphics[width=\textwidth]{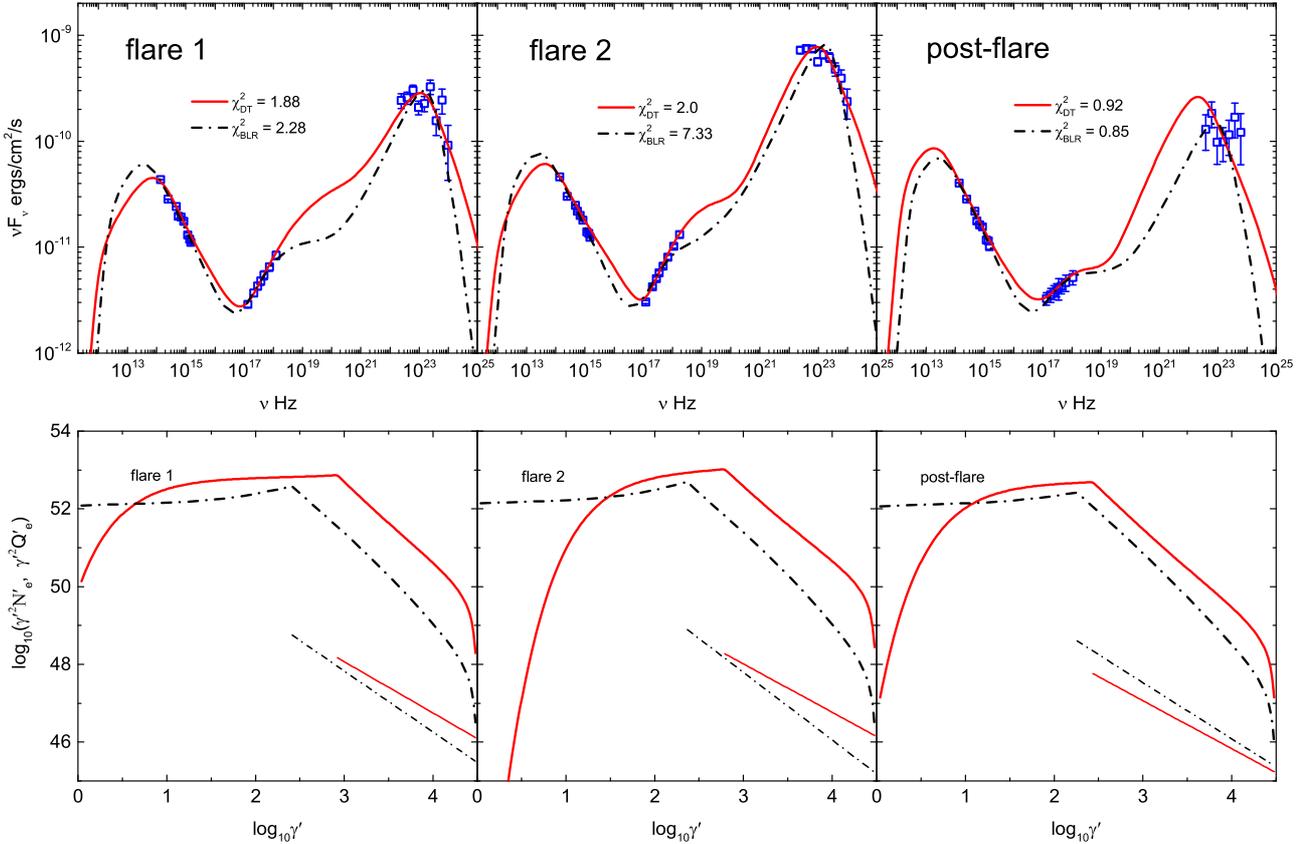}\\
  \caption{Same as Fig.\ref{figure2}, but for the SEDs reported in Paliya et al.(2015).}\label{figure3}%\cite{Paliya2015}
\end{figure*}

\begin{figure*}
  \centering
  \includegraphics[width=\textwidth]{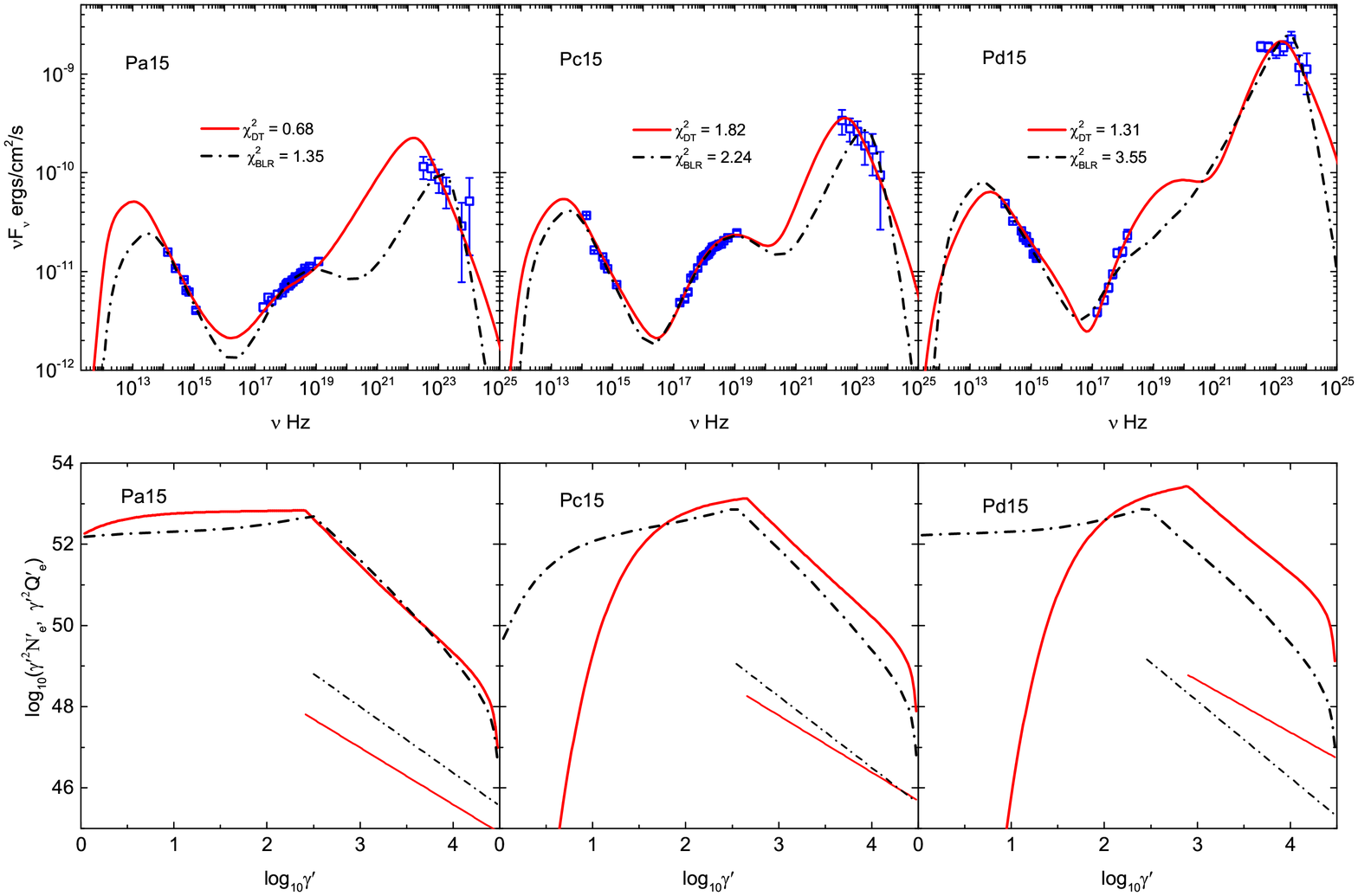}\\
  \caption{Same as Fig.\ref{figure2}, but for the SEDs reported in Hayashida et al. (2015). %\cite{Hayashida2015}.
  }\label{figure4}
\end{figure*}

\subsection{Fitting the SEDs}

In the upper panels of Figures \ref{figure1}-\ref{figure4}, we show the best-fitting results for the 14 SEDs.
Each SED is fitted with two models: the model with the BLR photons and the model with the DT photons.
The corresponding reduced $\rm\chi^2_{DT/BLR}$ is reported in each panel.
%For each state, the red-solid-thick line and black-dash-dotted-thick
%lines represent the SED fitting with the seed photons from the DT and BLR, respectively.
One can see that all the fittings with the DT photons, except for Period C in \cite{Hayashida2012}, are better than that with the BLR photons.
The highest-energy X-ray/gamma-ray data can be fitted better with the model of SSC+EC-DT.
%The X-ray emission is attributed to SSC, and the $\gamma$-ray emission is produced by EC.
In the SSC+EC-BLR model, the Klein-Nishina (KN) effect becomes important and suppresses the gamma-ray emission,
leading to the mismatch between the data and  the model (e.g., the first panel in Fig.~\ref{figure1}).
In addition, the high-energy hump in the SSC+EC-BLR model locates at higher energies, which causes the worse fitting to the X-ray data (e.g., the third  panel in Fig.~\ref{figure4})

The one-dimensional (1D) probability distributions of the free parameters are shown in Figures \ref{figure9}-\ref{figure13} in Appendix \ref{appA}.
The uncertainties on the parameters in the two cases are given in Tables \ref{MT_para} and \ref{BLR_para}, respectively.
We also report the marginalized 95\% confidence intervals (CIs) of the parameters.

One can see that  all parameters except for $\eta_{\rm esc}$ are well constrained.
In the EC-DT model, $\eta_{\rm esc}$ in four states are well constrained (see Table~\ref{MT_para}).
In the EC-BLR model, $\eta_{\rm esc}$ in all states are poorly constrained (see Table~\ref{BLR_para}).
The constraint on $\eta_{\rm esc}$ arises from the X-ray data.
$\eta_{\rm esc}$ determines the minimum Lorenz factor in the emitting EED $\gamma_{\rm c}'$,
i.e., $\gamma'_{\rm c}/|\dot{\gamma'}(\gamma'_{\rm c})|=\eta_{\rm esc}R'/c$.
$\gamma'_{\rm c}$ has a big impact on the low energy part (X-ray band) of the EC component.
If the EC emission contributes to the observed X-ray emission, $\gamma'_{\rm c}$ or $\eta_{\rm esc}$ could be well constrained.
In the EC-BLR model, the EC component peaks at higher energies, and the X-ray data are dominated by SSC emission.
Therefore, $\eta_{\rm esc}$ is poorly constrained.

In addition, it is worth pointing out that our model fails to fit the $\gamma$-ray spectrum in the Period C in \cite{Hayashida2012},
likely due to the simplification of the external photon fields.
Complex external photon fields \citep{Cerruti2013} may correct the discrepancy between the model and the data.

\subsection{Injected EEDs}

The injected EEDs obtained in the fittings are shown in the lower panels of Figures~\ref{figure1}-\ref{figure4}.
It can be seen that the parameters describing the injected EEDs, i.e., $P'_e$, $\gamma_{\rm min}'$ and $n$,  are well constrained.
In the EC-DT model, $\gamma_{\rm min}'$ is in the range from 248 to 877,
and $P'_e$ varies from $3.7\times10^{41}$ to $3.9\times10^{42}$ erg/s,
and $n$ is in the range of $\sim2.5-3.6$. It is noted that $n$ is larger than 3 except for the Periods A and B in \cite{Hayashida2012}.

Looking at the injected EEDs and the emitting EEDs in Figures~\ref{figure1}-\ref{figure4},
one can find that the radiative cooling of the electrons occurs in the fast-cooling regime,
i.e., $\gamma'_{\rm c}<\gamma'_{\rm min}$.
%where $\gamma'_{\rm c}$ is determined by the condition $\gamma'_{\rm c}/|\dot{\gamma'}(\gamma'_{\rm c})|=t'_{\rm esc}$.
%corresponding to the radiation at the peak frequency of the observed spectra
In the fast cooling regime, $\gamma'_{\rm min}$ is the break Lorentz factor of the emitting EED.
The spectral index $s$ between $\gamma'_{\rm c}$ and $\gamma'_{\rm min}$ in steady-state emitting EED depends on the cooling rate of the electrons with $\gamma'>\gamma'_{\rm min}$.
In the case of Thomson scattering or synchrotron energy-loss of the form $\dot{\gamma'}\sim\gamma'^2$, $s=2$.
If the dominant energy-loss rate is not the form of $\dot{\gamma'}\sim\gamma'^2$ (e.g., IC in KN regime),
$s$ should differ from 2 \citep[see][for a detailed investigation on $s$ in different energy-loss processes in 3C 279]{Yan2016b}.
For the electrons with $\gamma'>\gamma_{\rm min}'$, the index of the distribution changes to be $n+1$, 
when the form of $\dot\gamma'\propto\gamma'^2$ holds. %%If $n>2$, $\gamma_{min}'$ is responsible for the peak frequency of all radiation components.

It is noted that the EC-BLR model requires a larger injected electron power $P'_e$,
which is several times of that obtained in the EC-DT model.
This is caused by the KN effect.
%%\textbf{This is mainly due to the equilibrium EED in the fast-cooling regime, and the IC scattering off the BLR photons happening in the Klein-Nishina (KN) regime (see discussion in Section \ref{about_eed}).}

\subsection{Properties of the $\gamma$-ray emission region}

The magnetic field strength $B'$ and the Doppler factor $\delta_{\rm D}$ are two important physical quantities.
In Figures \ref{figure9}-\ref{figure13},
it is found that the two quantities are constrained very well with the current data.
In Table \ref{MT_para}, one can see that $B'$ varies in the range of [0.5-2.0] G,
and $\delta_D$ varies in the range of [34.6-47.5].
With the values of $\delta_D$,
we find that the values of $R'$ are in the range of $\sim(4.8-6.6)\times10^{15}$ cm.
They are consistent with that derived in previous works  \citep[e.g.,][]{Dermer2014,Yan2016a} where static emitting EEDs were used.
The large values of $\delta_D$ are also suggested by the VLBI study of the kinematics of the jet in 3C 279 \citep{Lister1997,Jorstad2004}.

\subsection{Correlations between model parameters and observed $\gamma$-rays}

\begin{figure}
  \centering
  \includegraphics[width=0.45\textwidth]{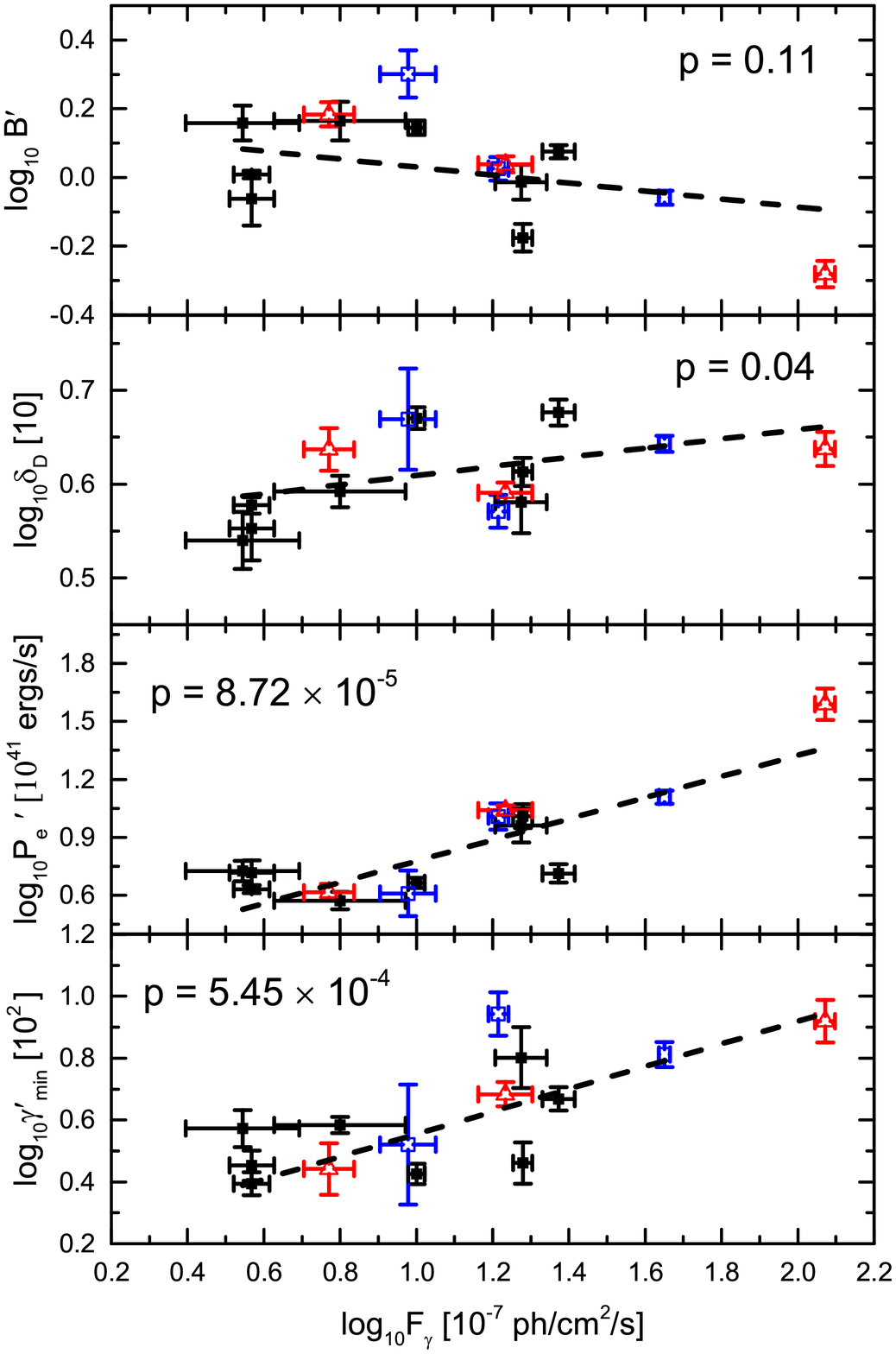}
  \caption{Evolutions of the model parameters ($B'$, $\delta_D$, $\gamma_{\rm min}'$, and $P_e'$)
   as a function of the observed $\gamma$-ray flux $F_\gamma$.
 The black-dashed line is the linear best-fitting  to the data.
  The red triangles, blue open squares and black filled squares are the results derived by fitting the three SEDs in Hayashida et al.(2015),
  the three SEDs in Paliya et al.(2015), and the eight SEDs in Hayashida et al.(2012), respectively.
  }\label{corr}
\end{figure}

%\textbf{The stringently constrained parameters enable us to examine what model parameters are correlated with source activity,
%which may be associated with acceleration and/or radiative cooling.}
The model parameters as a function of the observed $\gamma$-ray flux $F_\gamma$ \citep{Hayashida2012,Hayashida2015,Paliya2015} are showed in Figure~\ref{corr}.
We calculate the Pearson's probability for a null correlation, namely the p-value,
which is reported in the corresponding panel of Figure \ref{corr}.

Our results show that the $\gamma$-ray activity is tightly correlated with $P_e'$ and $\gamma'_{\rm min}$,
with the p-values of $p=8.72\times10^{-5}$ and $p=5.45\times10^{-4}$, respectively.
It indicates that the $\gamma$-ray activity is associated with the injection of the accelerated electrons.

In addition, there is a weak correlation between $F_\gamma$ and $\delta_D$, with $p=0.04$.
No correlation between $F_\gamma$ and $B'$ is found.

\subsection{Jet powers}
\label{jetpower}

Using the model parameters, we can derived the powers carried by the jet in the form of radiation ($L_r$), magnetic field ($L_B$),
electrons ($L_e$) and protons ($L_p$) \citep[e.g.,][]{Celotti1993}.
However, the poorly constrained $\gamma'_{\rm c}$ leads to large uncertainties on $L_e$ and $L_p$.
Here, we calculate $L_r$ and $L_B$ using our well constrained parameters (Table \ref{t3}).
One can find  $L_r\sim L_B$ except for the Period D in \cite{Hayashida2015}.

The jet power $L_{\rm kin}$ can also be estimated by the $L_{\rm kin}-L_{151}$ relation obtained by \cite{Godfrey2013},
\begin{equation}
 L_{\rm kin} = 3\times10^{44}\left(\frac{L_{151}}{10^{25} \rm W/Hz/sr}\right)^{0.67}~\rm erg/s,
\end{equation}
where $L_{151}$ is the 151 MHz radio luminosity from the extended jet.
The scaling relationship is roughly consistent with the theoretical relation presented in \cite{Willott1999}.
%%under the assumption of minimum energy arguments.
This approach is widely used to estimate the jet kinetic energy in AGNs.

Using the relation $L_{151}=d_L^2F^{151}$, we have $L_{\rm kin}=3\times10^{44}(9.23F^{151})^{0.67}$ erg/s, where $F^{151}$ is in the units of Jy.
With $F^{151}=22.08$ Jy \citep{Arshakian2010}, we obtain $L_{\rm kin}=1.05\times10^{46}$ erg/s, which is dozens times of $L_r$.
%The result indicates that the jet power estimated from the extended radio emission is lower than that from SED fitting assuming one proton per electron.
%It implies that some fraction of $e^\pm$ pairs may present in the jets of 3C 279.

\section{Discussions}\label{discussion}

At first, we would like to stress that our model implicitly assumes a small acceleration zone which cannot contribute significant photons to the observed radiations.

\subsection{On the acceleration mechanism}

Obviously, the values of $n$ significantly depart from the canonical $n\simeq2$ predicted by the non-relativistic shock acceleration,
and also differ from $n\simeq2.2$ expected by the classic relativistic shock acceleration \citep[e.g.,][]{Kirk2000,Baring1999,Achterberg2001,Ellison2004}.
%Note that the two typical values result from the case of test-particle approximation.
Although a steeper distribution ($n\simeq2.5$) can be produced considering the modification of shock by the
back-reaction of the accelerated particles \citep[e.g.,][]{Kirk1996}, it still fails to account for the large values of $n$ we obtained.

Note that the above discussions are given in the frame of (quasi-)parallel shocks.
Relativistic oblique shocks could produce much softer injection EED with $n>2.5$ \citep{Ellison2004,Niemiec2004,Sironi2009,Summerlin2012}.
In relativistic shocks, \citet{Summerlin2012} showed that the
spectral index $n$ varies dramatically from 1 to $>3$ with the changes of obliquity and magnetic turbulence.
Steep electron distribution with $n\sim3$ can be produced in relativistic shocks with large obliquity and low turbulence.
Therefore, our results indicate that the relativistic shocks with large obliquity and low turbulence may be responsible for the acceleration of electrons in 3C 279.

In relativistic shocks, the minimum Lorentz factor of the distribution is \citep[e.g.,][]{Sari1998,Piran1999}
\begin{equation}
\gamma_{\rm min}'\simeq\frac{m_p}{m_e}\frac{n-2}{n-1}\epsilon_e\Gamma_{\rm sh},~n>2
\end{equation}
where $\Gamma_{\rm sh}$ is the bulk Lorentz factor across the shock front, and $\epsilon_e$ is the fraction of shock energy that goes into the electrons.
\iffalse
\footnote{It should be pointed out that the formula mentioned above is suitable for parallel shocks.
For relativistic oblique shock, exact calculation of the injection minimum energy only could be made through kinetic particle-in-cell simulation that beyond the scope of this work.
 However, it should be valid for the purpose of our analysis, if we assume that for relativistic oblique shock the injection minimum energy is larger than that expected by the parallel shock \citep{Niemiec2004,Summerlin2012}.
}.
\fi
From our results, we use $n=3.4$, $\gamma'_{\rm min}=300$, and assume $\Gamma_{\rm sh}=10$ \citep{Ushio2010}, then we obtain $\epsilon_e\simeq0.03$.
This indicates that the acceleration is low-efficiency, which is consistent with the numerical simulation result in \citet{Sironi2009}.

The strongest evidence for an oblique shock can be found in the polarization maps of the jet emissions \citep{Lind1985,Cawthorne1990,Cawthorne2006,Nalewajko2009,Nalewajko2012}.
\cite{Abdo2010} discovered a dramatic change in the optical polarization associated with the $\gamma$-ray flare in Period D in \cite{Hayashida2012}.
They suggested that the observed polarization behavior may be the result of the jet bending.
Jet bending has been observed in a number of AGNs \citep[e.g.,][]{Graham2014}.
In this scenario, an oblique shock could be formed due to the interaction of the jet with the external medium.
\cite{Denn2000} have organized an extensive VLBI monitoring.
They revealed the existent of the oblique shocks in the knots through the observed linear polarization behavior.
\cite{Lister2005} have shown that the distribution of the electric vector position angles (EVPA) offsets is similar to
that predicted by an ensemble of oblique shocks with random orientations \citep{Lister1998}.
\cite{Dulwich2009} have shown that the high-resolution data from the Very Large Array, Hubble Space Telescope and Chandra observatories
 support the presence of an oblique shock in the kiloparsec-scale jet of the powerful radio galaxy 3C 346.
Very recently, some authors proposed that radio-to-$\gamma$-ray variabilities may be caused by the oblique shocks in AGN jets \citep{Hughes2011,Hovatta2014,Aller2014,Hughes2015}.
In particular, Using a relativistic oblique shock acceleration + radiation-transfer model, \citet{Bottcher2019} successfully explained the SEDs and variabilities of 3C 279 during flaring activity in 
the period December 2013 - April 2014 reported in \citet{Hayashida2015}.

Our results show that the $\gamma$-ray activities strongly correlated with the injection of electrons.
This indicates that the $\gamma$-ray activities could be caused by the acceleration of the electrons in the relativistic oblique shock.

\subsection{On the Magnetization and Radiative Efficiency}

The most promising scenario for launching blazar powerful jets involves the central accumulation of large magnetic
flux and the formation of magnetically arrested/choked accretion flows (MACF) \citep{Narayan2003,Igumenshchev2008,Tchekhovskoy2009,Tchekhovskoy2011,McKinney2012,Chen2018}.
In this scenario, the jet is powered by the Blandford-Znajek (BZ) mechanism that extracts BH rotational energy,
and the jet production efficiency for maximal BH spin is estimated by $\eta_j\simeq1.9(\phi_{\rm BH}/50)^2$ 
where $\phi_{\rm BH}$ is the dimensionless magnetic flux threading the BH \citep{Blandford1977,Tchekhovskoy2010,Sikora2013,Sikora2013b}.
The value of $\phi_{\rm BH}$ is typically on the order of 50 according to the numerical simulations by \cite{McKinney2012}, although it depends on the details of the model.

Assuming $L_{\rm jet}=L_{\rm kin}$, one can find $\eta_j\equiv \epsilon L_{\rm jet}/L_d\simeq1.6$ for
$\epsilon=0.3$\footnote{$\epsilon\equiv L_d/\dot{M}c^2$ is the radiation efficiency of an accretion disk with $\dot{M}$ denoting the mass accretion rate.} \citep{Thorne1974}
and $L_d\sim2\times10^{45}$ erg/s \citep{Pian1999}.
It is in good agreement with $\eta_j=1.9$ in the MCAF scenario for a typical value of $\phi_{\rm BH}=50$.
Therefore, our result supports the BZ mechanism for jet  launching

The magnetization and radiative efficiency are usually considered to be the indicator of acceleration mechanism occurring in blazar jets.
We derive the magnetization parameter $\sigma_B$ and radiative efficiency $\eta_r$ \citep{Kang2014,Sikora2016,Fan2018},

\begin{eqnarray}
  \sigma_B &=& L_B/(L_{\rm kin}-L_B), \\
  \eta_r &=& L_r/(L_{\rm kin}+L_r) .
 % n_e/n_p &=& L_p/(L_{kin}-L_B-L_e),
\end{eqnarray}

Since $L_{\rm kin}$ is the time-averaged kinetic power of a source with the radio flux $F^{151}$,
we use the average values of $L_B$ and $L_r$ and get $\sigma_B\simeq\eta_r\simeq0.02$.
\citet{Baring2017} showed that electrons would be efficiently accelerated by relativistic shocks in blazar jets with $\sigma_B$ changing from $\sim10^{-4}$ to 0.06.

\section{Summary}\label{summary}

Using a time-dependent one-zone SSC+EC model and the MCMC fitting technique, we analyzed 14 high-quality SEDs of 3C 279.
We assume that the $\gamma$-ray emission region is either in the BLR or in the DT.
The results show that the SEDs are better fitted in the latter case.
The injected EED is well constrained in each state.
The index of the injected EED is large, ranging from 2.7 to 3.8, which cannot be produced in (quasi-)parallel shocks.
We argue that the steep injected EED may be the result of the acceleration of relativistic oblique shocks.
According to the correlations of $F_{\gamma}$ and $\gamma_{\rm min}'$, $P'_e$, the $\gamma$-ray flares are caused by the acceleration.
\section*{Acknowledgements}

We thank the referee for helpful suggestions.
We acknowledge the National Natural Science Foundation of China (NSFC-11803081, NSFC-U1738124) and
the joint foundation of Department of Science and Technology of Yunnan Province and Yunnan University [2018FY001(-003) and 2018FA004].
BZD acknowledges funding support from National Key R\&D Program of China under grant No. 2018YFA0404204.
WH acknowledges funding supports from Key Laboratory of Astroparticle Physics of Yunnan Province (No. 2016DG006)
and the Scientific and Technological Research Fund of Jiangxi Provincial Education Department (No. GJJ180584).
DHY is also supported by the CAS ``Light of West China'' Program and Youth Innovation Promotion Association.

%%%%%%%%%%%%%%%%%%%%%%%%%%%%%%%%%%%%%%%%%%%%%%%%%%



%\bibliographystyle{mnras}


%%%%%%%%%%%%%%%%%%%%%%%%%%%%%%%%%%%%%%%%%%%%%%%%%%

%%%%%%%%%%%%%%%%% APPENDICES %%%%%%%%%%%%%%%%%%%%%

\appendix
\section{One-dimensional probability distributions of the free model parameters}
\label{appA}

\begin{figure*}
\includegraphics[width=0.45\textwidth]{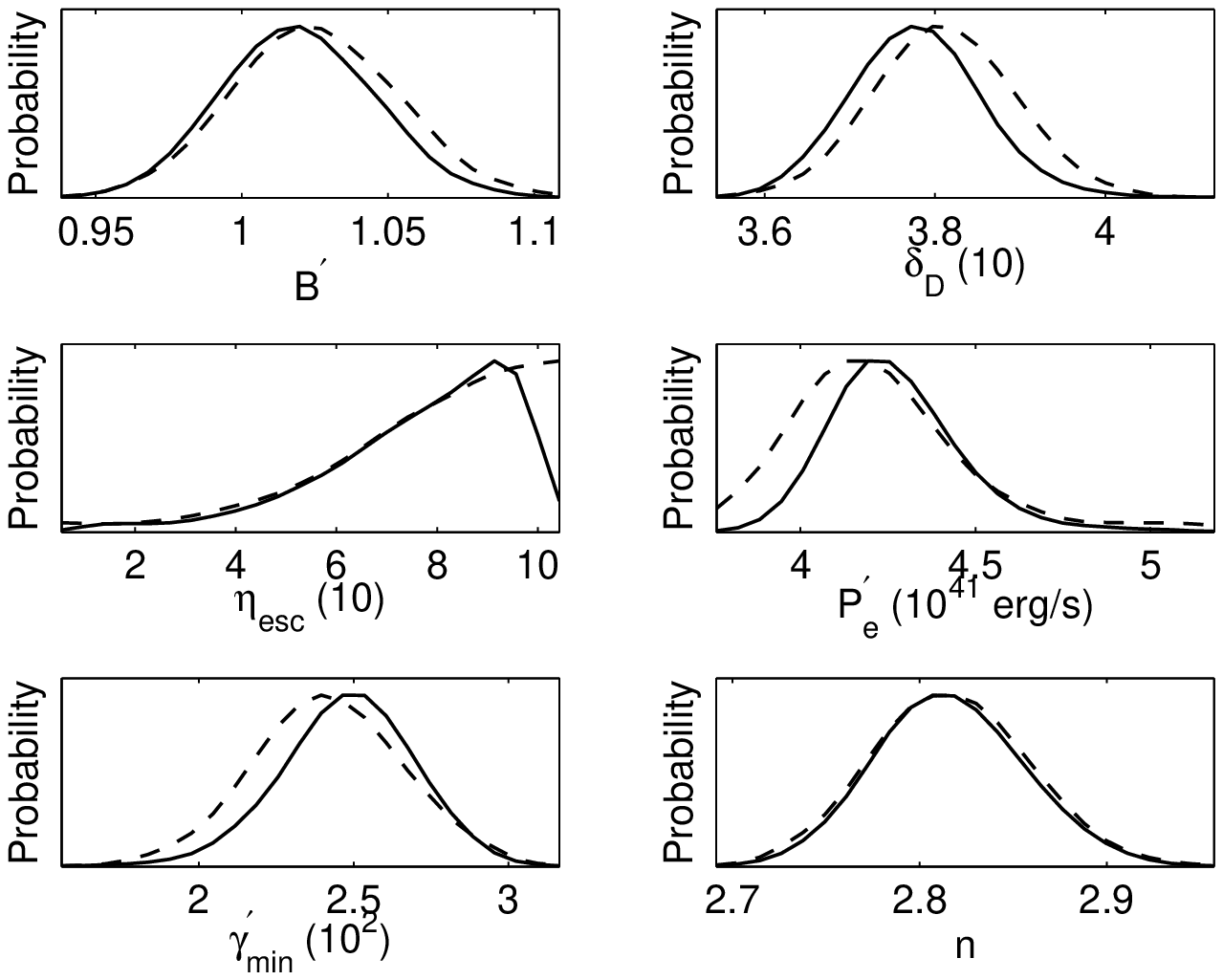}\includegraphics[width=0.45\textwidth]{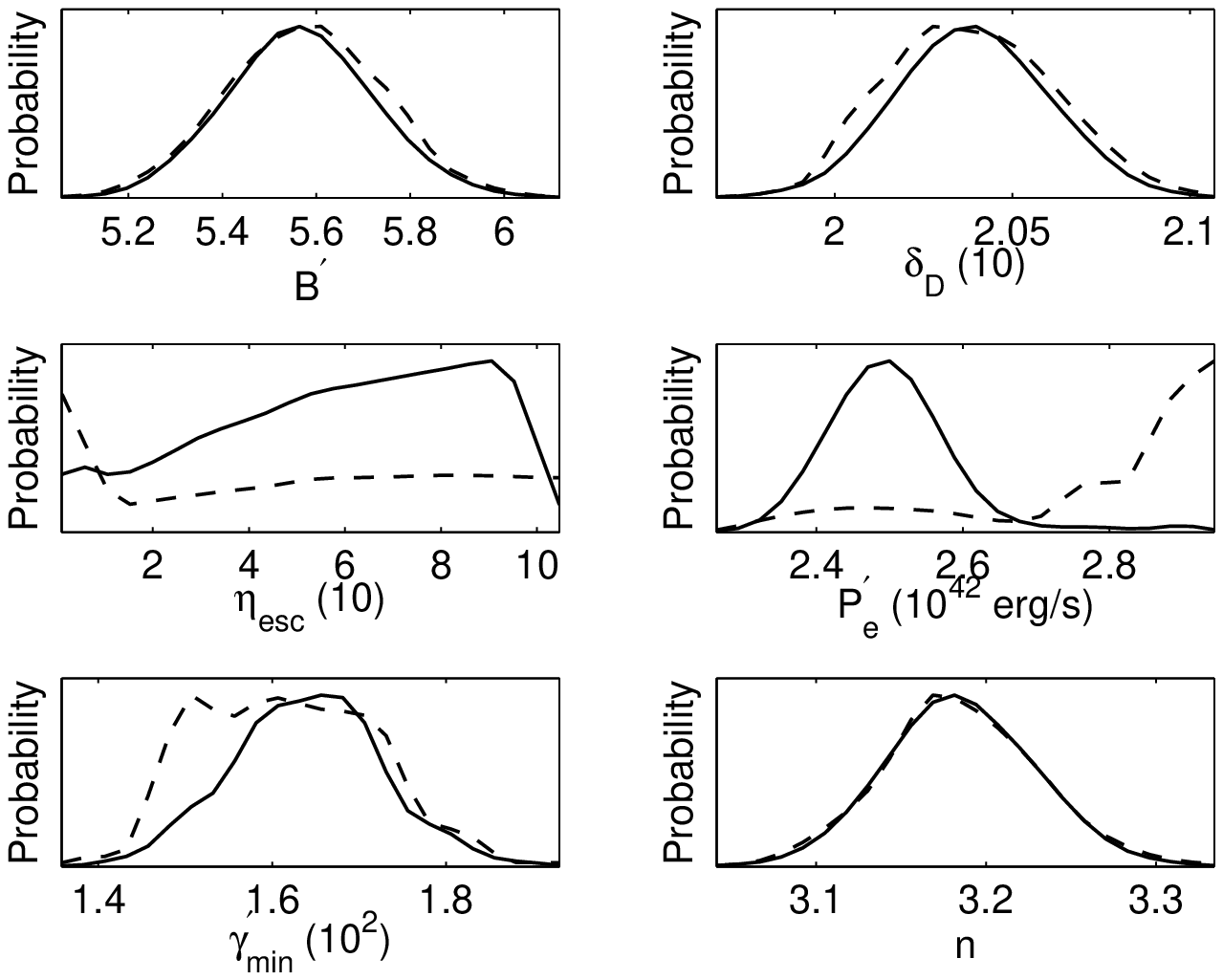}
\includegraphics[width=0.45\textwidth]{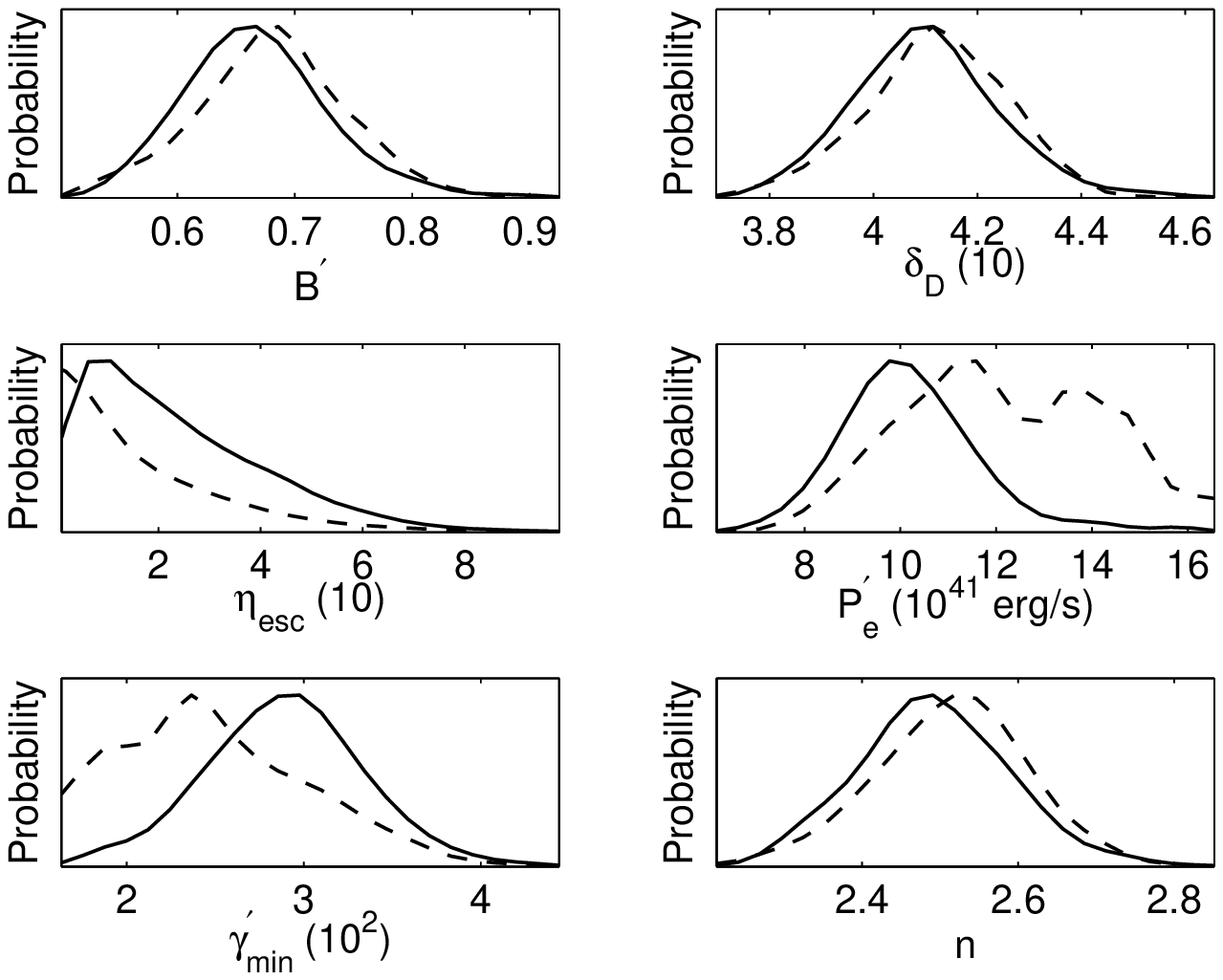}\includegraphics[width=0.45\textwidth]{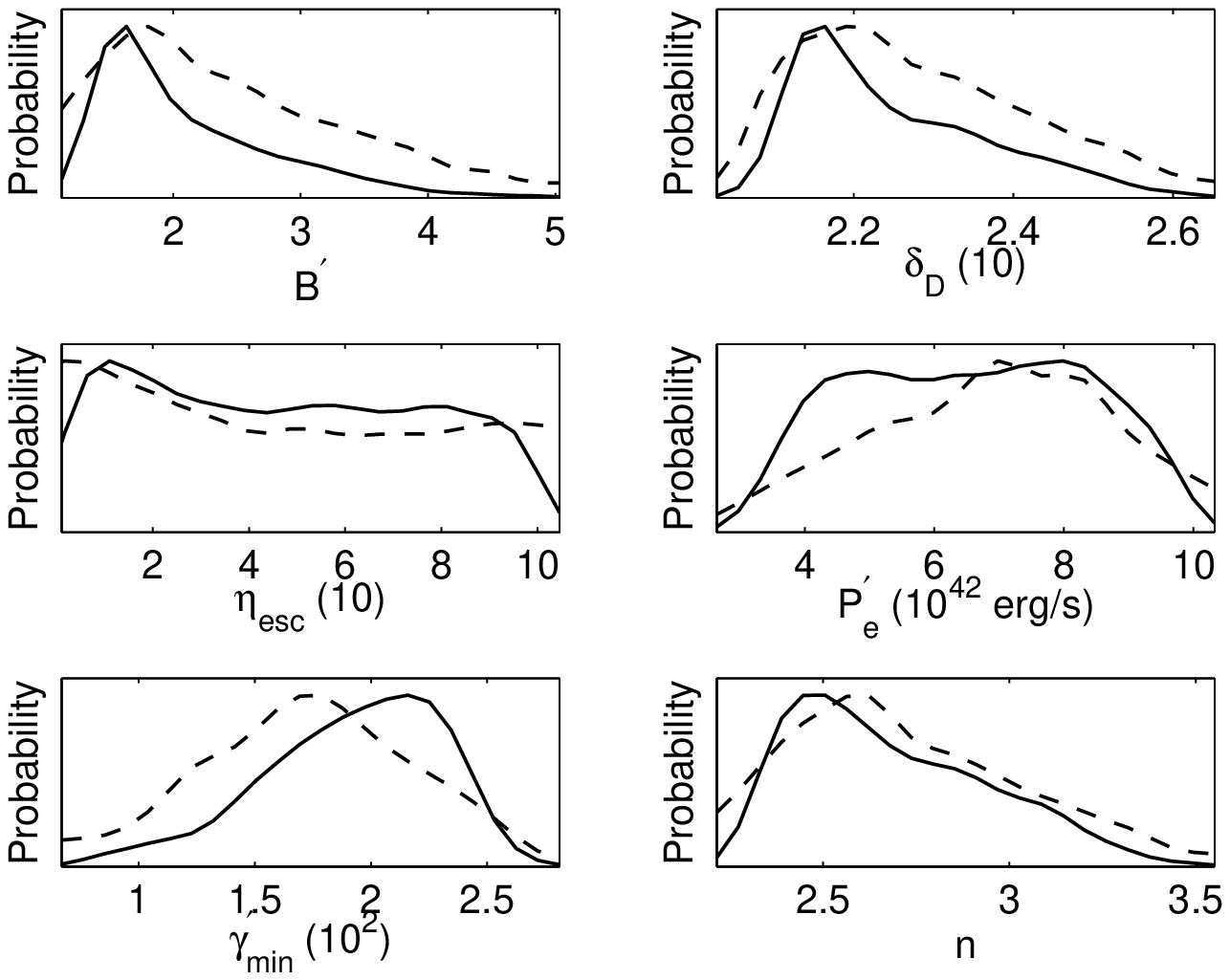}
\includegraphics[width=0.45\textwidth]{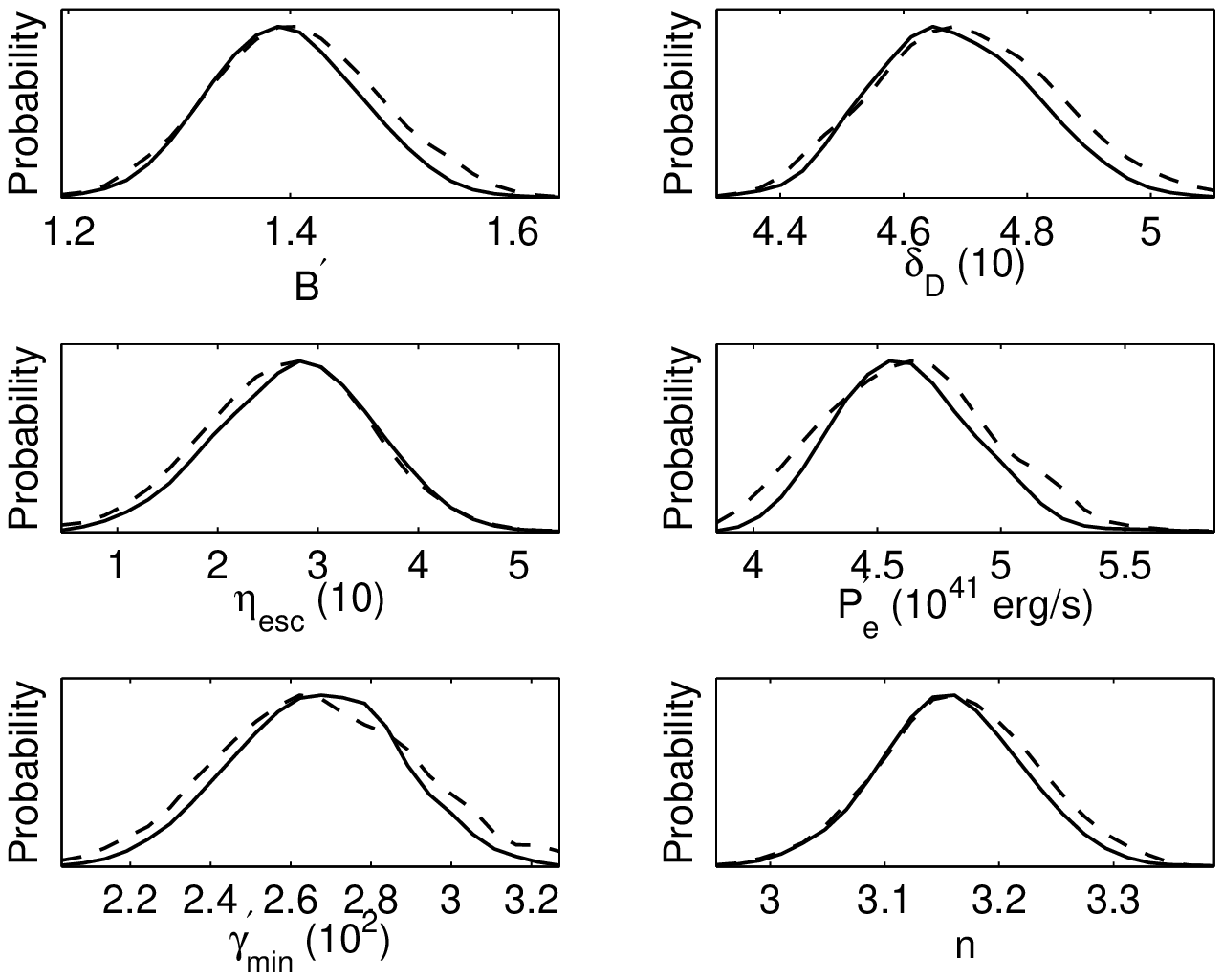}\includegraphics[width=0.45\textwidth]{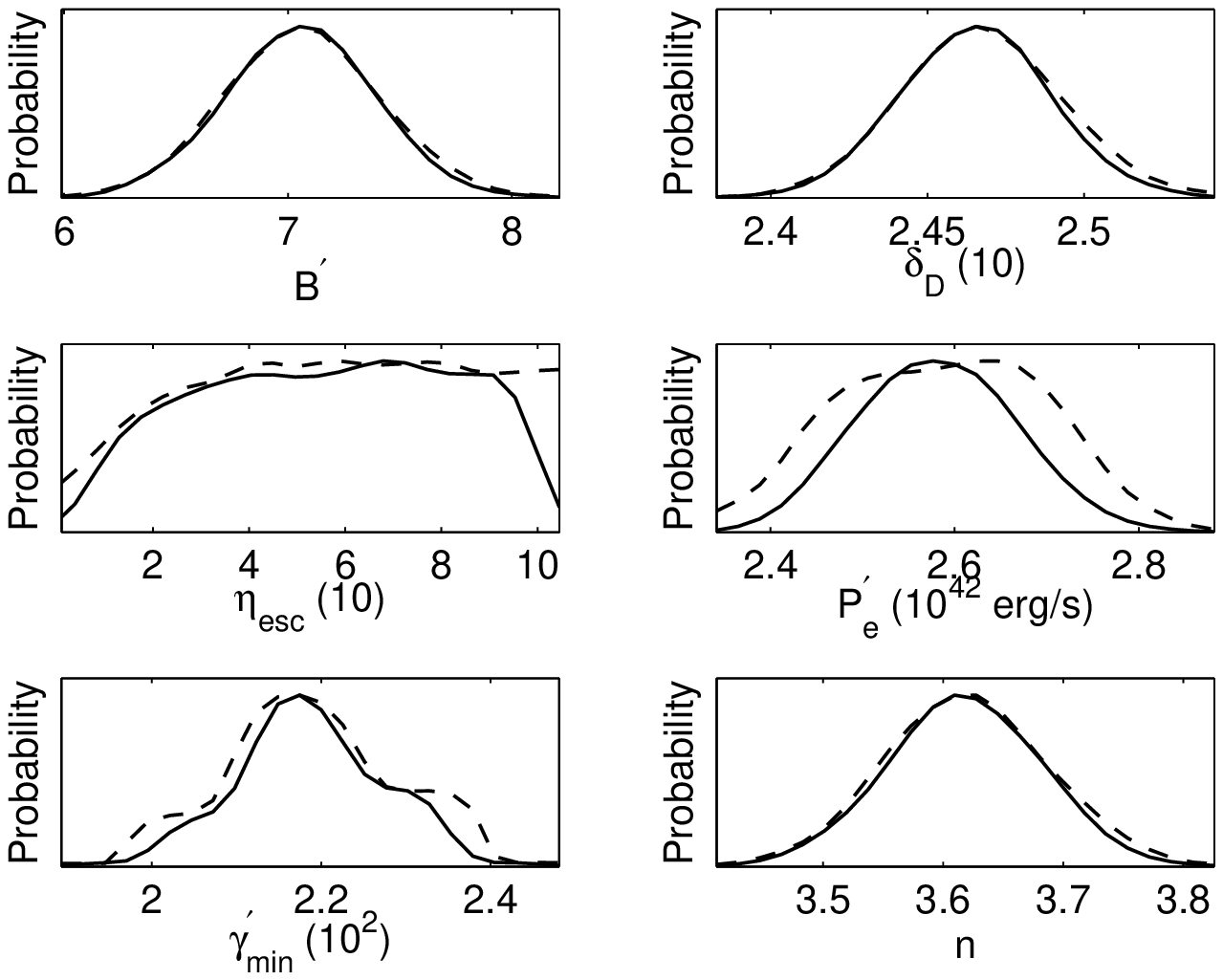}
\caption{1D probability distributions of the free model parameters obtained from SED fittings with the seed photons from DT (left panels) and BLR (right panels).
The dashed lines show the maximum likelihood distributions and solid lines show the marginalized probability distributions.
From top to bottom, the plots are the results obtained from fitting the SEDs in Periods A, B, and C reported in Hayashida et al.(2012), respectively.
%%The plots are arranged as the following Figure \ref{figure1}
 \label{figure9}}
\end{figure*}

\begin{figure*}
\includegraphics[width=0.45\textwidth]{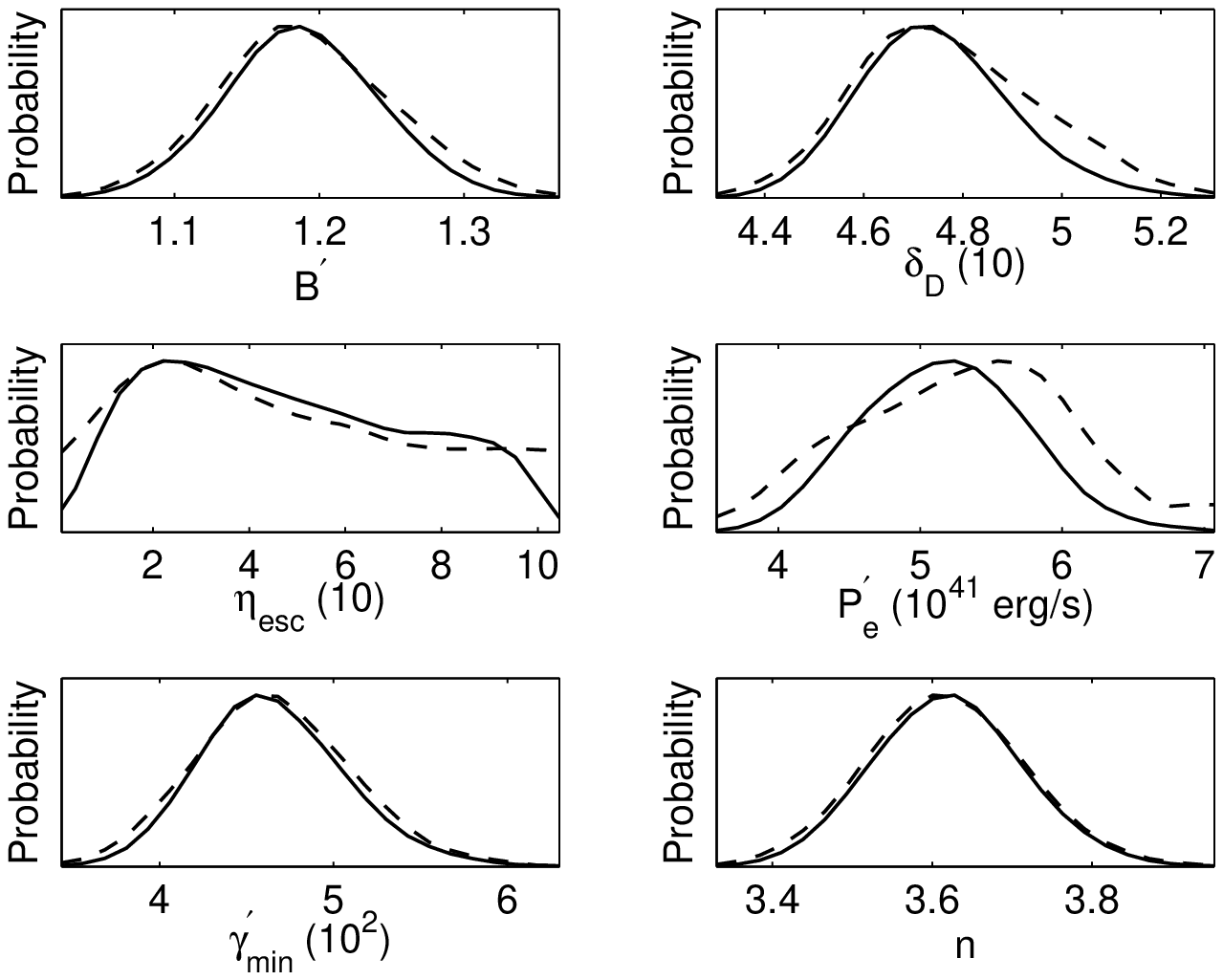}\includegraphics[width=0.45\textwidth]{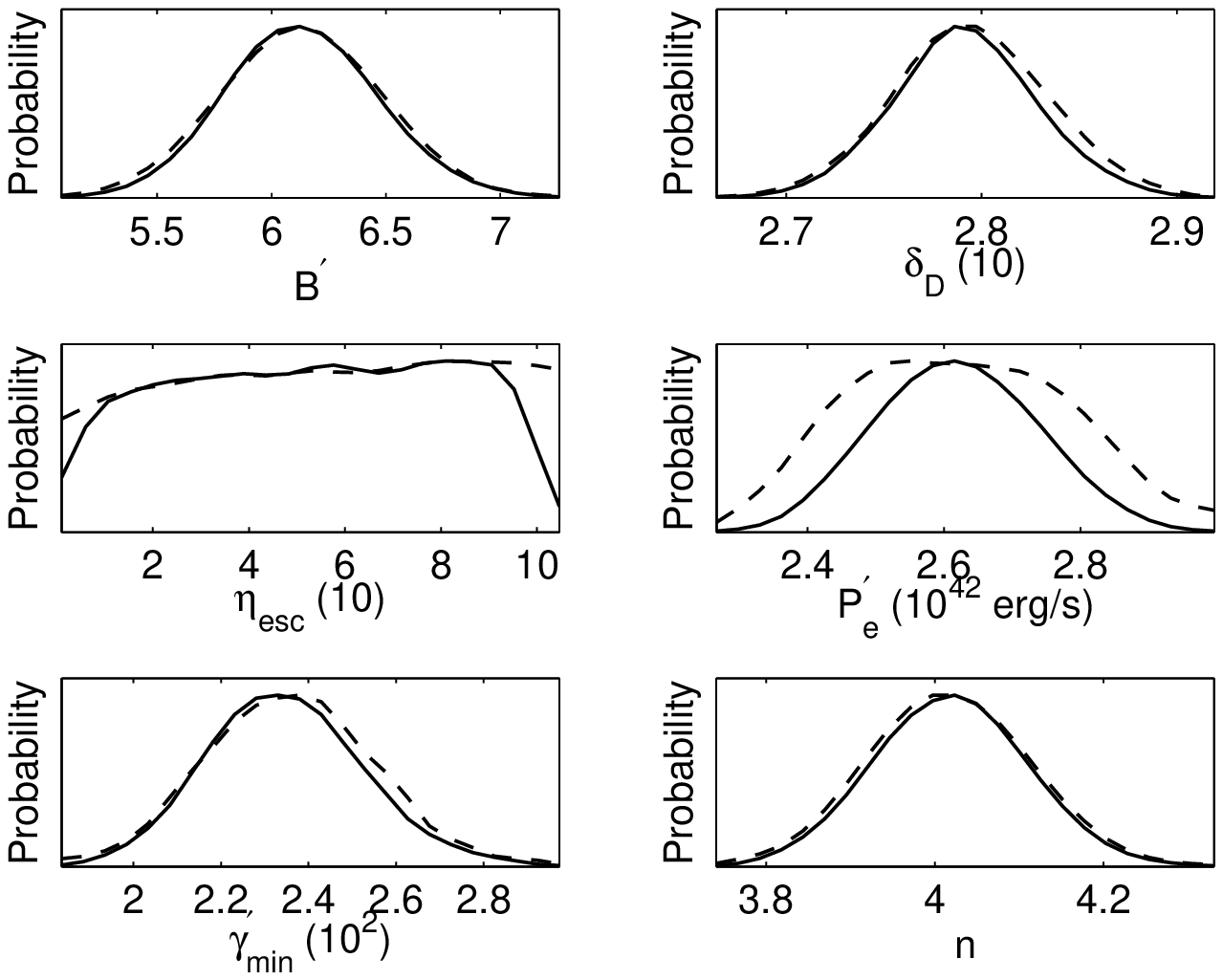}
\includegraphics[width=0.45\textwidth]{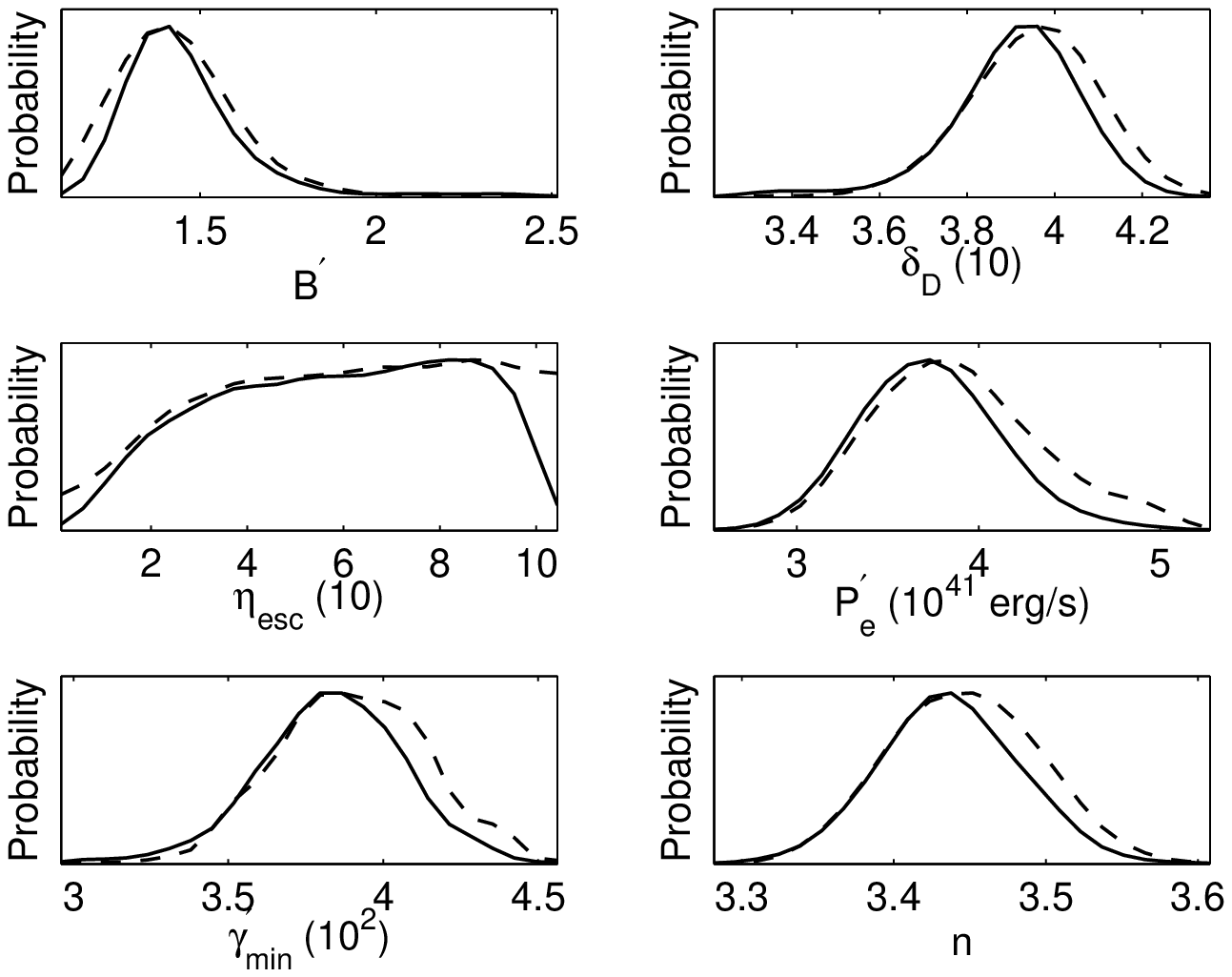}\includegraphics[width=0.45\textwidth]{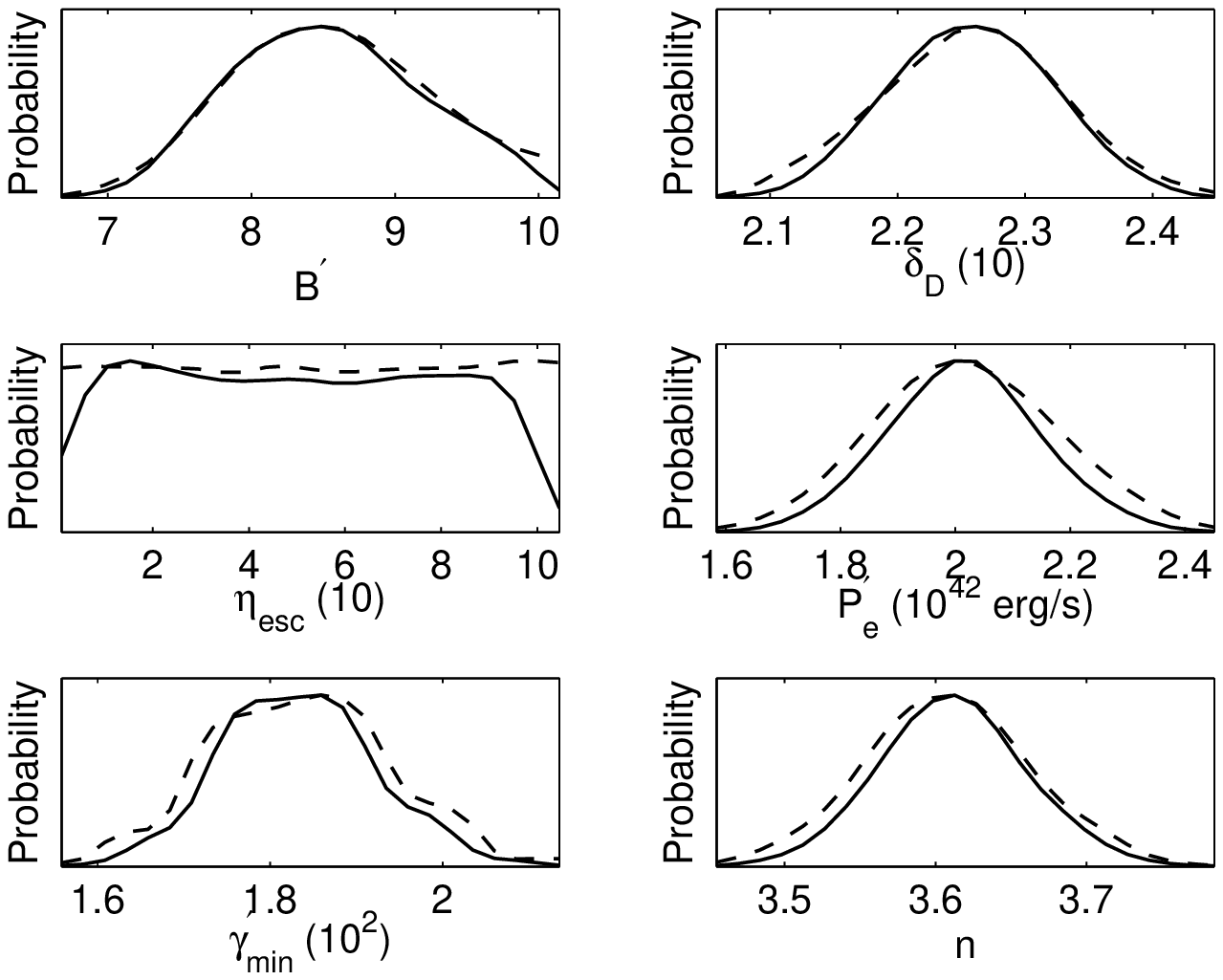}
\includegraphics[width=0.45\textwidth]{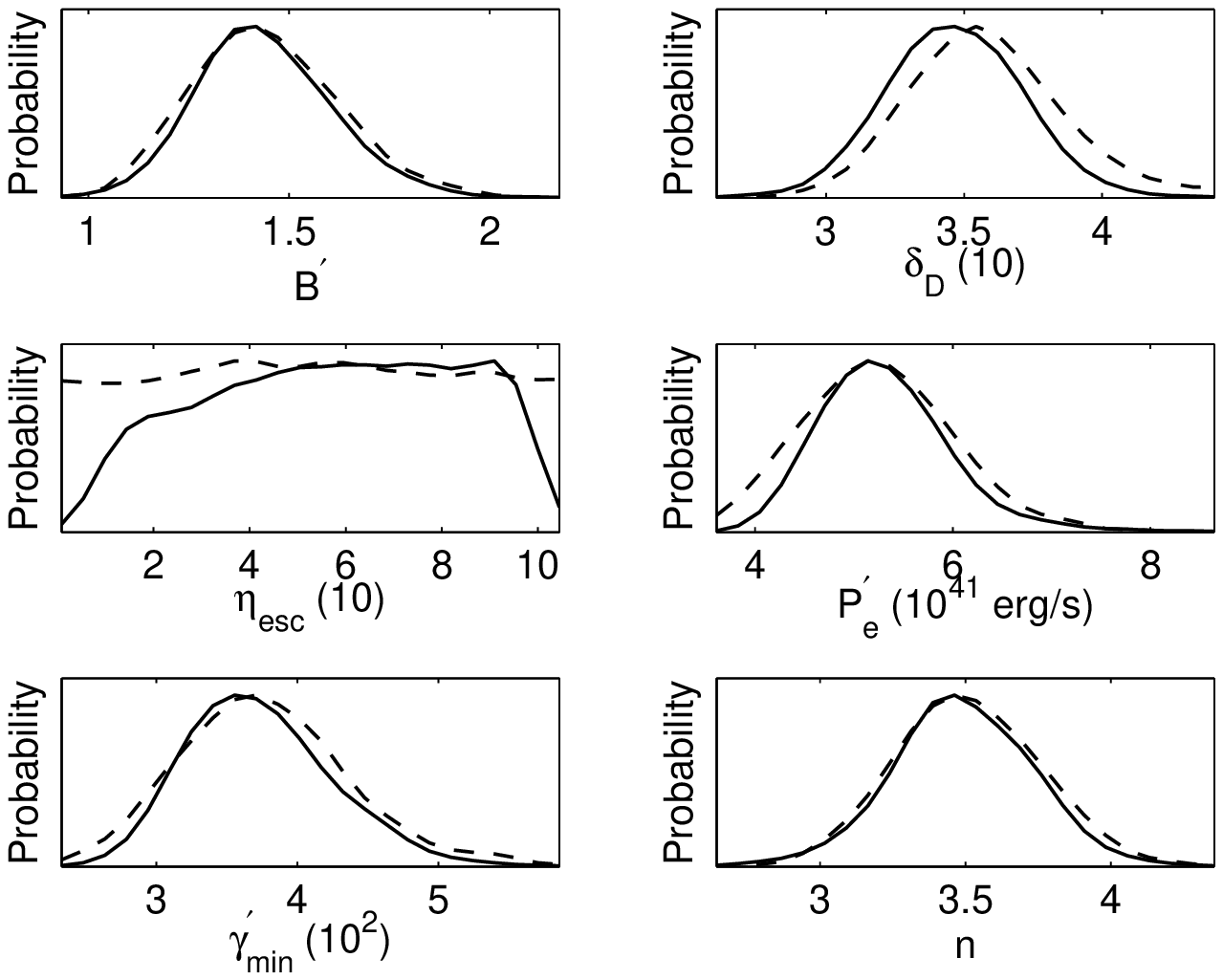}\includegraphics[width=0.45\textwidth]{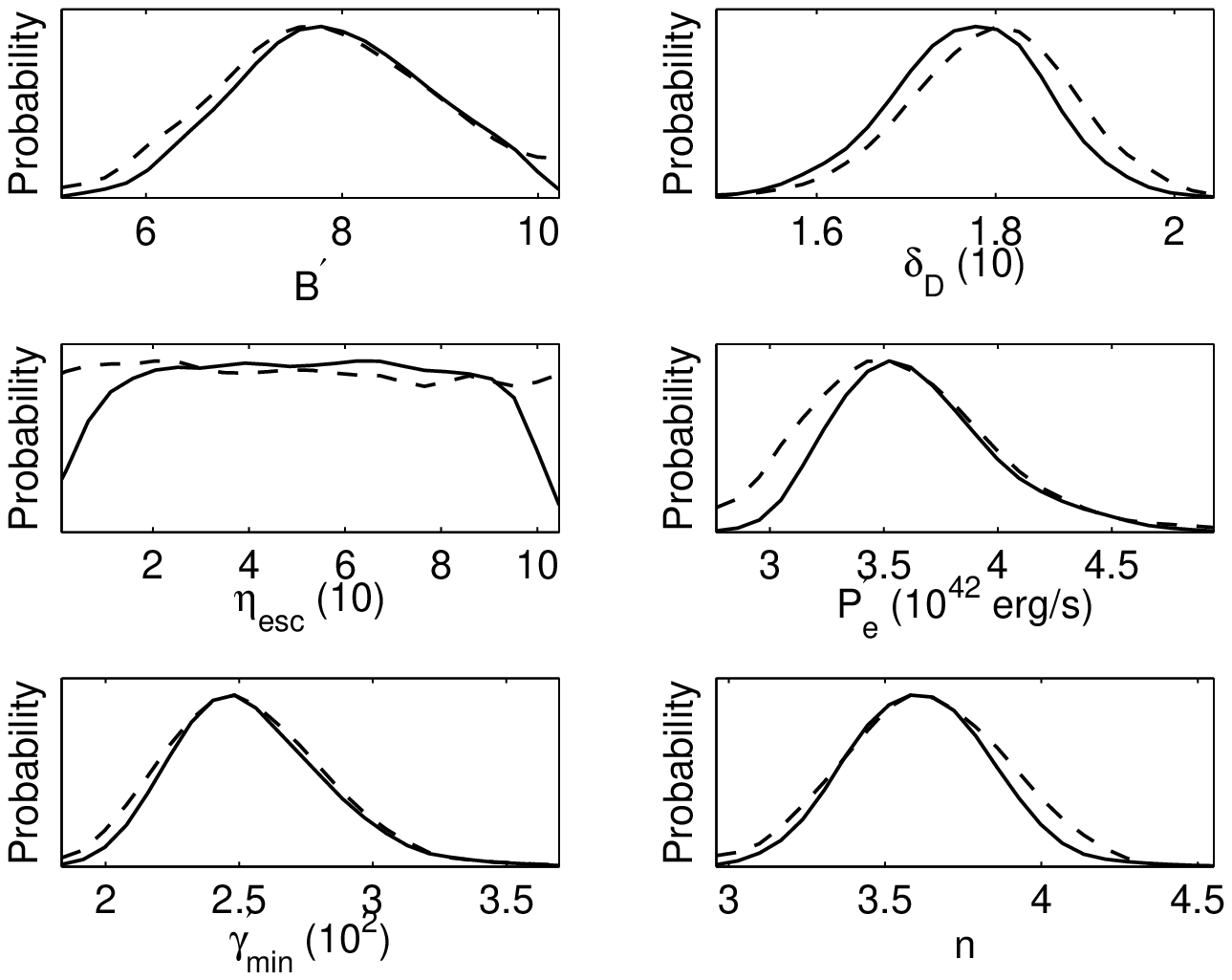}
\caption{Same as Figure~\ref{figure9}, but for the SEDs in Periods D (Top), E(Middle) and F(Bottom) reported in Hayashida et al.(2012).
%%The plots are arranged as the following Figure \ref{figure1}
 \label{figure10}}
\end{figure*}

\begin{figure*}
\includegraphics[width=0.45\textwidth]{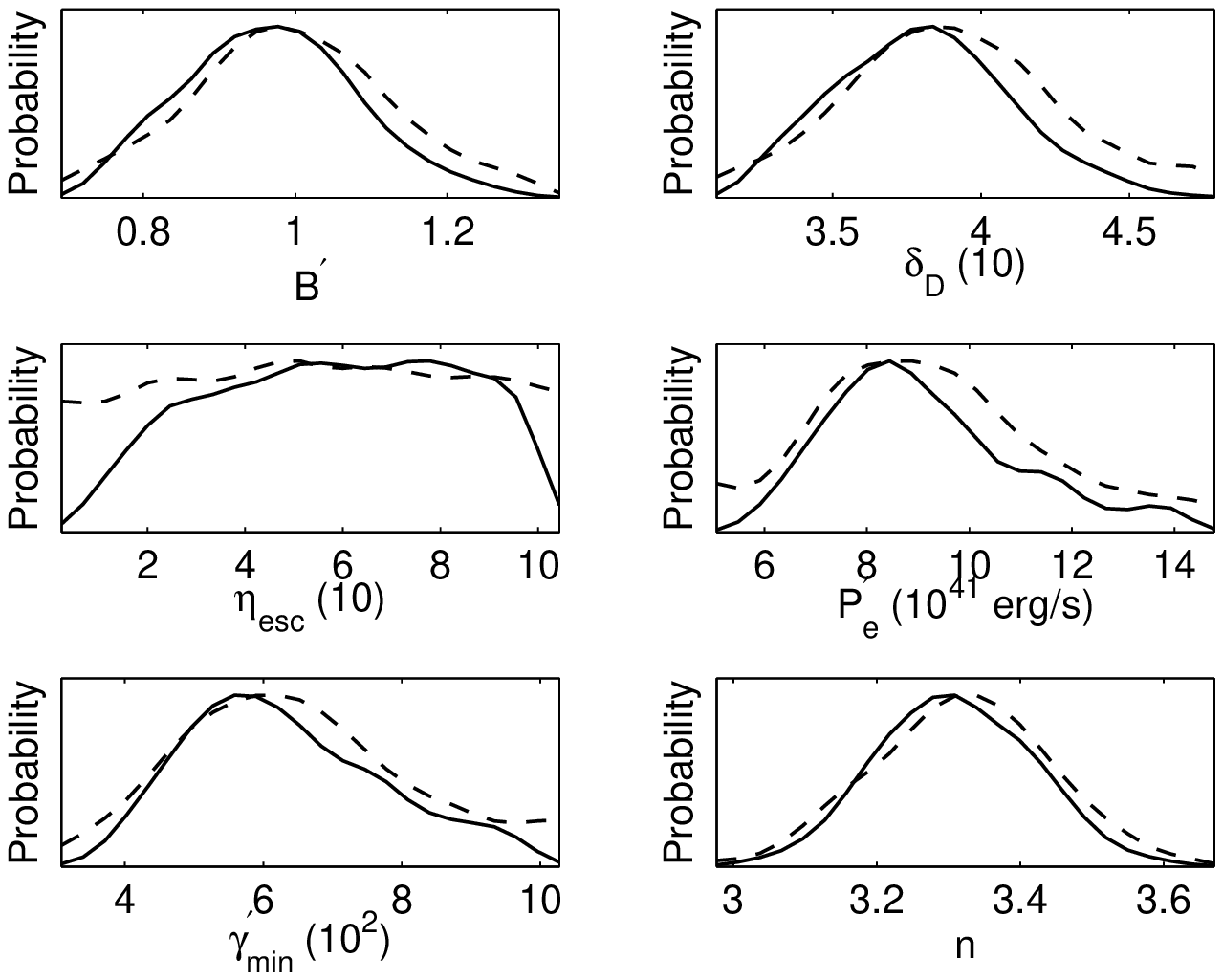}\includegraphics[width=0.45\textwidth]{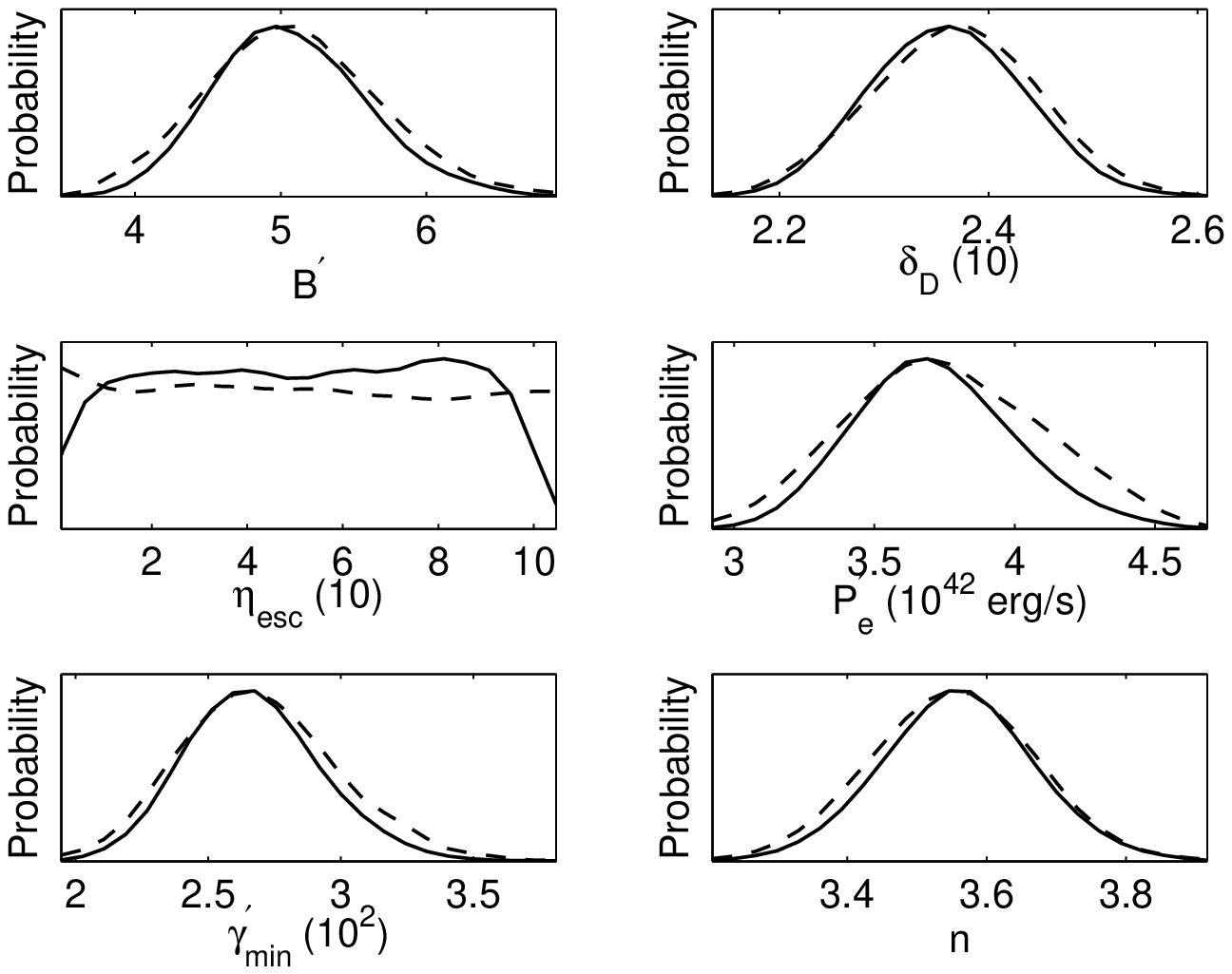}
\includegraphics[width=0.45\textwidth]{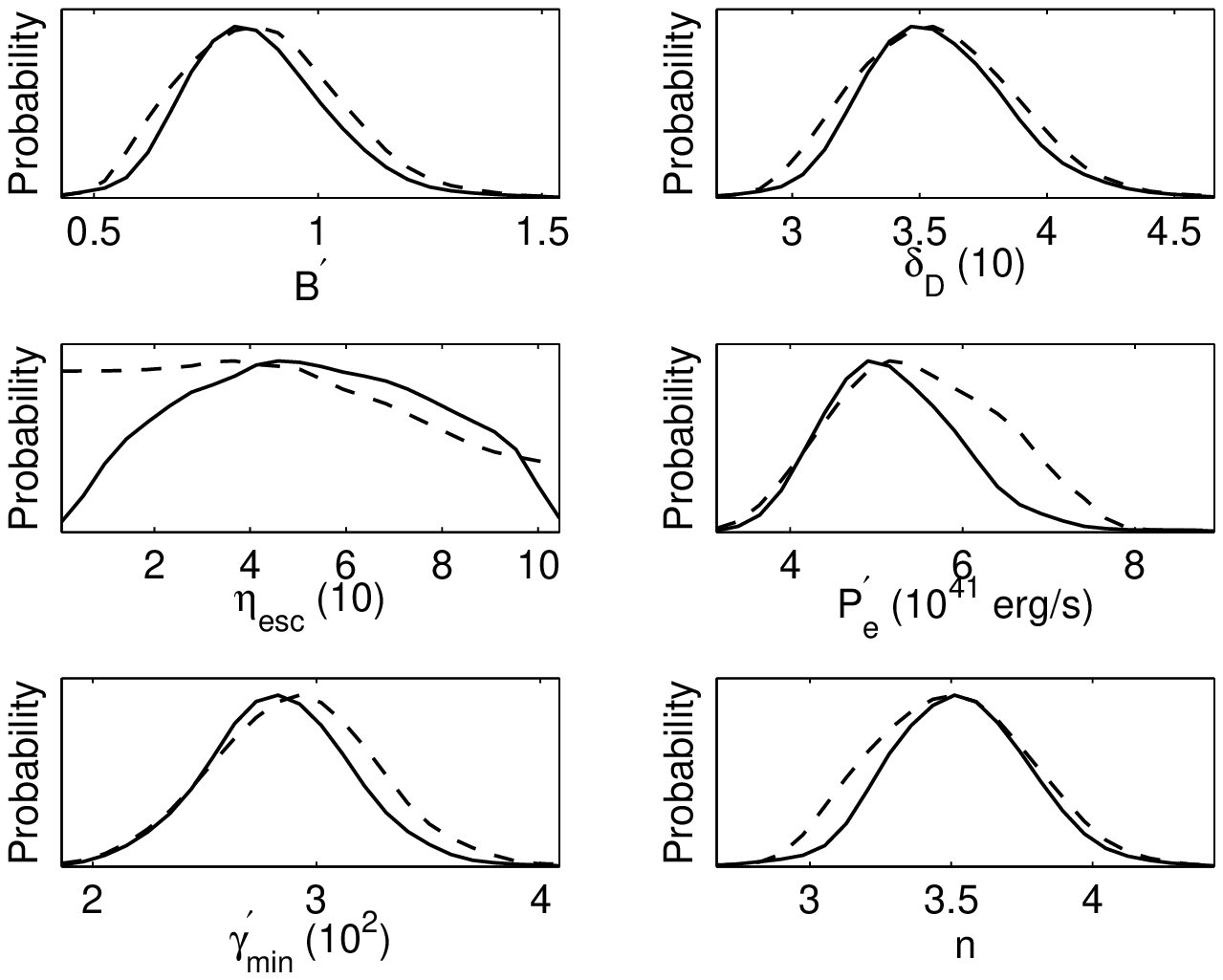}\includegraphics[width=0.45\textwidth]{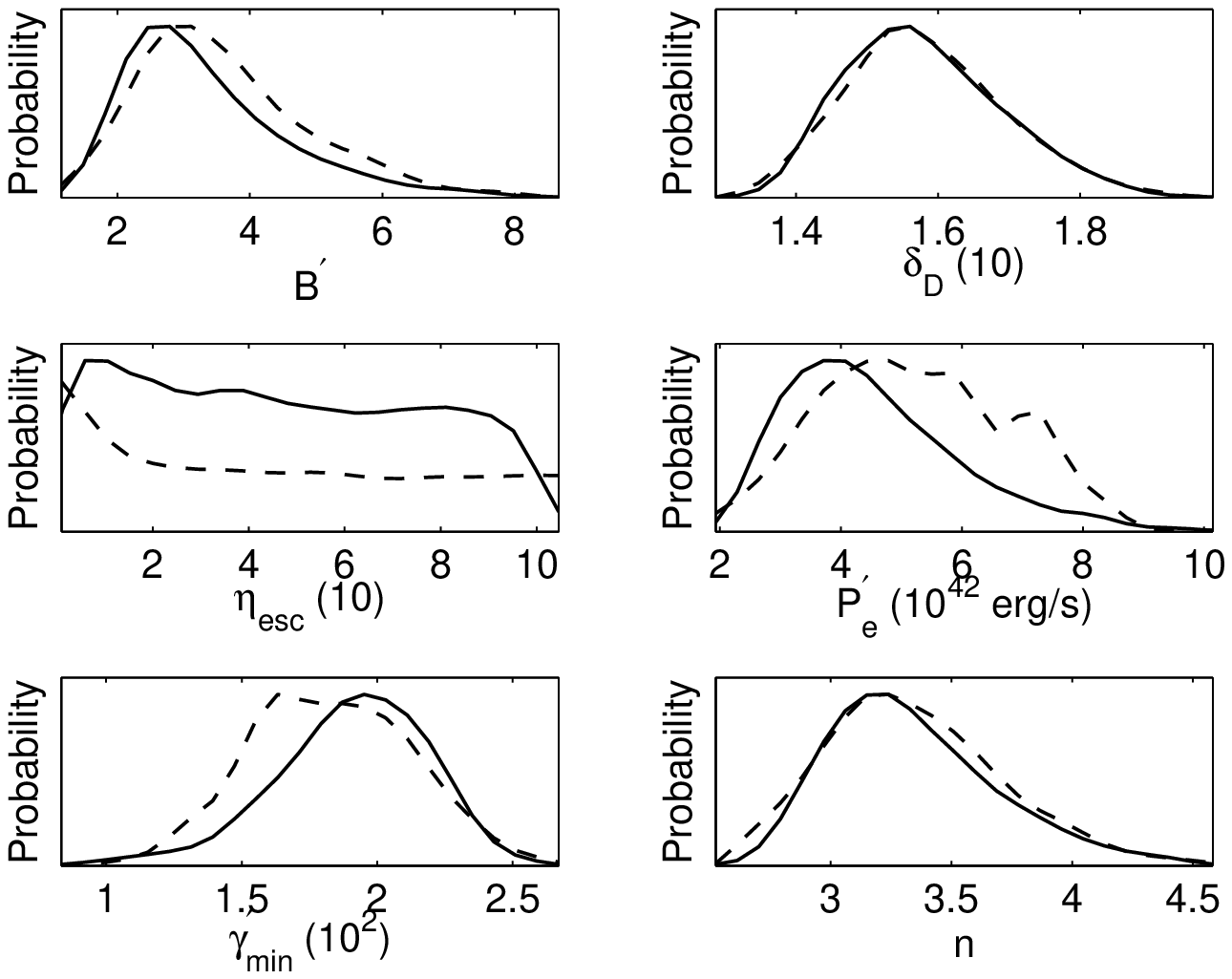}
\caption{Same as Figure~\ref{figure9}, but for the SEDs Periods G (Top) and H (Middle) reported in Hayashida et al.(2012).\label{figure11}}
\end{figure*}

\begin{figure*}
\includegraphics[width=0.45\textwidth]{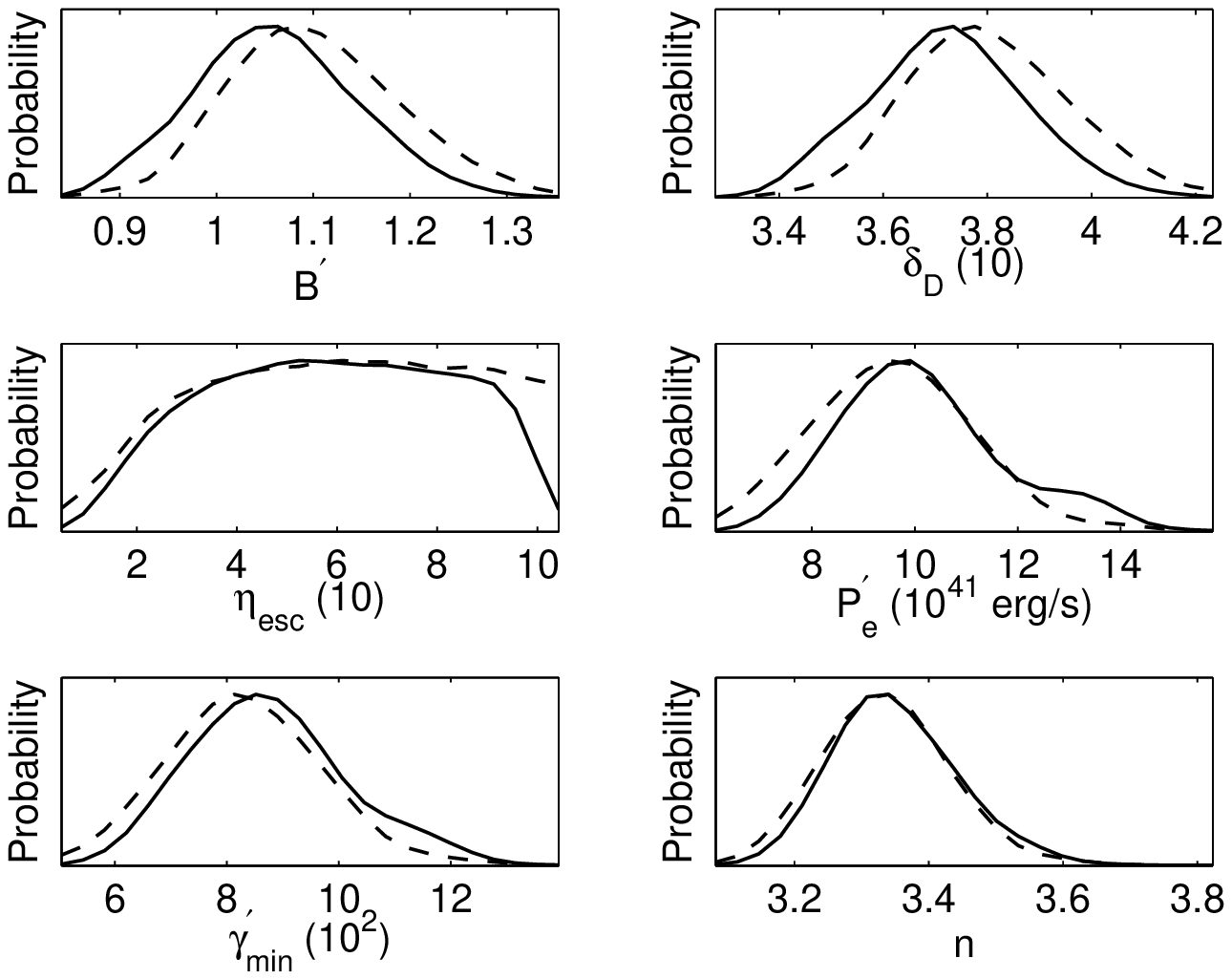}\includegraphics[width=0.45\textwidth]{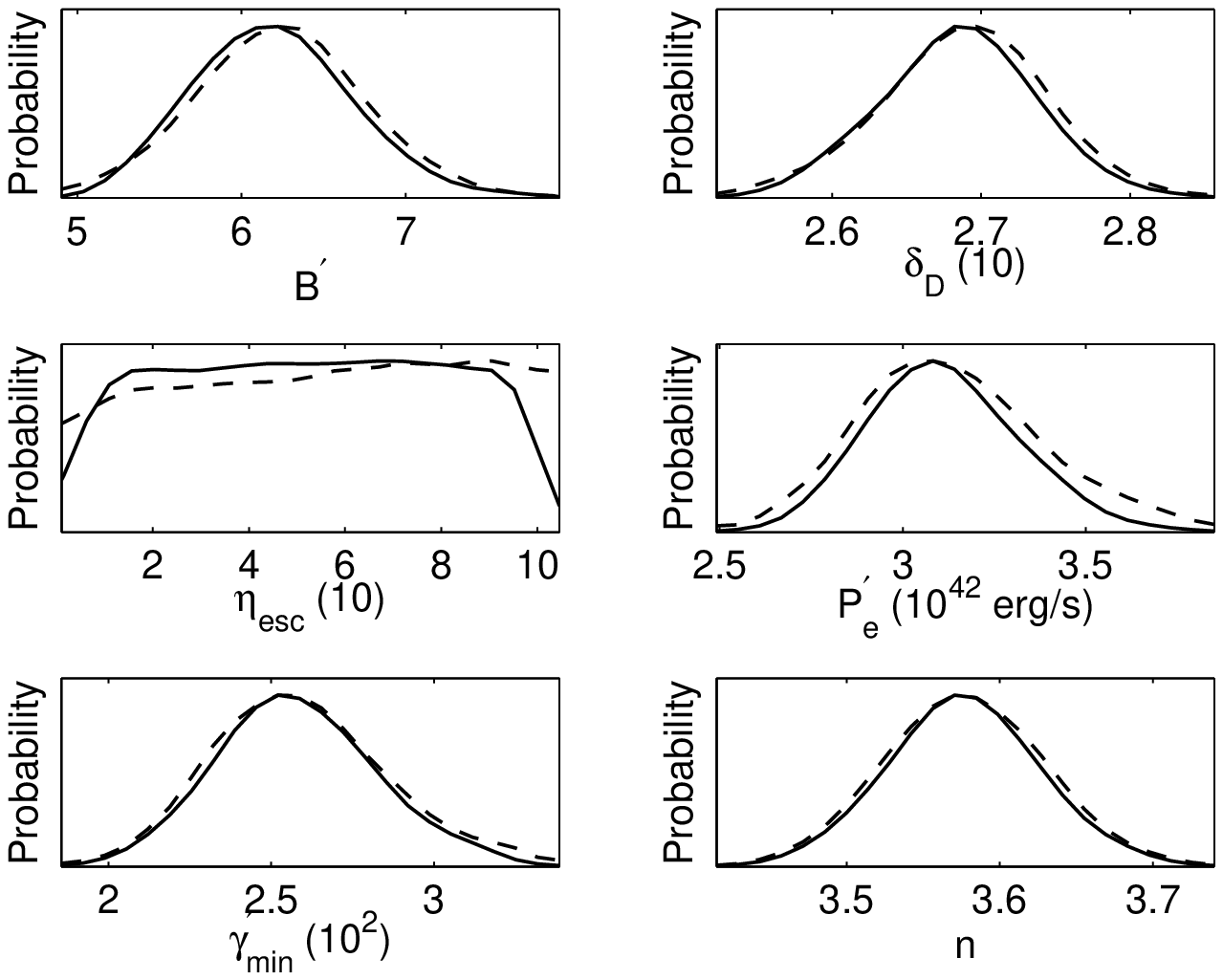}
\includegraphics[width=0.45\textwidth]{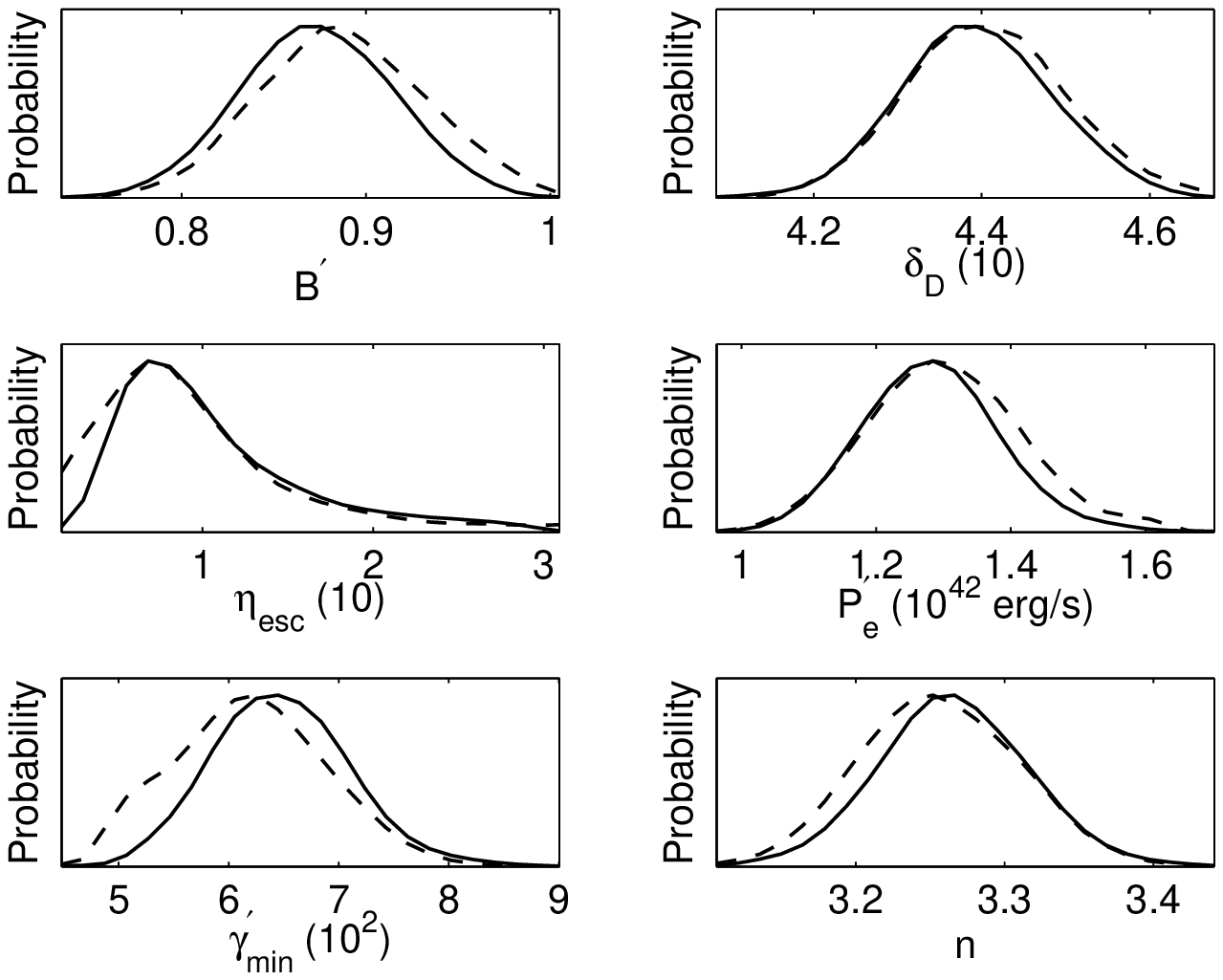}\includegraphics[width=0.45\textwidth]{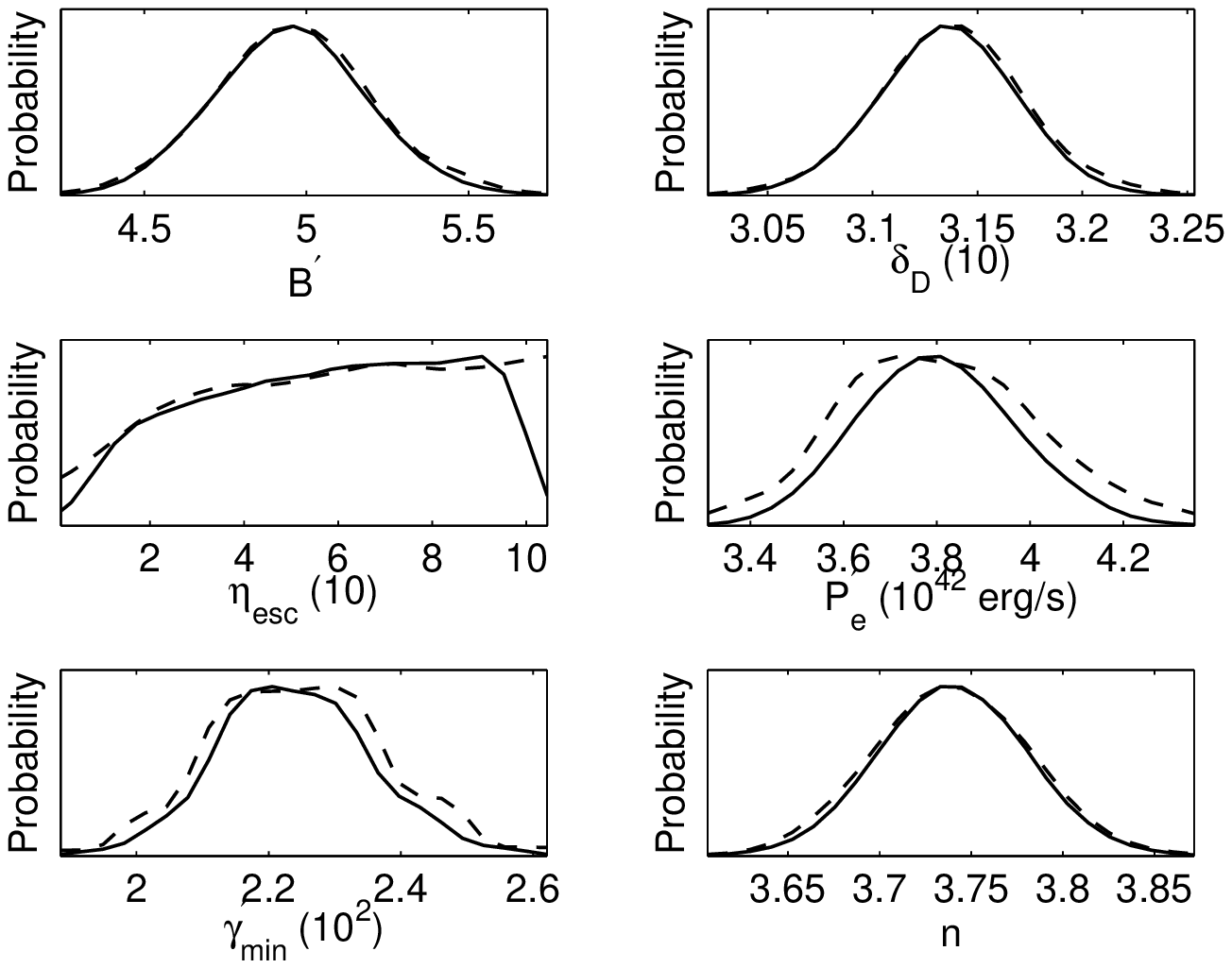}
\includegraphics[width=0.45\textwidth]{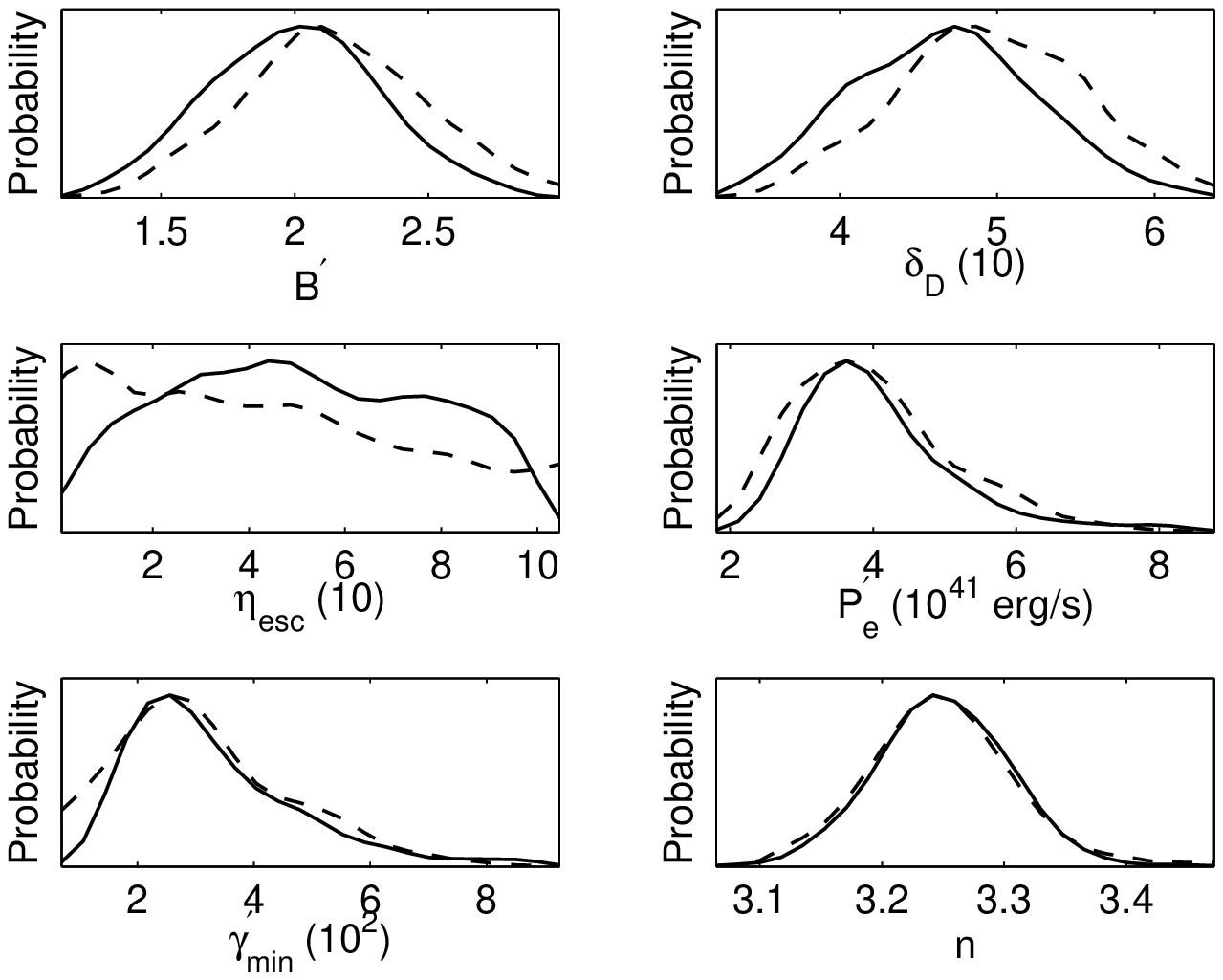}\includegraphics[width=0.45\textwidth]{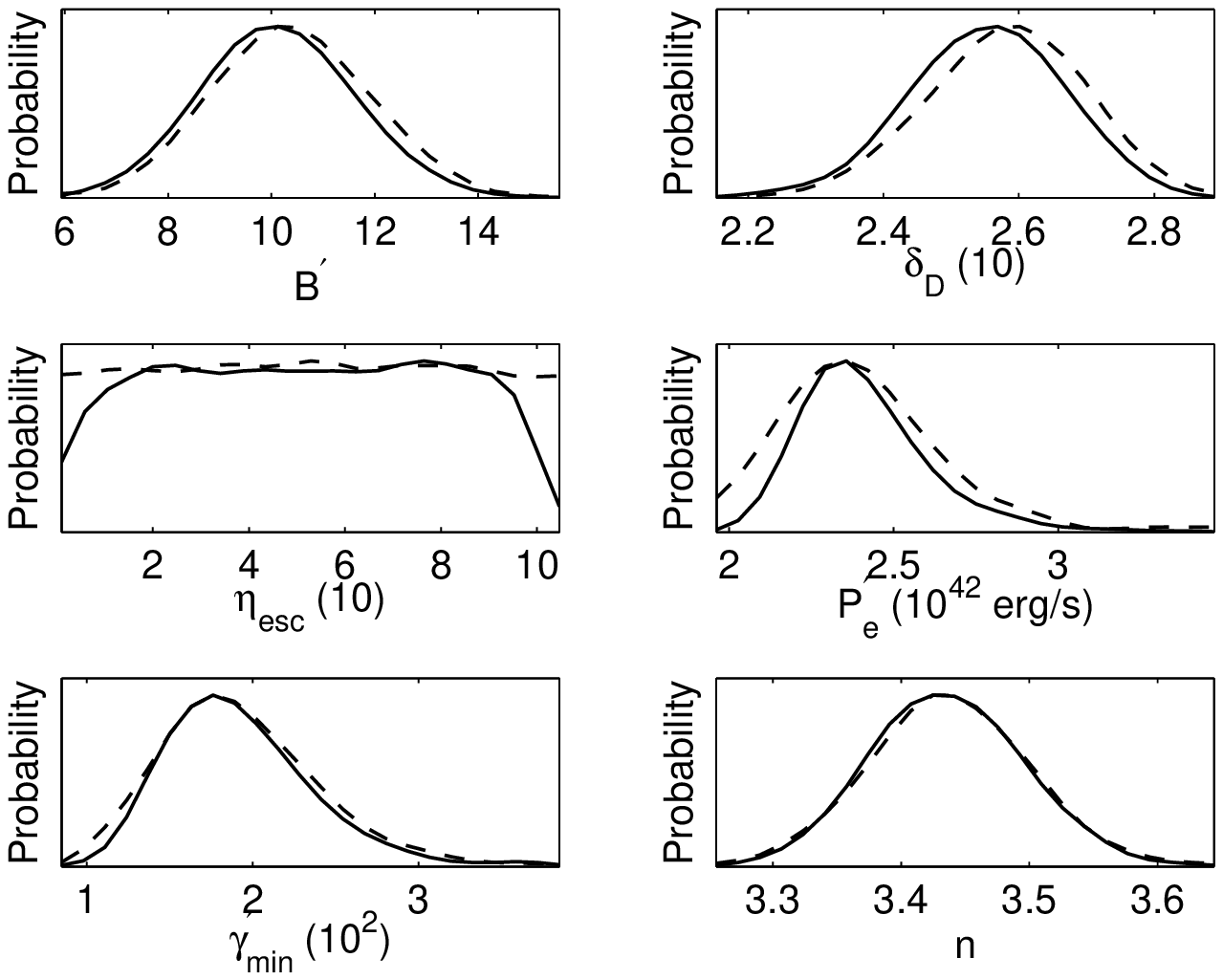}
\caption{Same as Figure~\ref{figure9}, but for the SEDs reported by Paliya et al.(2015).
The plots from top to bottom are Flare1, Flare2 and post-flare, respectively. \label{figure12}}
\end{figure*}

\begin{figure*}
\includegraphics[width=0.45\textwidth]{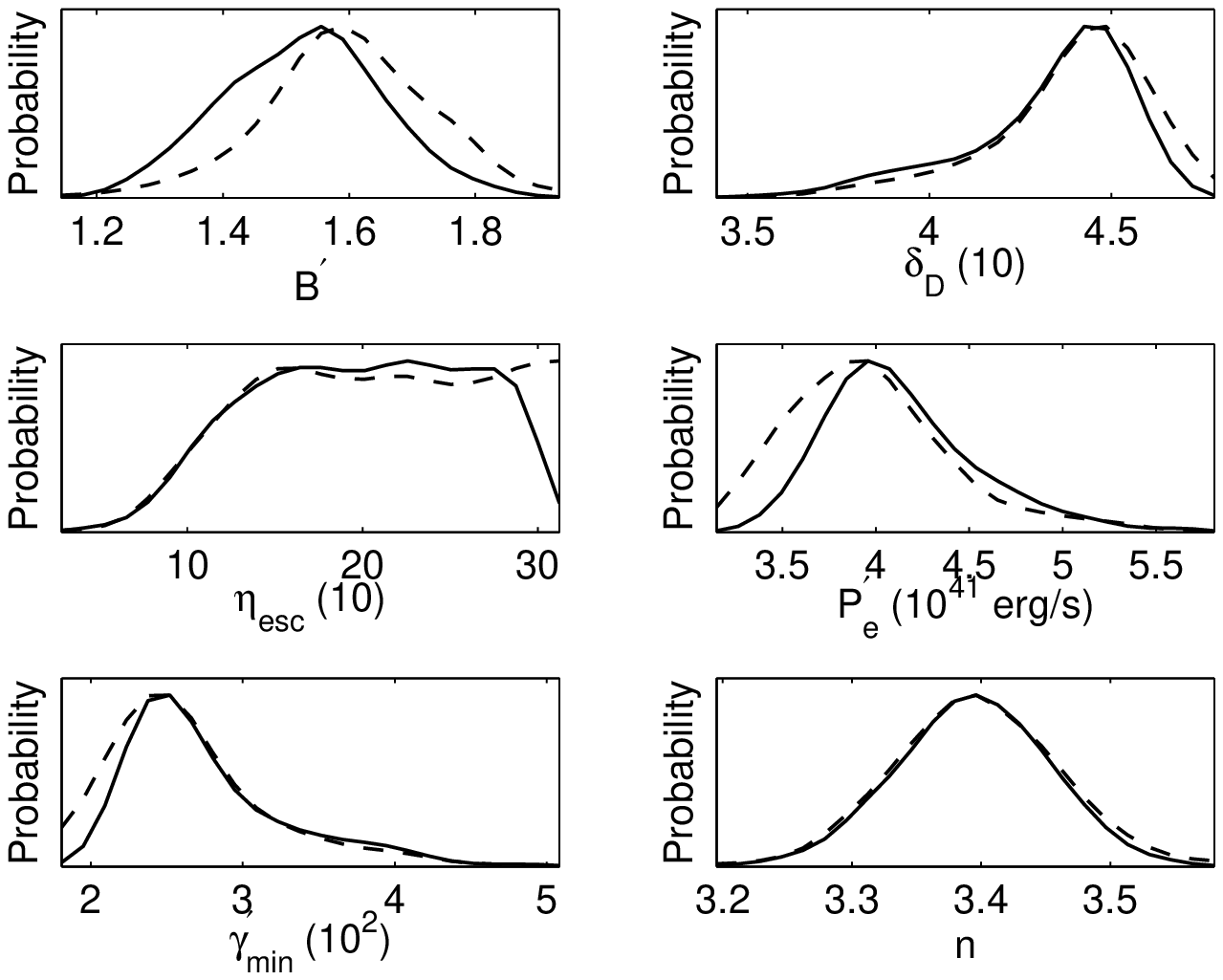}\includegraphics[width=0.45\textwidth]{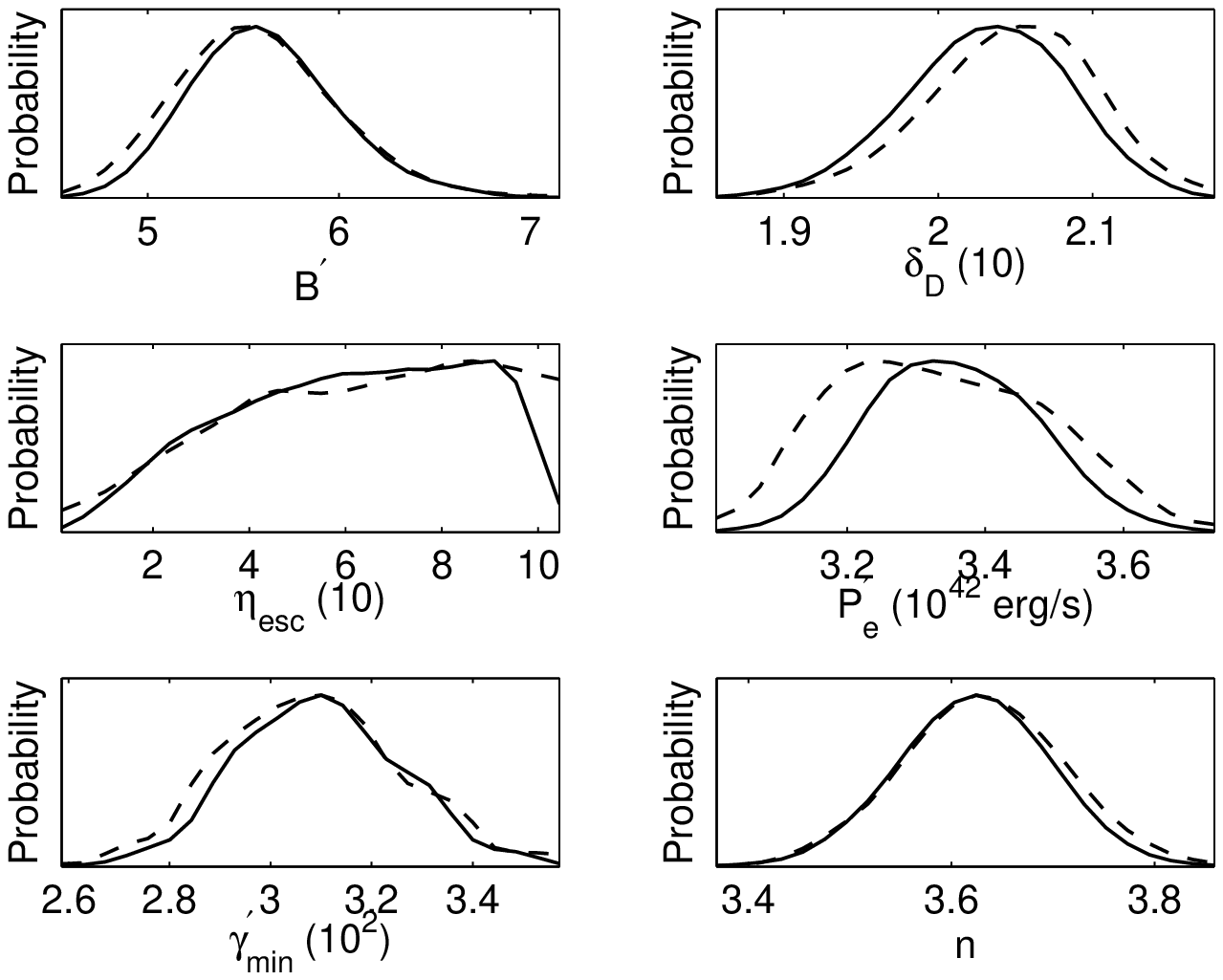}
\includegraphics[width=0.45\textwidth]{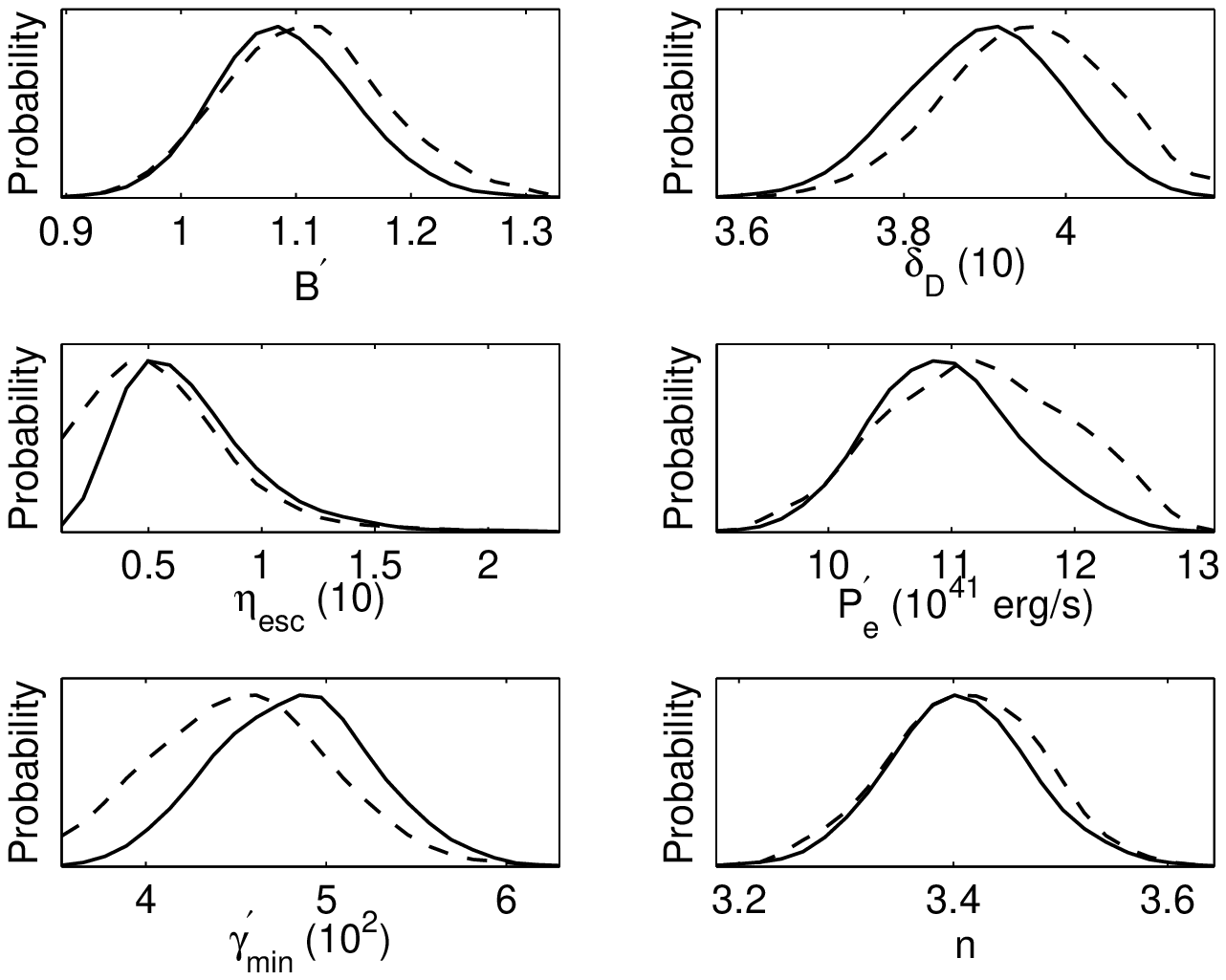}\includegraphics[width=0.45\textwidth]{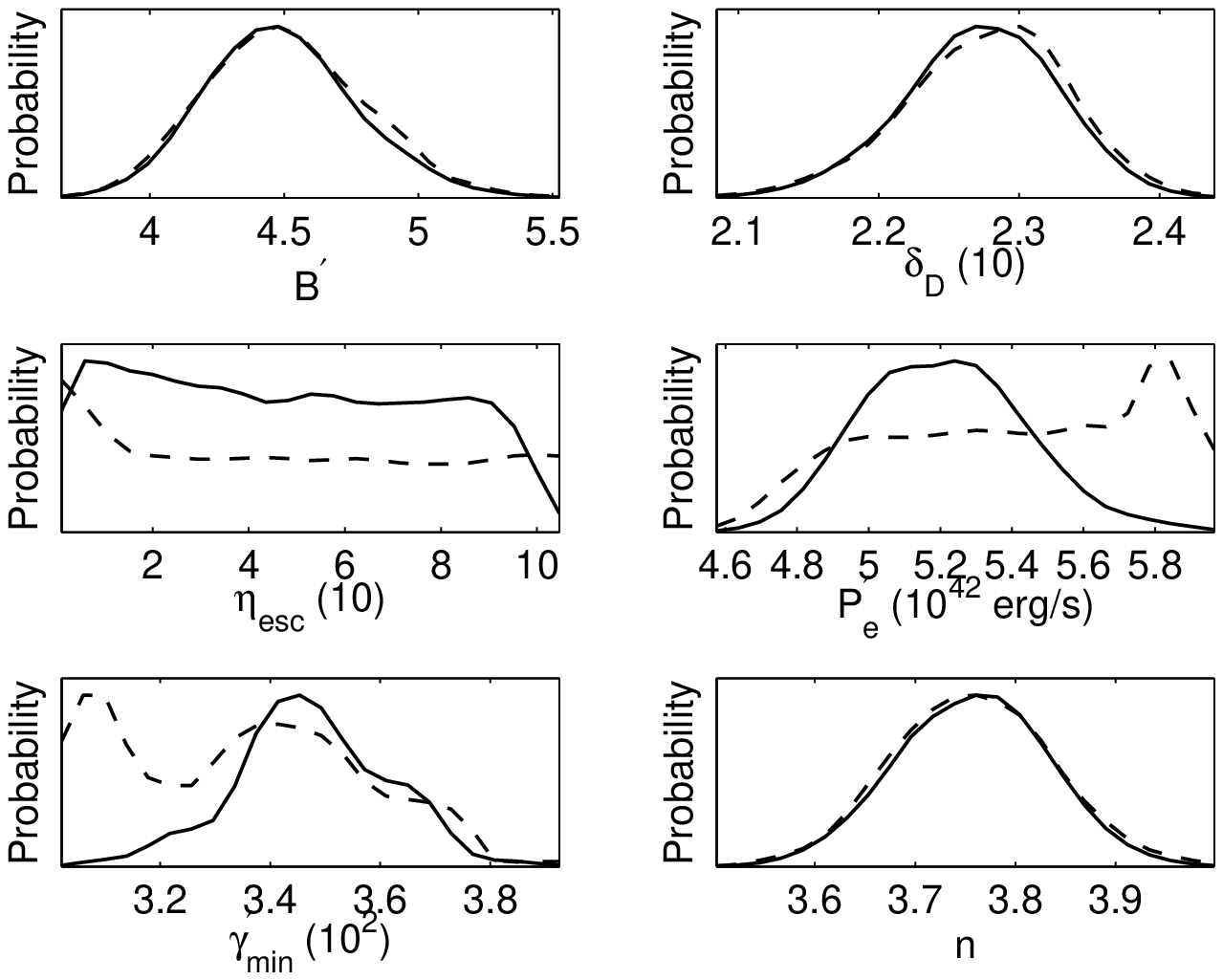}
\includegraphics[width=0.45\textwidth]{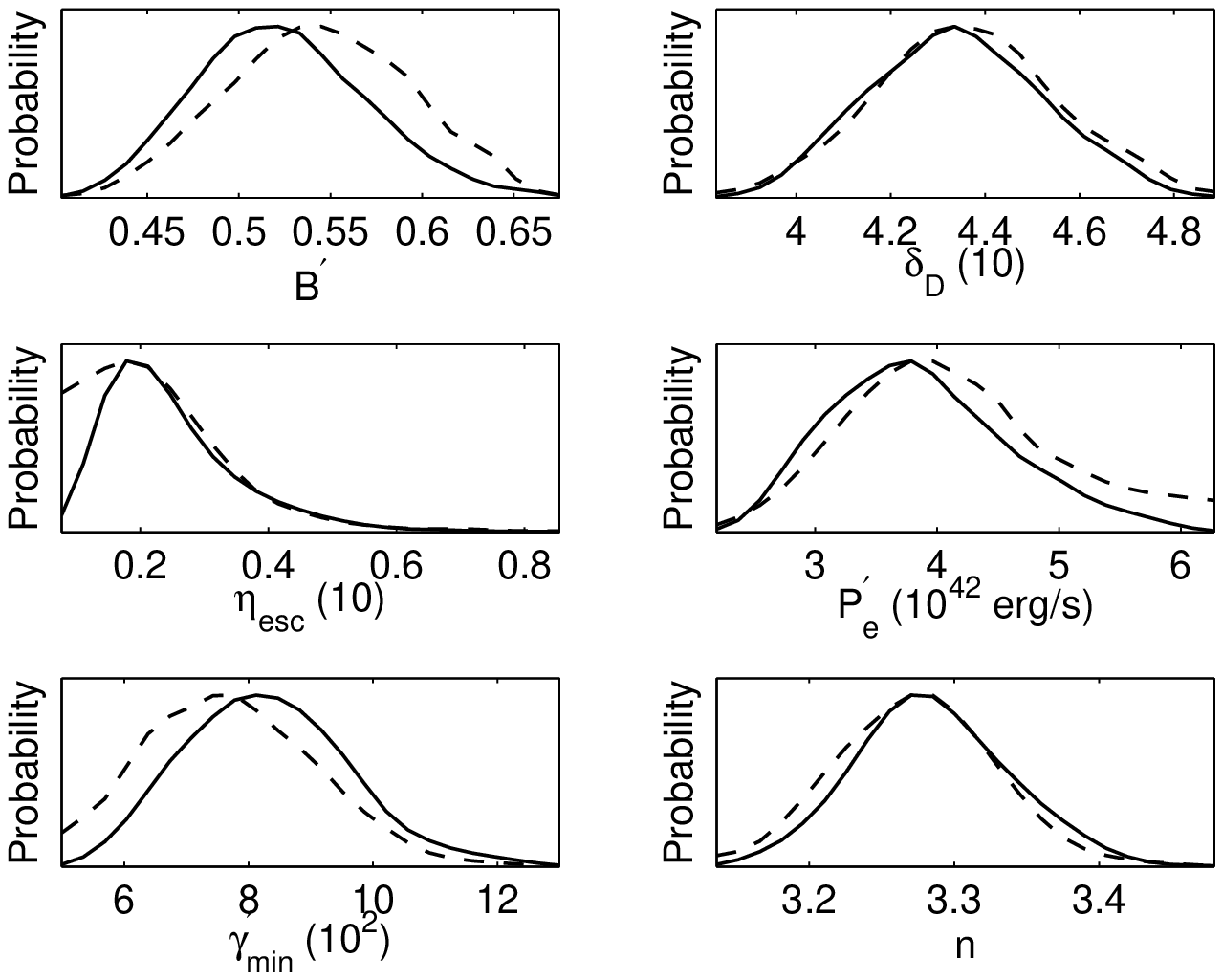}\includegraphics[width=0.45\textwidth]{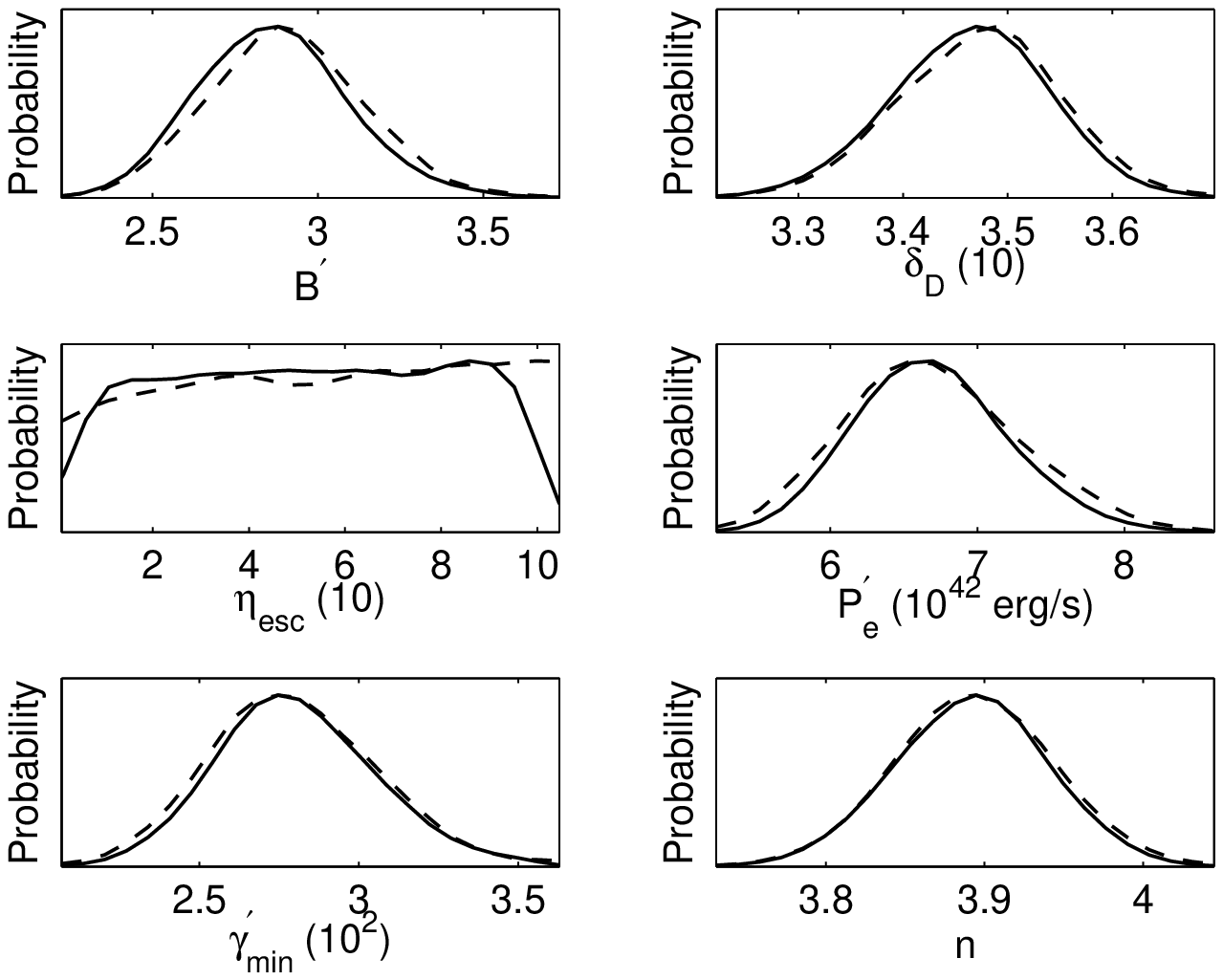}
\caption{Same as Fig.\ref{figure9}, but for the SEDs reported by Hayashida et al.(2015).
The plots from top to bottom are Periods A, C and D, respectively.\label{figure13}}
\end{figure*}

\section{One-dimensional probability distributions of the derived parameters obtained from SED fittings with the DT photons}

\label{appB}
\begin{figure*}
\includegraphics[width=0.4\textwidth]{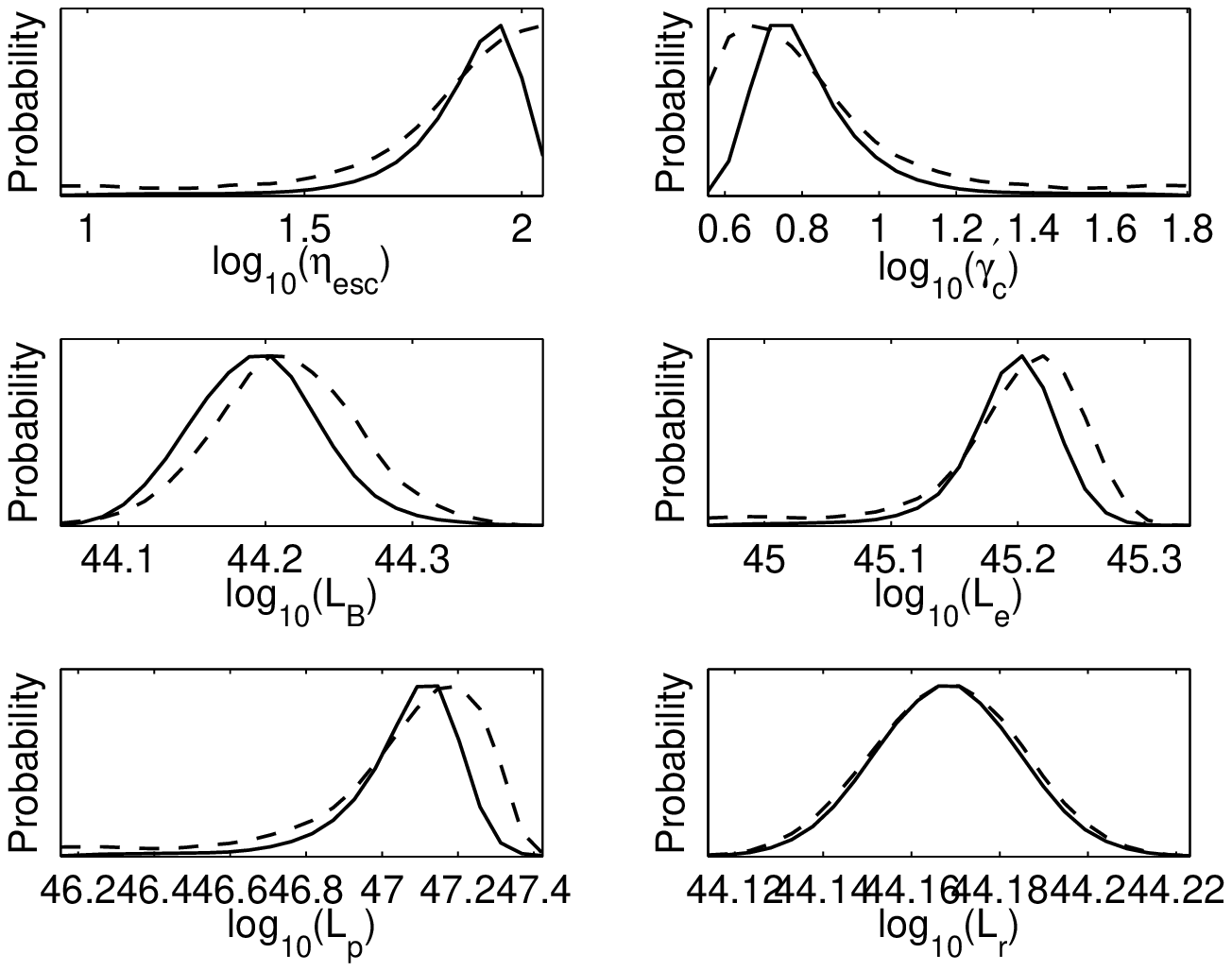}\includegraphics[width=0.4\textwidth]{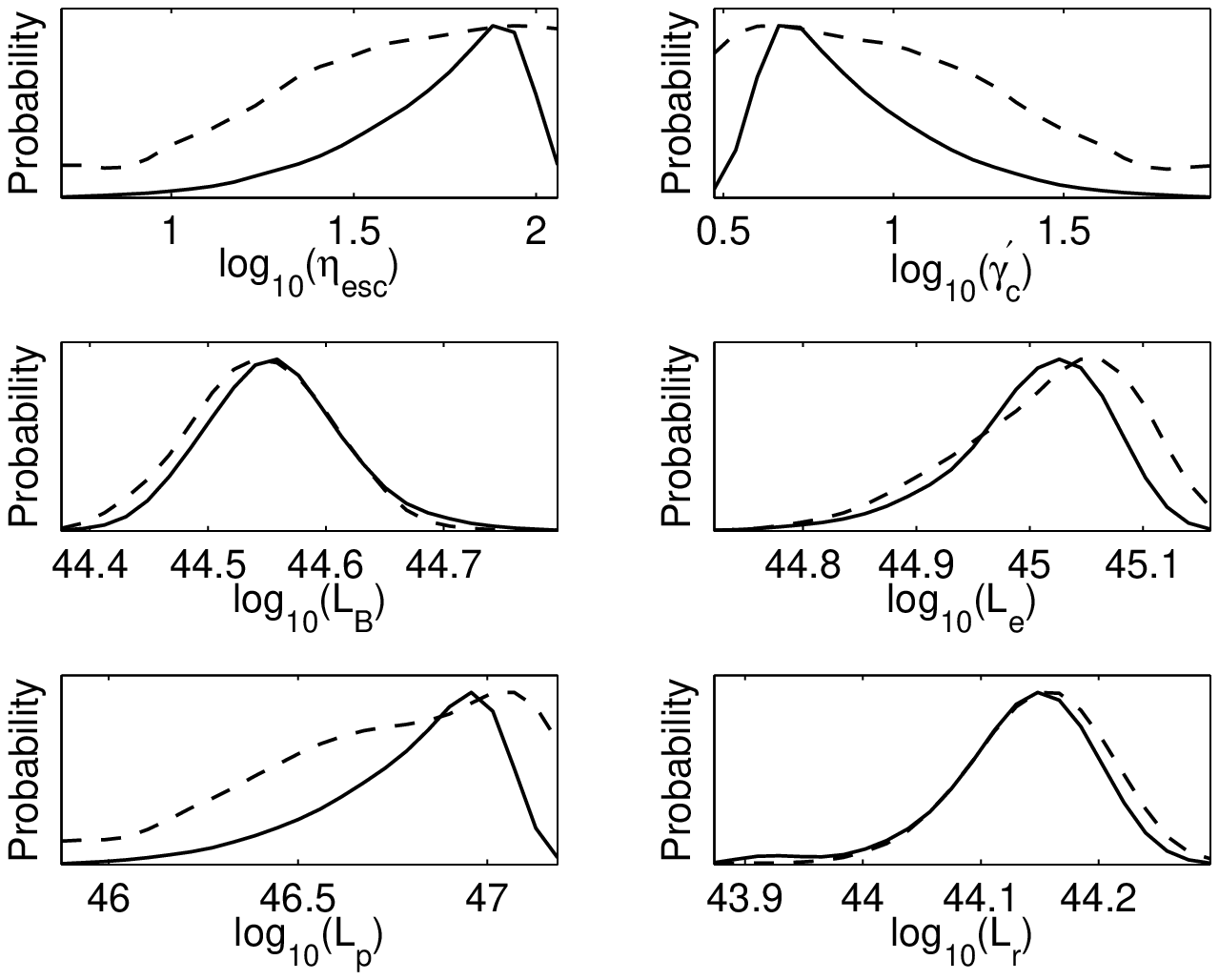}
\includegraphics[width=0.4\textwidth]{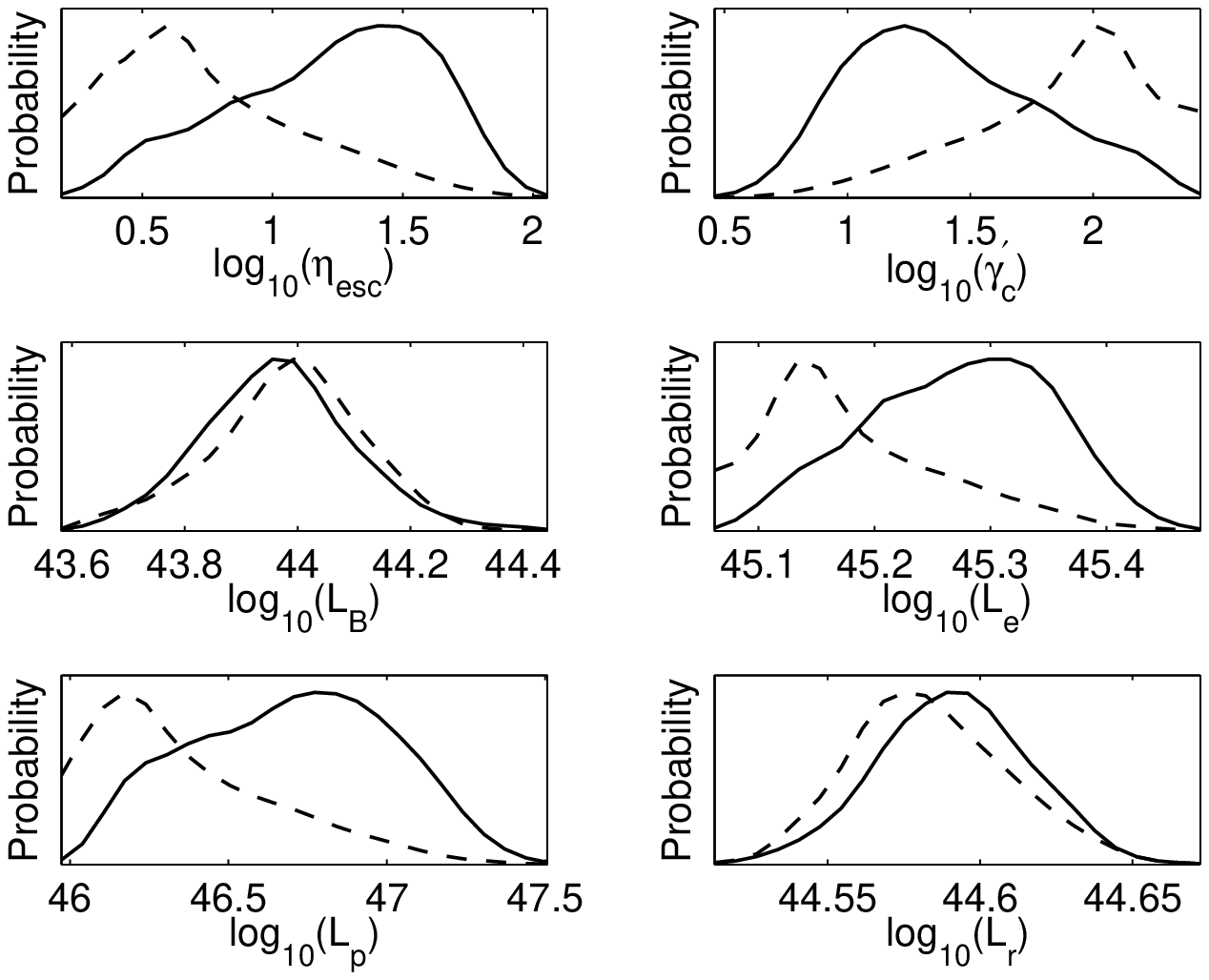}\includegraphics[width=0.4\textwidth]{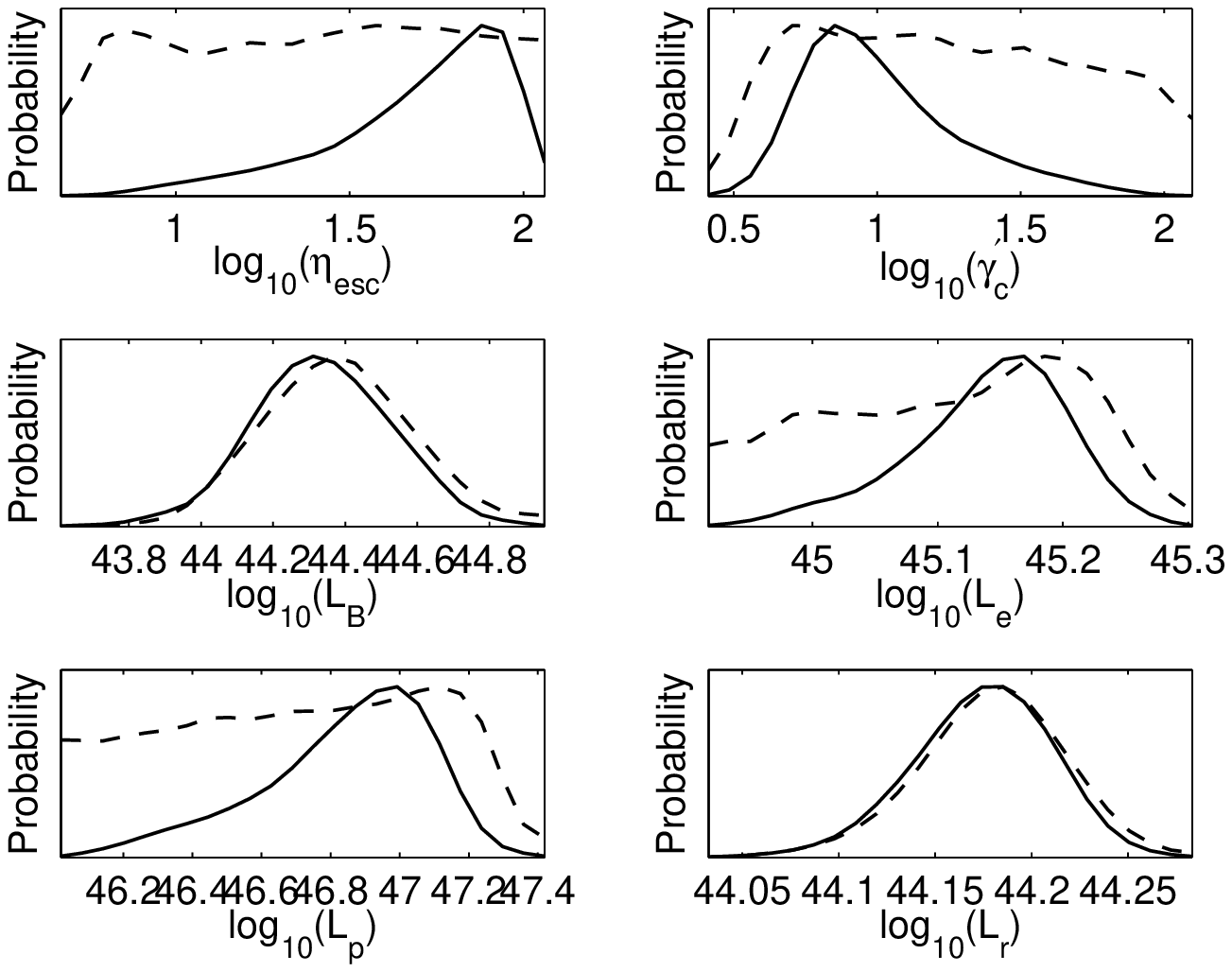}
\includegraphics[width=0.4\textwidth]{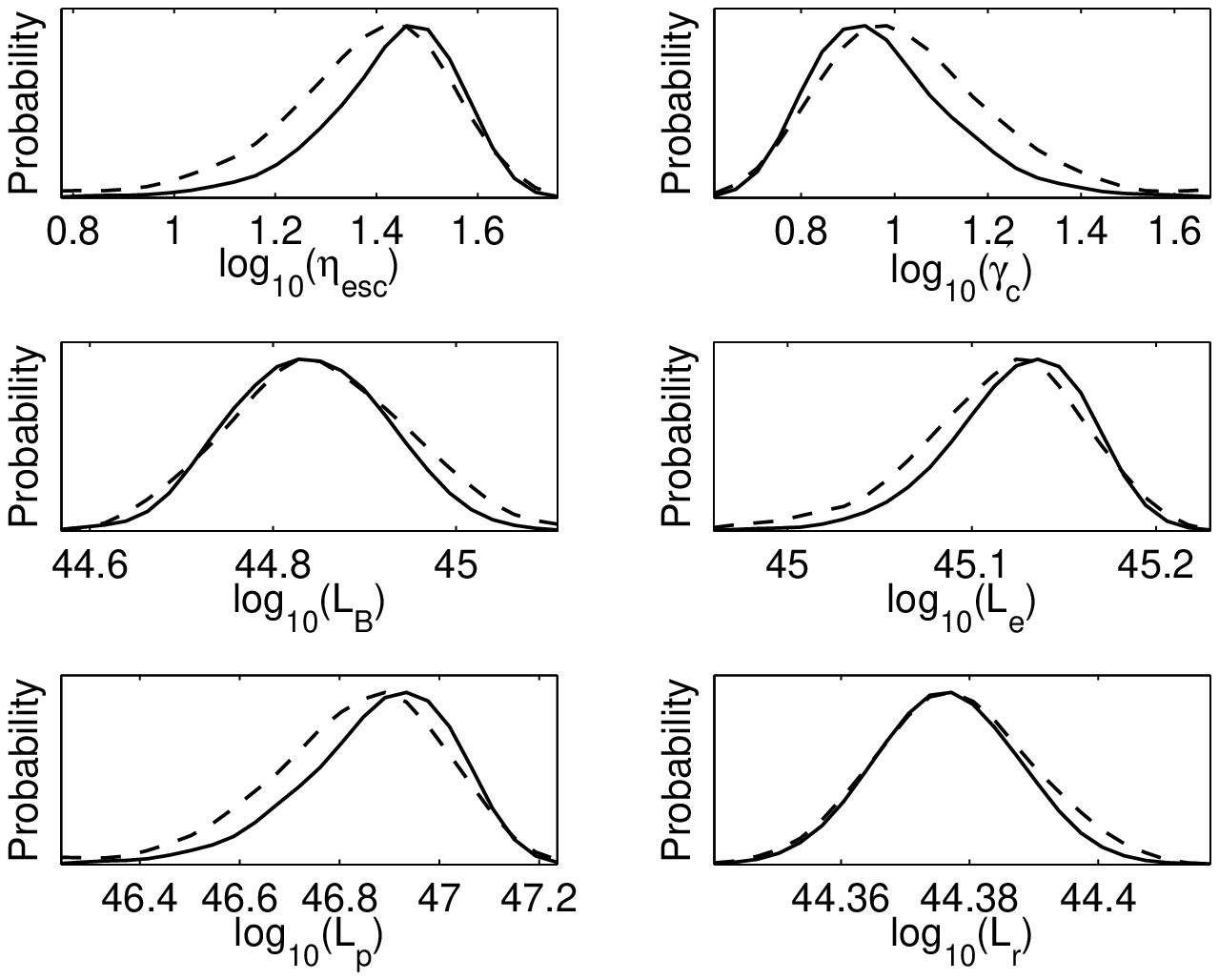}\includegraphics[width=0.4\textwidth]{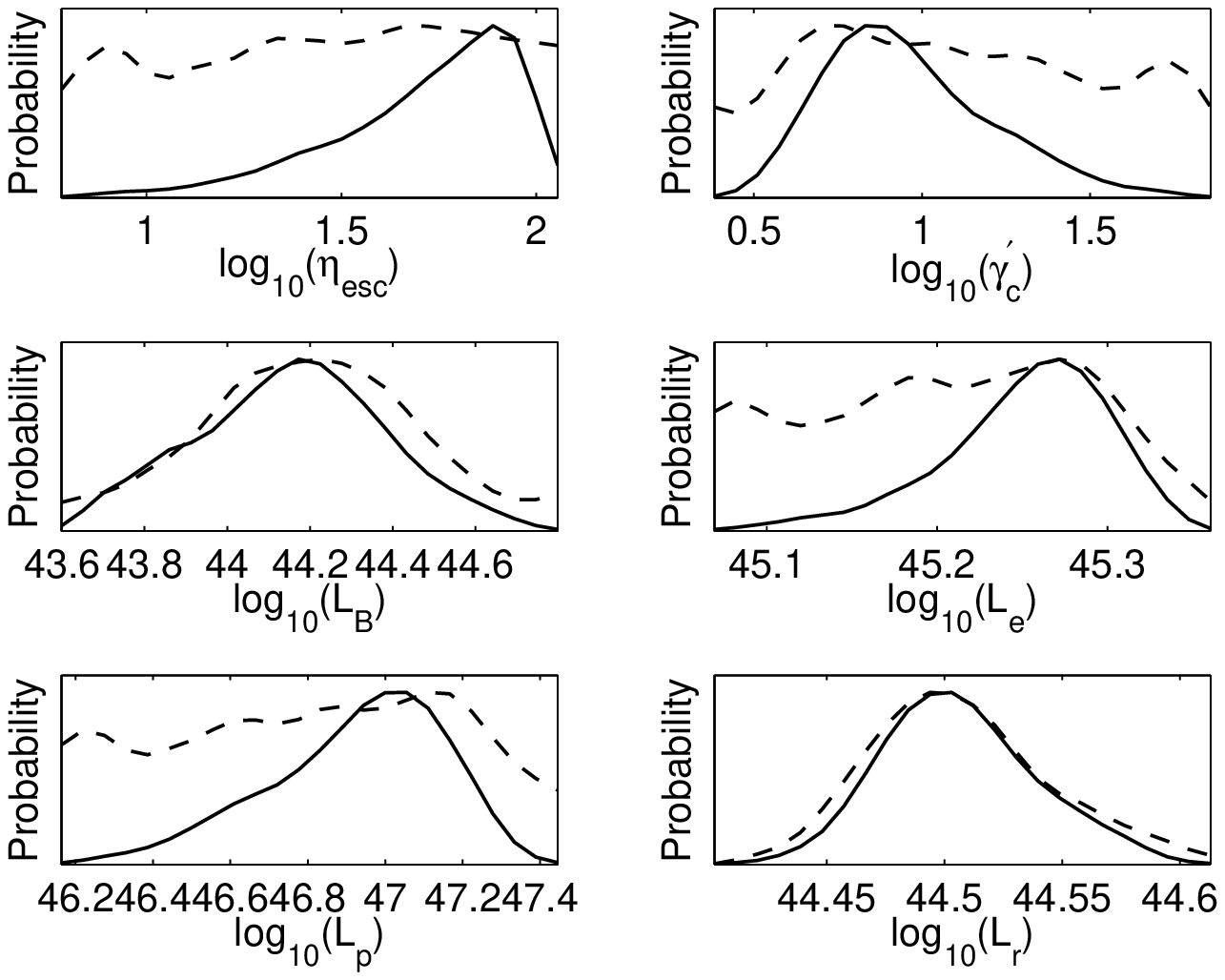}
\includegraphics[width=0.4\textwidth]{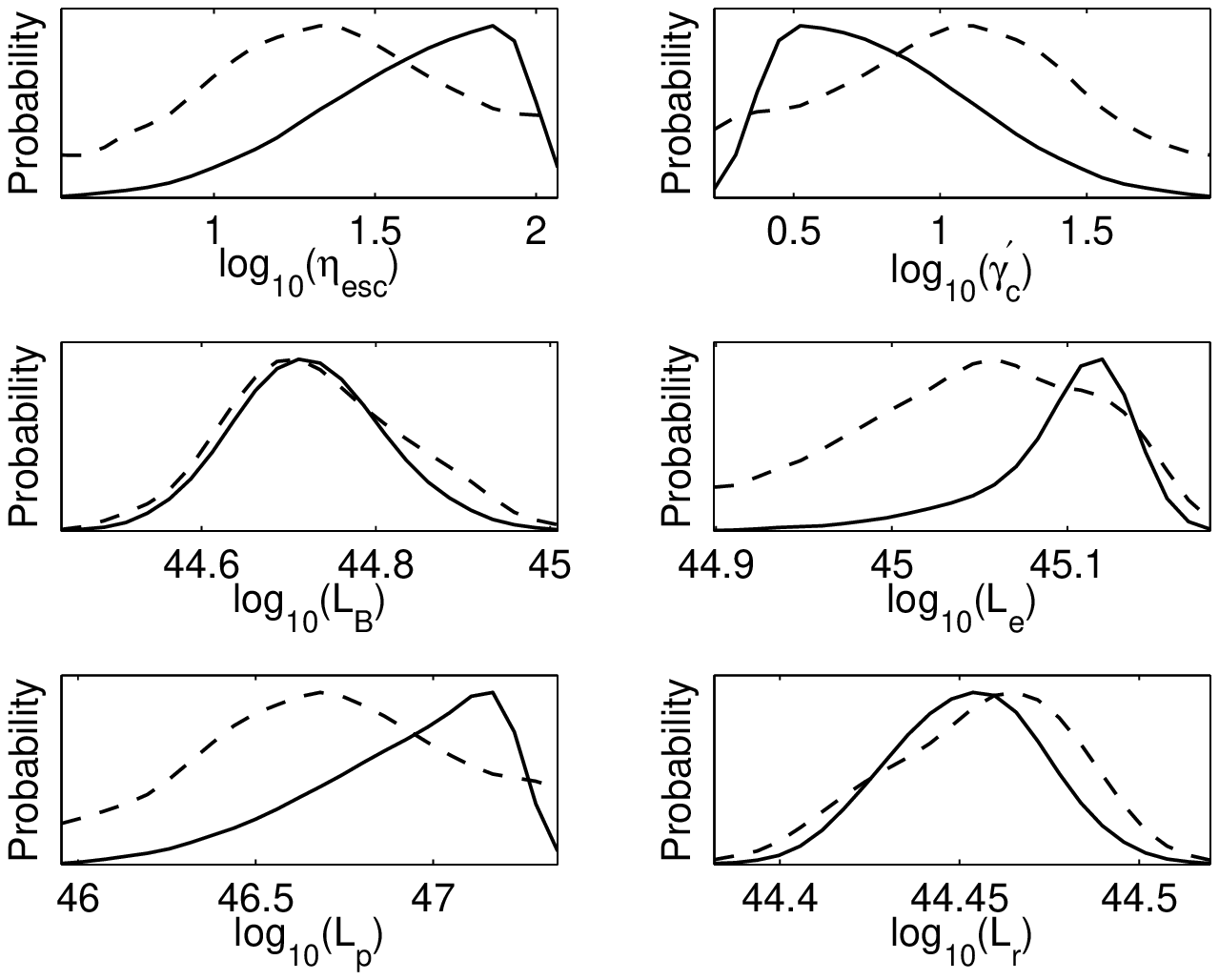}\includegraphics[width=0.4\textwidth]{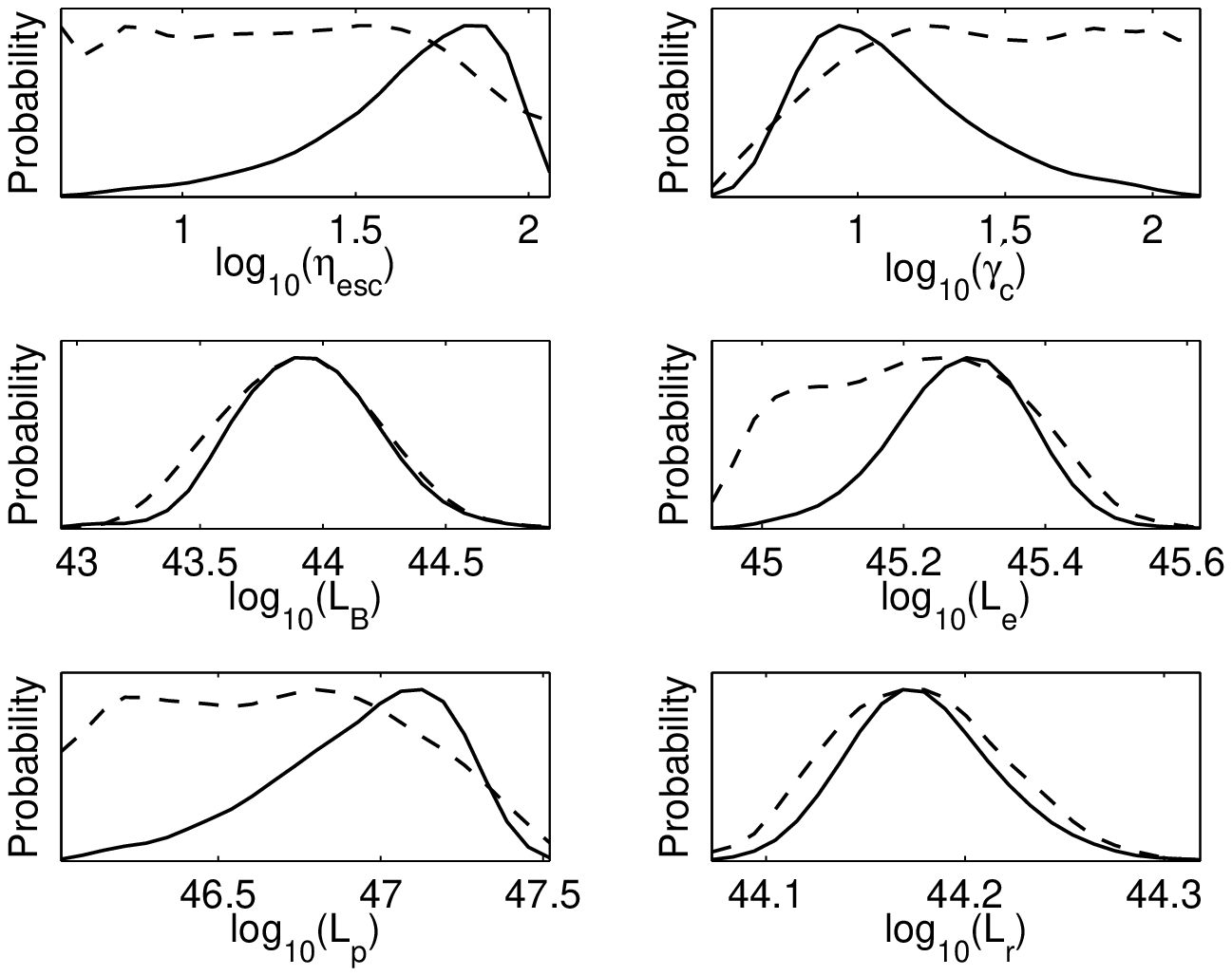}
\caption{1D probability distributions of the derived parameters in the EC-DT model.
%%The dashed lines show the maximum likelihood distributions and solid lines show the marginalized probability distributions.
In the left panel, the plots from top to bottom are arranged in the following order Periods A, B, C and D reported in Hayashida et al.(2012), and Periods E, F, G and H are shown in the right panel.
%%The plots are arranged as the following Figure \ref{figure1}
\label{figure14}
}
\end{figure*}

\begin{figure*}
\includegraphics[width=0.45\textwidth]{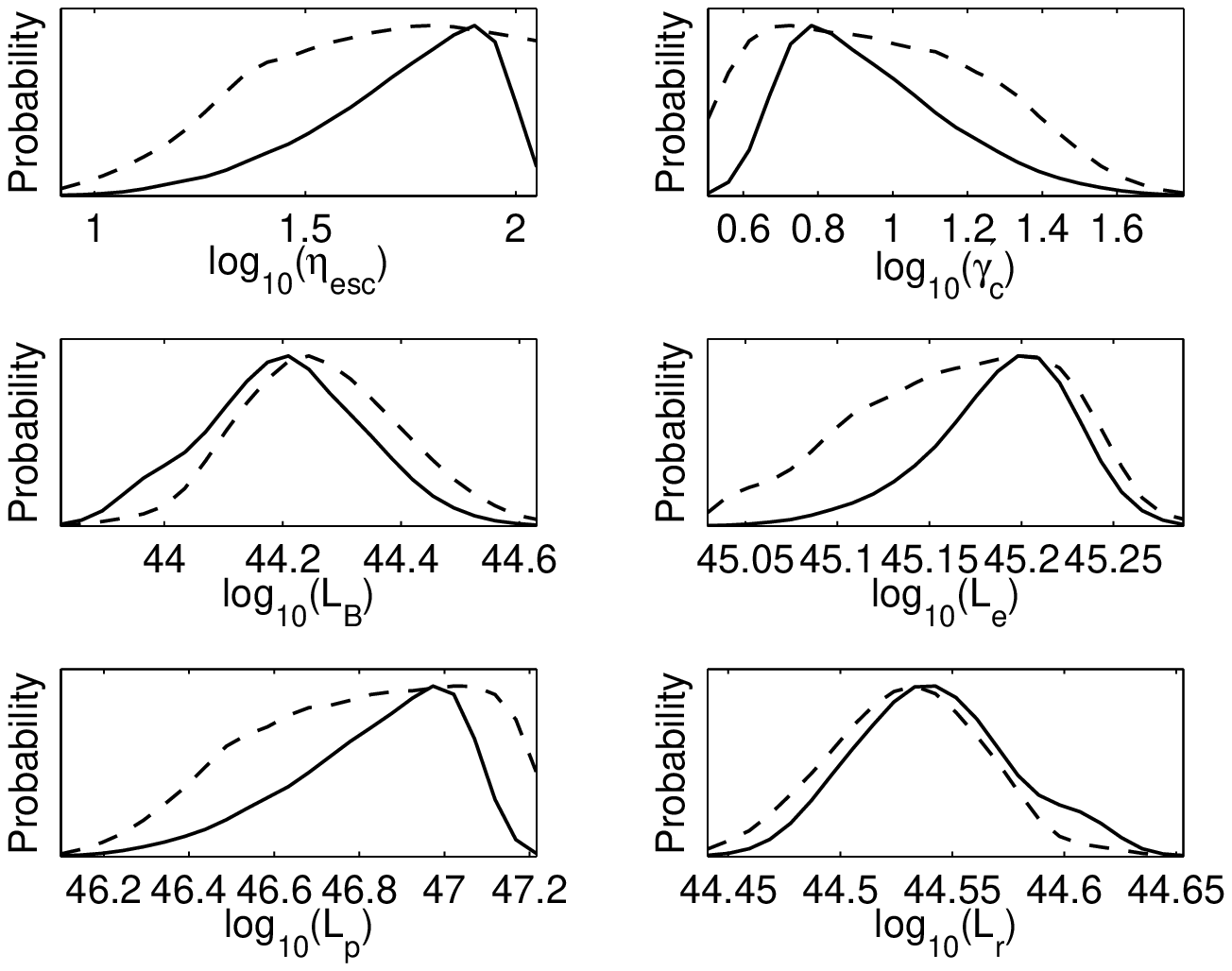}\includegraphics[width=0.45\textwidth]{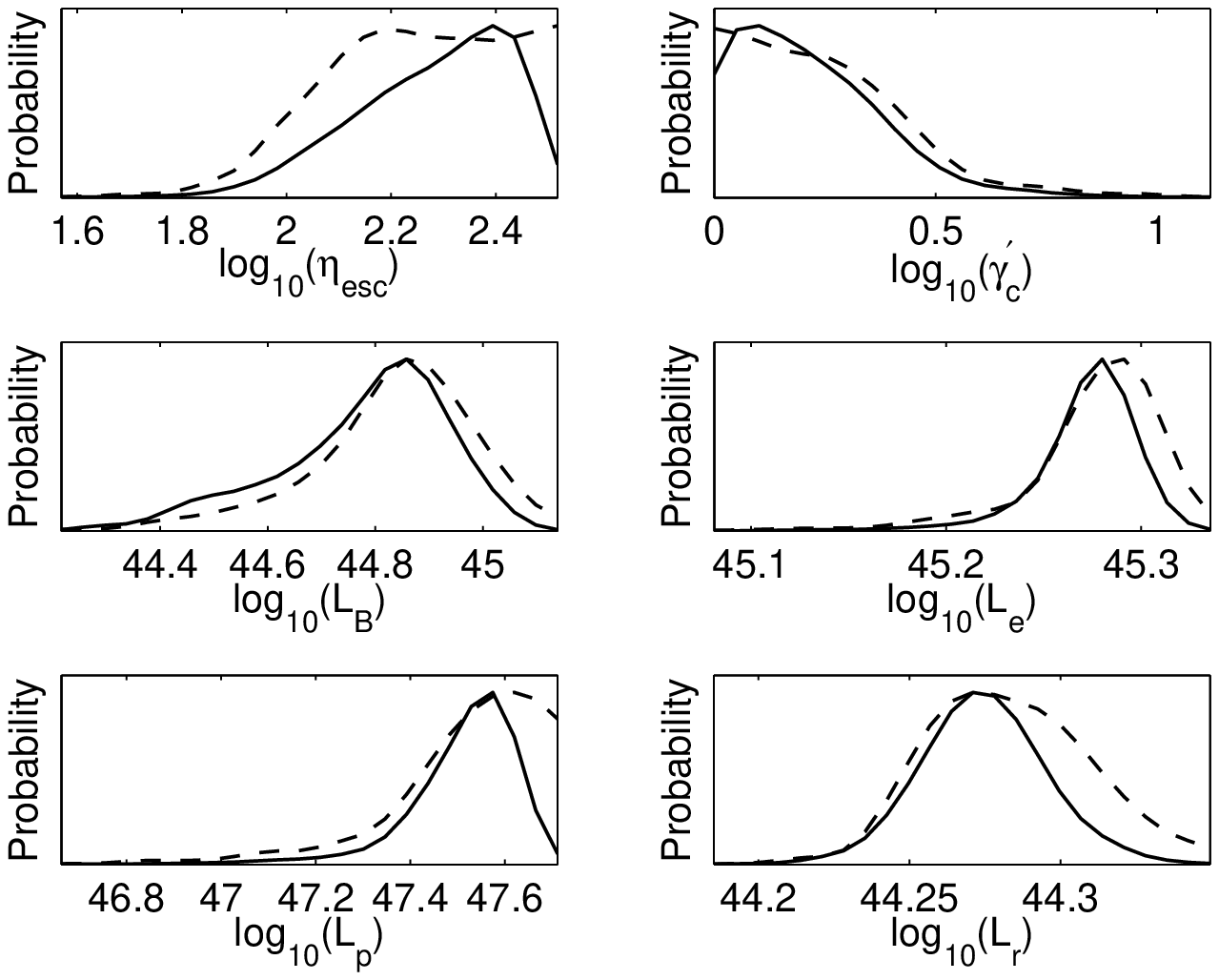}
\includegraphics[width=0.45\textwidth]{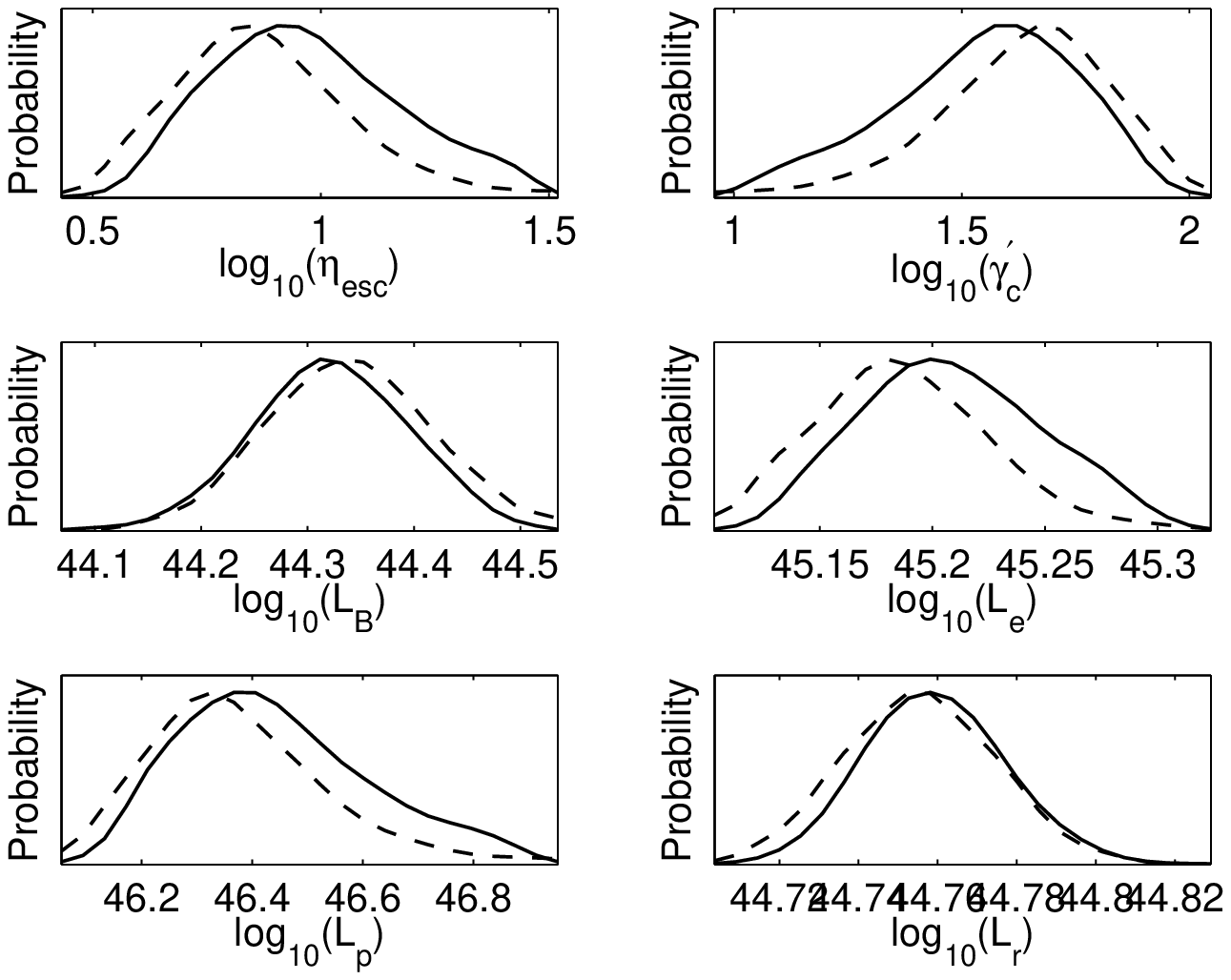}\includegraphics[width=0.45\textwidth]{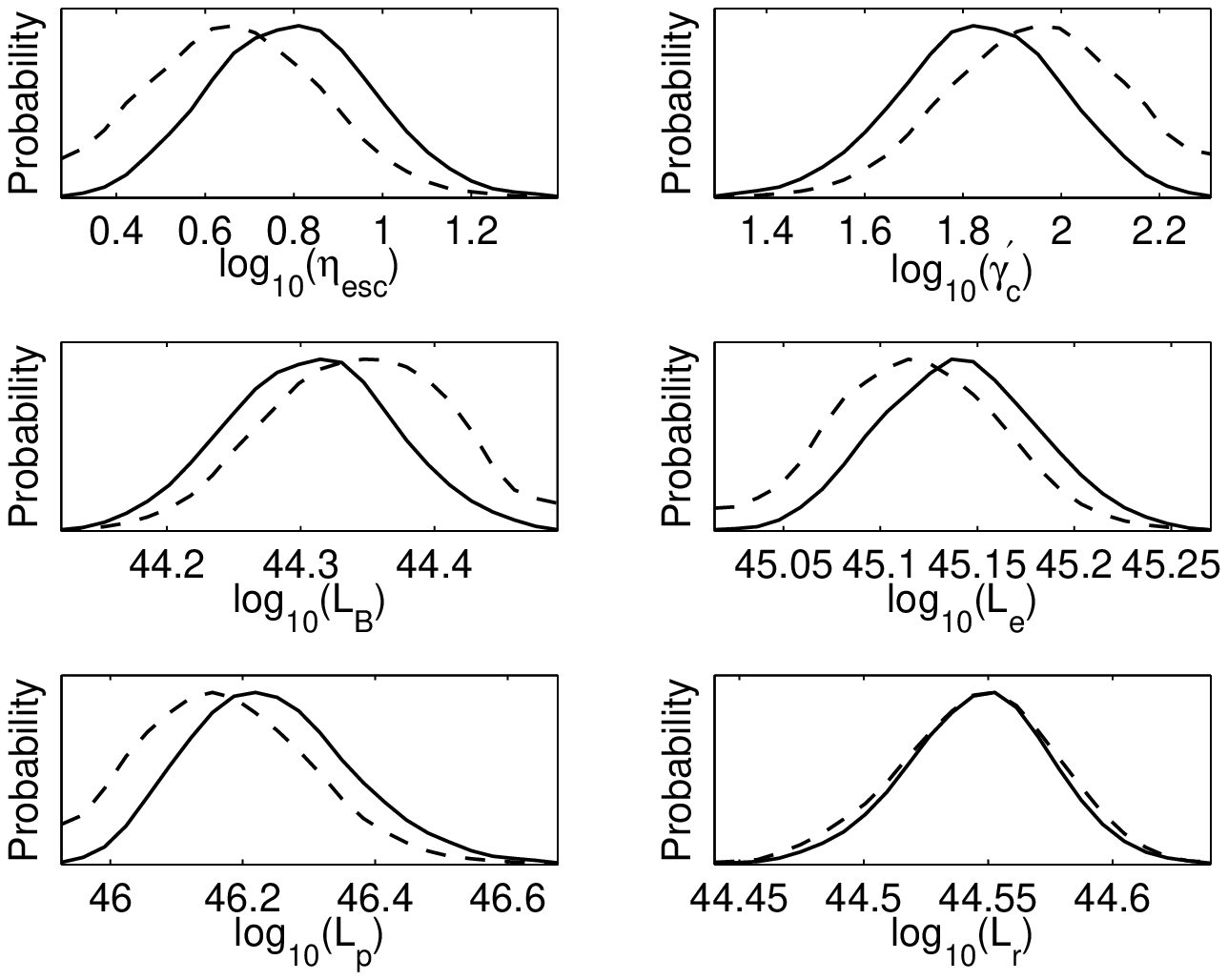}
\includegraphics[width=0.45\textwidth]{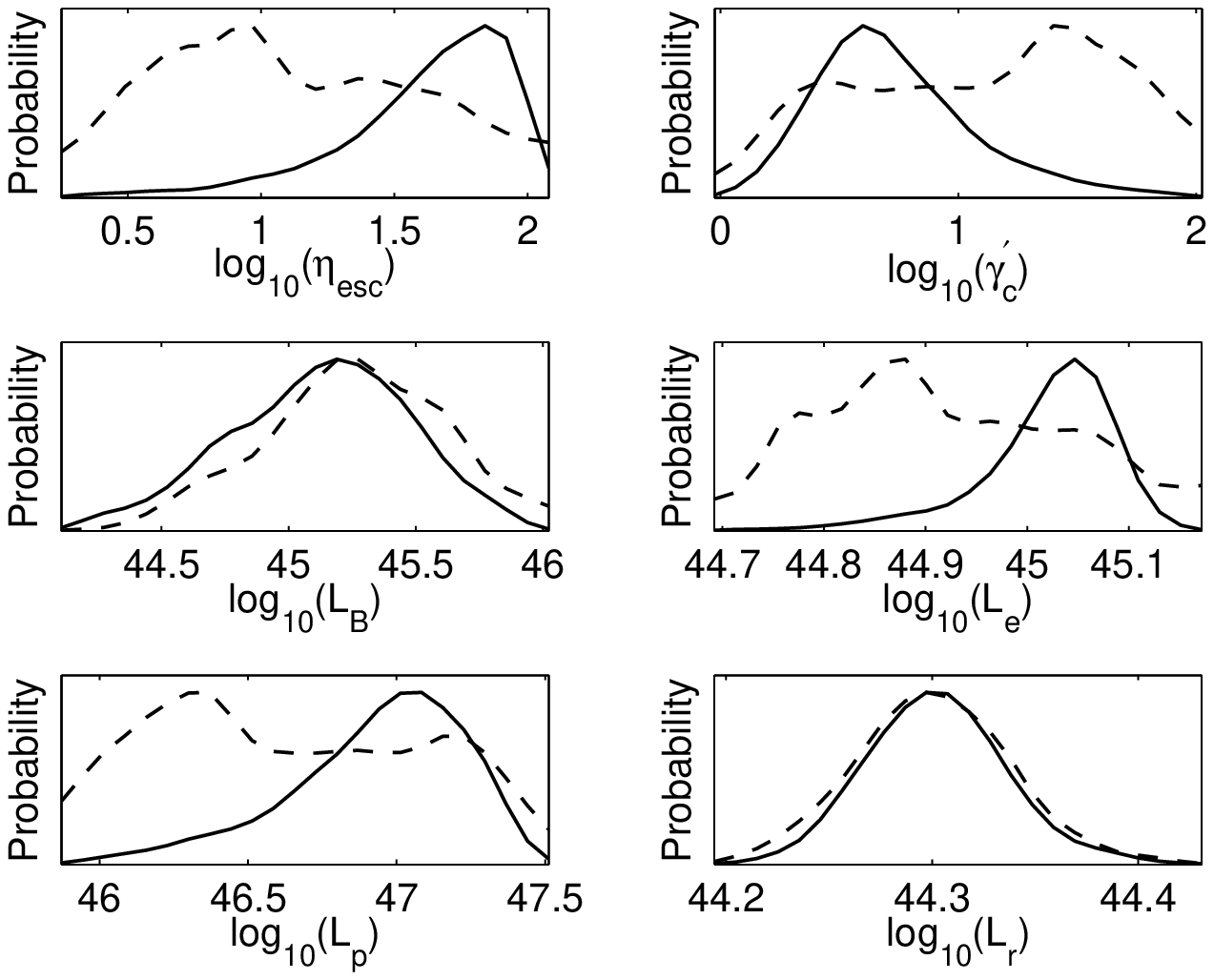}\includegraphics[width=0.45\textwidth]{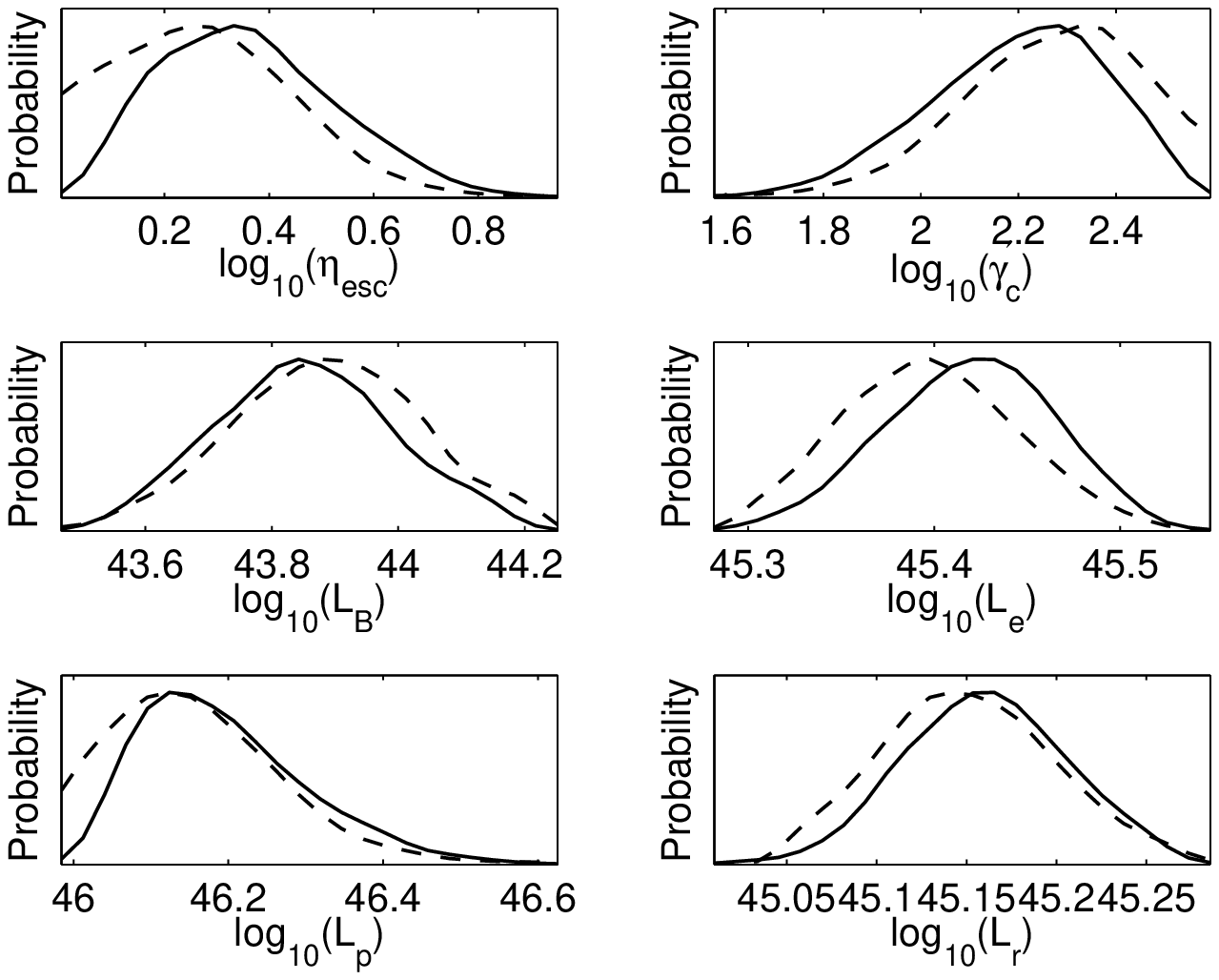}
\caption{Same as Figure~\ref{figure14}, but for the SEDs reported by Paliya et al.(2015) (Left) and the SEDs reported by Hayashida et al.(2015) (Right).
In the left panel, the plots from top to bottom are Flare1, Flare2 and Post-Flare, respectively.
In the right panel, the plots from top to bottom are Periods A, C and D, respectively. \label{figure15}}
\end{figure*}

%\bsp	% typesetting comment
%\label{lastpage}
\end{document}